hep-th/9805023

\def\Ad            {\mathrm{Ad}}
\newcommand\aef[3] {\alpha_{({#1},{#2},{#3})}}
\newcommand\aefx[4]{\alpha_{({#1},{#2},{#3})}^{(#4)}}

\newcommand\aefxb[4]{\overline{\alpha}_{({#1},{#2},{#3})}^{(#4)}}
\def\ala           {\alpha_\lambda}

\def\amu           {\alpha_\mu}

\newcommand\alf[1] {\alpha_{\Lambda_{({#1})}}}

\def\AIo           {{A(\Io)}}
\def\aro           {{\alpha_\rho}}
\newcommand\ASU[2] {\cA^{({#1}+{#2})}}
\newcommand\as[1]  {[\alpha_{{#1}}]}
\newcommand\asx[2] {[\alpha_{{#1}}^{({#2})}]}
\newcommand\asd[2] {[\alpha_{({#1},{#2})}]}
\newcommand\asdx[3]{[\alpha_{({#1},{#2})}^{({#3})}]}
\newcommand\asf[3] {[\alpha_{({#1},{#2},{#3})}]}
\newcommand\asfx[4]{[\alpha_{({#1},{#2},{#3})}^{(#4)}]}
\newcommand\asfb[3]{[\overline{\alpha}_{({#1},{#2},{#3})}]}
\newcommand\asfxb[4]{[\overline{\alpha}_{({#1},{#2},{#3})}^{(#4)}]}
\newcommand\asN[2] {[\alpha^{({#1})}_{#2}]}
\newcommand\asv[2] {[\alpha_{{#1},{#2}}]}

\def\bbC           {\mathbb{C}}

\def\bbN           {\mathbb{N}}

\def\bbZ           {\mathbb{Z}}
\def\be            {\begin{equation}}
\def\bearl         {\begin{array}{l}}
\def\bearll        {\begin{array}{ll}}
\def\bearlll       {\begin{array}{lll}}
\def\bearrl        {\begin{array}{rl}}
\def\bea           {\begin{eqnarray}}
\def\beaa          {\begin{eqnarray*}}
\def\bfe           {{\bf1}}

\newcommand\bproof {\noindent {\it Proof. }}
\def\can           {\gamma}
\def\cani          {\gamma^{-1}}
\def\canr          {\theta}

\def\cA            {\mathcal{A}}
\def\cC            {\mathcal{C}}

\def\cG            {\mathcal{G}}
\def\cH            {\mathcal{H}}

\def\cM            {\mathcal{M}}
\def\cN            {\mathcal{N}}
\def\chib          {\bar{\chi}}
\newcommand\chid[2]{\chi_{({#1},{#2})}}
\newcommand\chif[3]{\chi_{({#1},{#2},{#3})}}

\def\cM            {\mathcal{M}}
\newcommand\co[1]  {\overline{{#1}}}

\def\CA            {\cC_\cA}

\def\cS            {\mathcal{S}}
\def\cT            {\mathcal{T}}
\def\cV            {\mathcal{V}}
\def\cW            {\mathcal{W}}
\def\CA            {\cC_\cA}

\def\CM            {\cC_\cM}

\newcommand\dafx[4]{d_{({#1},{#2},{#3})}^{(#4)}}

\def\DelAI         {\Delta_\cA(I)}
\def\DelAIo        {\Delta_\cA(\Io)}

\def\DelNIo        {\Delta_\cN(\Io)}
\def\DelMIo        {\Delta_\cM(\Io)}
\def\DelMIoO       {\Delta_\cM^{(0)}(\Io)}

\def\DiffS         {\mathrm{Diff}(S^1)}
\def\DiffIS        {\mathrm{Diff}_I(S^1)}

\def\dim           {\mathrm{dim}}
\newcommand\del[2] {\delta_{{#1},{#2}}}

\def\E             {\mathrm{e}}
\def\ee            {\end{equation}}
\def\eear          {\end{array}}
\def\eea           {\end{eqnarray}}
\def\eeaa          {\end{eqnarray*}}
\def\Eig           {\mathrm{Eig}}
\def\End           {\mathrm{End}}

\newcommand\eproof {\hspace*{\fill}\nolinebreak\hspace*{\fill}
                   {\sl Q.E.D.}
                   \par\vspace{3mm}}
\newcommand\eps[2] {\varepsilon({#1},{#2})}

\newcommand\epsp[2]{\varepsilon^+({#1},{#2})}
\newcommand\epspm[2]{\varepsilon^\pm({#1},{#2})}
\newcommand\epso[2]{\epsilon({#1},{#2})}
\newcommand\epsom[2]{\epsilon^-({#1},{#2})}

\newcommand\epsopm[2]{\epsilon^\pm({#1},{#2})}

\newcommand\erf[1] {Eq.\ (\ref{#1})}
\def\Exp           {\mathrm{Exp}}

\def\fB            {\mathfrak{B}}

\def\for           {\quad\mbox{for}\,\,\,\,}
\def\fUI           {f_{U_I}}
\def\fUIt          {f_{\tilde{U}_I}}

\def\Gtwo          {\mathrm{G}_2}

\def\Hom           {\mathrm{Hom}}
\def\I             {{\rm i}}
\def\Ic            {I^\mathrm{c}}

\def\Bild          {\mathrm{Im}\,}

\def\id            {\mathrm{id}}
\def\Io            {I_\circ}
\def\Jz            {\mathcal{J}_z}
\newcommand\K[2]   {K^{({#1})}_{#2}}
\newcommand\Kb[2]  {\bar{K}^{({#1})}_{#2}}

\def\la            {\langle}
\newcommand\lad[2] {\lambda_{({#1},{#2})}}
\newcommand\labl[1]{\label{#1}\ee}
\newcommand\lablsec[1]{\label{#1}}
\newcommand\lablth[1]{\label{#1}}

\newcommand\lasd[2]{[\lambda_{({#1},{#2})}]}
\newcommand\lasf[3]{[\lambda_{({#1},{#2},{#3})}]}
\newcommand\lasN[2]{[\lambda^{({#1})}_{#2}]}
\newcommand\laN[2] {\lambda^{({#1})}_{#2}}
\newcommand\lasv[2]{[\lambda_{{#1},{#2}}]}
\newcommand\lav[2] {\lambda_{{#1},{#2}}}
\def\LG            {\mathit{LG}}
\def\LH            {\mathit{LH}}
\def\LIG           {\mathit{L}_I\mathit{G}}
\def\LIcG          {\mathit{L}_{\Ic}\mathit{G}}
\def\LIH           {\mathit{L}_I\mathit{H}}

\def\LISUn         {\mathit{L}_I\mathit{SU}(n)}

\newcommand\ls[1]  {[\lambda_{{#1}}]}

\def\LSUd          {\mathit{LSU}(3)}
\def\LSUn          {\mathit{LSU}(n)}
\def\LSUz          {\mathit{LSU}(2)}

\def\LTSM          {[\Delta]_\cM(\Io)}

\def\LTSN          {[\Delta]_\cN(\Io)}
\def\min           {\mathrm{min}}
\def\MIo           {{M(\Io)}}

\newcommand\mm[3]  {Z_{{#1},{#2}}^{({#3})}}
\def\mod           {\mathrm{mod}}

\newcommand\N[3]   {N_{{#1},{#2}}^{{#3}}}
\def\NIo           {{N(\Io)}}

\def\Nres          {\tilde{N}}

\def\pio           {\pi_0}
\def\pioi          {\pi_0^{-1}}
\newcommand\psiv[2]{\psi^{#1}_{#2}}
\newcommand\psivs[2]{(\psi^{#1}_{#2})^*}
\def\PSLZ          {\mathit{PSL}(2;\bbZ)}

\def\ra            {\rangle}

\def\rmc           {\mathrm{c}}
\def\rms           {\mathrm{s}}
\def\rmv           {\mathrm{v}}
\def\RotS          {\mathrm{Rot}(S^1)}
\def\Sect          {\mathrm{Sect}}
\newcommand\Sex[2] {S^\mathrm{ext}_{{#1},{#2}}}
\newcommand\Sexm[2]{\cS^\mathrm{ext}_{{#1},{#2}}}
\def\sib           {{\sigma_\beta}}

\def\sio           {\sigma_0}

\def\sioh          {\hat{\sigma}_0}
\def\sioI          {\sigma_{0;I}}
\def\SLZ           {\mathit{SL}(2,\bbZ)}
\newcommand\Sm[2]  {S_{{#1},{#2}}}
\def\SOf           {\mathit{SO}(5)}

\def\SUd           {\mathit{SU}(3)}
\def\SUn           {\mathit{SU}(n)}
\def\SUz           {\mathit{SU}(2)}
\def\SUf           {\mathit{SU}(4)}

\def\suzh          {\widehat{\mathfrak{su}}(2)}
\def\tr            {\mathrm{tr}\,}

\def\Usiom         {U_{\sio,-}}
\def\Usiop         {U_{\sio,+}}

\newcommand\V[3]   {V_{{#1};{#2}}^{#3}}
\def\Vir           {\mathfrak{Vir}}
\newcommand\Z[3]   {Z^{({#1})}({#2},{#3})}

\def\ADEs          {$\mathcal{ADE}_7$}
\def\per           {positive energy representation}
\newcommand\ddA[1] {$\mathrm{A}_{#1}$}
\newcommand\ddD[1] {$\mathrm{D}_{#1}$}
\newcommand\ddE[1] {$\mathrm{E}_{#1}$}
\def\ddAg          {$\mathcal{A}$}
\def\ddDg          {$\mathcal{D}$}
\def\ddEg          {$\mathcal{E}$}
\def\ddEgms        {$\mathcal{E}^{(12)}_\mathrm{MS}$}
\newcommand\ddAgx[1]{$\mathcal{A}^{({#1})}$}
\newcommand\ddDgx[1]{$\mathcal{D}^{({#1})}$}
\newcommand\ddEgx[1]{$\mathcal{E}_{{#1}}$}
\def\Deven         {$\mathrm{D}_{\mathrm{even}}$}
\def\Dodd          {$\mathrm{D}_{\mathrm{odd}}$}

\documentclass[11pt]{article}
\usepackage{amssymb,amsfonts,epic,eepic}
\begin{document}


\title{Modular Invariants, Graphs and $\alpha$-Induction for
Nets of Subfactors II}
\author{{\sc Jens B\"ockenhauer} and {\sc David E. Evans}\\ \\
University of Wales Swansea\\
Department of Mathematics\\Singleton Park\\Swansea SA2 8PP}
\maketitle
\nopagebreak

\begin{abstract}
We apply the theory of $\alpha$-induction of sectors which we elaborated
in our previous paper to several nets of subfactors arising from conformal
field theory. The main application are conformal embeddings and orbifold
inclusions of $\SUn$ WZW models. For the latter, we construct the extended
net of factors by hand. Developing further some ideas of F.\ Xu, our
treatment leads canonically to certain fusion graphs, and in all our
examples we rediscover the graphs Di Francesco, Petkova and Zuber
associated empirically to the corresponding $\SUn$ modular invariants.
We establish a connection between exponents of these graphs and the
appearance of characters in the block-diagonal modular invariants,
provided that the extended modular S-matrices diagonalize the endomorphism
fusion rules of the extended theories. This is proven for many cases,
and our results cover all the block-diagonal $\SUz$ modular invariants,
thus provide some explanation of the A-D-E classification.
\end{abstract}

\tableofcontents


\newtheorem{definition}{Definition}[section]
\newtheorem{lemma}[definition]{Lemma}
\newtheorem{corollary}[definition]{Corollary}
\newtheorem{theorem}[definition]{Theorem}
\newtheorem{proposition}[definition]{Proposition}


\section{Introduction}

\subsection{Background}

The $\SUn_q$ subfactors of Wenzl \cite{wenz} can be understood from
the viewpoint of statistical mechanics \cite{evka2},
the IRF models of \cite{djmo} or from the viewpoint of conformal
field theory, irreducible highest weight \per s
of the loop groups of $\SUn$ \cite{wass3}.
These viewpoints are also related to the study and classification of
modular partition functions on a torus. The statistical mechanical
models of \cite{djmo} are generalizations of the Ising model. The
configuration space of the Ising model, distributions of symbols
$+,-$ on the vertices of the square lattice $\bbZ^2$,
can also be thought of as distributions of edges of the Dynkin
diagram \ddA 3 on the edges of a square lattice, where the
end vertices are labelled by $+$ and $-$. This model can be
generalized by replacing \ddA 3 by other graphs $\Gamma$ such as Dynkin
diagrams or indeed the Weyl alcove \ddAgx m \ of the level $k$
integrable representations of $\SUn$, where $m=k+n$ is the
altitude. The vertices of \ddAgx {n+k} are given by weights
$\{\Lambda=\sum_{i=1}^{n-1} m_i \Lambda_{(i)} \,:\, m_i\in\bbN_0
\,,\,\, \sum_{i=1}^{n-1} m_i\le k \}$
where the $\Lambda_{(i)}$ are the $n-1$ weights of the fundamental
representation, and the oriented edges are given by the
vectors $e_i$ defined by $e_1=\Lambda_{(1)}$,
$e_i=\Lambda_{(i)}-\Lambda_{(i-1)}$, $i=1,2,...,n-1$,
$e_n=\Lambda_{(n-1)}$. We can also label our states by
partitions or Young tableaux  $(p_j)_{j=1}^{n-1}$,
$k \ge p_1 \ge p_2 \ge \cdots \ge p_{n-1} \ge p_n \equiv 0$
obtained by the transformation
$(m_i)_{i=1}^{n-1}\mapsto (p_j)_{j=1}^{n-1}$, where
$p_j = \sum_{i=j}^{n-1} m_i$.
The unoccupied state corresponds to $(0,0,\ldots,0)$ or the
empty Young tab\-leau in the two descriptions,
which we often denote by $\ast$ or $0$.

A configuration is then a distribution of the edges of $\Gamma$
over $\bbZ^2$, and associated to each local configuration is a
Boltzmann weight (see Figure \ref{figBW})
\setlength{\unitlength}{1pt}
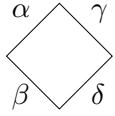
\begin{figure}[tb]
\begin{center}
\begin{picture}(50,45)
\path(0,20)(20,40)(40,20)(20,0)(0,20)
\put(2,2){$\beta$}
\put(2,35){$\alpha$}
\put(32,35){$\gamma$}
\put(32,2){$\delta$}
\end{picture}
\end{center}
\caption{Boltzmann weight where $\alpha,\beta,\gamma,\delta$
are edges of $\Gamma$}
\label{figBW}
\end{figure}
satisfying the Yang-Baxter equation of Figure \ref{figYBE}.
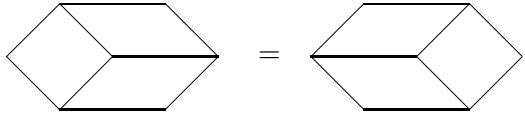
\begin{figure}[tb]
\begin{center}
\begin{picture}(215,60)
\path(10,30)(30,50)(50,30)(30,10)(10,30)
\put(30,50){\line(1,0){40}}
\put(30,10){\line(1,0){40}}
\put(50,30){\line(1,0){40}}
\put(70,50){\line(1,-1){20}}
\put(70,10){\line(1,1){20}}
\put(105,28){$=$}
\multiput(125,30)(40,0){2}{\line(1,1){20}}
\multiput(125,30)(40,0){2}{\line(1,-1){20}}
\put(125,30){\line(1,0){40}}
\multiput(145,10)(0,40){2}{\line(1,0){40}}
\put(185,50){\line(1,-1){20}}
\put(185,10){\line(1,1){20}}
\end{picture}
\end{center}
\caption{Yang-Baxter equation}
\label{figYBE}
\end{figure}
The justification of the term $\SUn$ models is as follows. By
Weyl duality, the representation of the permutation group on
$\bigotimes M_n$ is the fixed point algebra of the product action
of $\SUn$. Deforming this, there is a representation of the
Hecke algebra in $\bigotimes M_n$ whose commutant is a representation
of a deformation of $\SUn$, the quantum group $\SUn_q$. The
Boltzmann weights lie in this Hecke algebra representation,
and at criticality reduce to the natural braid generators $g_i$, so
that the Yang-Baxter equation of Figure \ref{figYBE}
reduces to the braid
relation $g_ig_{i+1}g_i=g_{i+1}g_ig_{i+1}$. The labels of
the irreducible representations of either the Hecke algebra
(e.g.\ the permutation group when $q=1$) or the quantum group
(e.g.\ $\SUn$ when $q=1$) are generically given by \ddAg,
a Young tableaux of at most $n-1$ rows. However when $q$
is a root of unity $\E^{2\pi \I/m}$ we have the further
constraint of at most $k=m-n$ columns, where
$k$ is the level, i.e.\ \ddAgx m \ (e.g.\ when $n=2$
the vertices of the Dynkin diagram \ddA {k+1}).

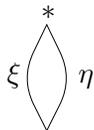
\begin{figure}[tb]
\begin{center}
\begin{picture}(40,60)
\spline(20,10)(10,30)(20,50)
\spline(20,10)(30,30)(20,50)
\put(5,28){$\xi$}
\put(32,28){$\eta$}
\put(18,52){$*$}
\end{picture}
\end{center}
\caption{Matrices of partition functions $[T_{\xi\eta}]$ where
$\xi,\eta$ are paths on $\Gamma$ with fixed initial
vertex $*$ and same terminal vertex generate a von Neumann
algebra ${\cal N}$}
\label{figPF}
\end{figure}

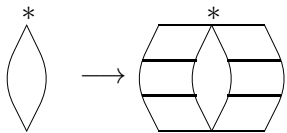
\begin{figure}[tb]
\begin{center}
\begin{picture}(130,60)
\spline(20,10)(10,30)(20,50)
\spline(20,10)(30,30)(20,50)
\put(18,52){$*$}
\put(40,28){$\longrightarrow$}
\spline(90,10)(80,30)(90,50)
\spline(90,10)(100,30)(90,50)
\spline(70,10)(60,30)(70,50)
\spline(110,10)(120,30)(110,50)
\put(70,10){\line(1,0){40}}
\put(70,50){\line(1,0){40}}
\put(70,10){\line(1,0){40}}
\put(70,50){\line(1,0){40}}
\put(64,23.33){\line(1,0){20}}
\put(96,23.33){\line(1,0){20}}
\put(64,36.66){\line(1,0){20}}
\put(96,36.66){\line(1,0){20}}
\put(88,52){$*$}
\end{picture}
\end{center}
\caption{Embedding of ${\cal N}\subset{\cal M}$ by $T\to{\rm
Ad}(V)(T)$ where $V=g_1g_2\cdots$ is the product of
Boltzmann weights at criticality}
\label{figNM}
\end{figure}

The Boltzmann weights involve paths of length two in the
Bratteli diagram using the embedding graph $\Gamma$. As we
look at larger and larger partition functions (based on some
fixed initial vertex $\ast$) then we can complete with
respect to a natural trace and obtain a von Neumann algebra
as in Figure \ref{figPF}.
A subfactor $N\subset M$
can be obtained with the aid of the initial Boltzmann
weights placed on the boundary as in Figure \ref{figNM}.
For the $\SUn_q$ subfactors, this just amounts to
$\{ g_i\,:\,\,i=1,2,3,...\}''\subset\{ g_i\,:\,\,i=0,1,2,...\}''$
because of the braid relations
$\Ad(g_1g_2\cdots)(g_i)=g_{i+1}$. The center $\bbZ_n$ of
$\SUn$ acts on \ddAgx m leaving the Boltzmann weights
invariant and hence induces an action on $M$, leaving $N$
globally invariant, yielding the orbifold subfactor
$N^{\bbZ_n}\subset M^{\bbZ_n}$. The action of the center
$\bbZ_n$ (corresponding to the simple currents)
on \ddAgx m is as follows. We set $A_0=\ast$,
and label the other end vertices of \ddAgx m by
$A_1=A_0+(m-n)e_1$, $A_2=A_1+(m-n)e_2$, ... ,
$A_{n-1}=A_{n-2}+(m-n)e_{n-1}$. Define a rotation
symmetry of the graph \ddAgx m by
$\sigma(A_j+\sum c_r e_r)=A_{j+1}+\sum c_r e_{r+1}$
where the indices are in $\bbZ_n$.

Let us now turn to the loop group picture. The loop group
$\LG$ is the group of smooth maps from $S^1$ into a compact
Lie group $G$ under pointwise multiplication.
We are interested in projective representations of
$\LG\rtimes\RotS$, where $\RotS$ is the rotation group, which
are highest weight representations in that the generator $L_0$
of the rotation group is bounded below. Such representations are
called \per s and are classified by irreducible representations
of $G$ and a level $k$. For unitary irreducible \per s,
the possibilities are severely restricted.
Indeed $k$ must be integral and, for a given value of the
level, there are only a finite number of admissible (vacuum vector)
irreducible representations of $G$. For the case of $G=\SUn$,
the admissible ones at level $k$ are the vertices
of \ddAgx m, where $m=n+k$. Restricting to loops
$\LIG$ concentrated on an interval $I\subset S^1$,
$\LIG=\{f\in \LG \,:\, f(z)=e\,,\,\,\,z\notin I\}$,
we get for each positive energy representation $\pi$
a subfactor $\pi(\LIG)''\subset\pi(\LIcG)''$ if
$\Ic$ is the complementary interval, of type III
and of finite index --- e.g.\ of index
$4\cos^2(\pi/(k+2))$ in the case of the
fundamental representation of $\SUz$ and level $k$.

We next turn to the modular invariant picture.
It is argued on physical grounds that the partition
function $Z(\tau)$ in a conformal field theory should
be invariant under re-parameterization of the torus by
$\SLZ$. In the string theory formulation, modular
invariance is essentially built into the definition
of the partition function (although Nahm \cite{nahm}
has argued the case for modular invariance in terms
of the chiral algebra and its representations rather
than a functional integral setting). In the transfer
matrix picture of statistical mechanics we can write
the partition function as an average over $\E^{-\beta H}$,
where $H$ is the Hamiltonian, now $L_0+\bar{L}_0-c/12$
(the shift by $c/24$ arising from mapping the Virasoro
algebra on the plane to a cylinder). We have a momentum
$P$ ($=L_0-\bar{L}_0$) describing evolution along the
closed string, so taking both evolutions into account,
we first compute
\be
Z(\tau) = \tr(\E^{-\beta H}\E^{\I\eta P}) =
\tr(\E^{2\pi \I\tau(L_0-c/24)}
\E^{-2\pi \I\bar{\tau}(\bar{L}_0-c/24)}) \,.
\labl{parf1}
Here $2\pi\I\tau=-\beta+\I\eta$ parameterizes the metric
of the torus, and we then have to average over
$\tau$. If we choose one $\tau$ from each orbit under the
action of $\PSLZ$ and integrate we implicitly assume that
$Z(\tau)$ is modular invariant.

>From a Hilbert space decomposition of the loop group
representation the partition function \erf{parf1}
decomposes as
\be
Z = \sum_{\Lambda,\Lambda'} Z_{\Lambda,\Lambda'} \,
\chi_\Lambda \, \bar{\chi}_{\Lambda'}
\labl{parf2}
where $\chi_\Lambda$ is the conformal character $\tr(q^{L_0-c/24})$,
$q=\E^{2\pi\I\tau}$ of the unitary positive energy irreducible
representation $\pi_\Lambda$, $\Lambda\in$\ddAgx m,
where $m=n+k$ for some fixed level $k$.

The problem then is to find or classify all expressions
of the form \erf{parf2} where $Z$ is $\SLZ$ invariant,
subject to the normalization $Z_{0,0}=1$ and
$Z_{\Lambda,\Lambda'}$ is a non-negative integer.
A simple argument of Gannon \cite{gann1} shows that
$\sum_{\Lambda,\Lambda'}Z_{\Lambda,\Lambda'}\le 1/S_{0,0}^2$
where $S_{0,0}$ is a matrix entry of the S-matrix action
of $\SLZ$ on characters, and hence for a given $G$ at a
fixed level, there are only finitely many possible modular
invariants. They have been completely classified in the
case $\SUz$ by \cite{caiz} and in the case $\SUd$ by \cite{gann2},
and the program of Gannon to the complete classification
is far advanced --- see e.g.\ the notes of Chapter 8
of \cite{evka} for a review. The Gannon program involves
identifying first a special class of modular invariants,
the \ADEs\ invariants which satisfy
$Z_{0,\Lambda}\neq 0$ implies that $\Lambda$ is (the weight
labelling) a simple current or equivalently $Z_{0,\Lambda}\neq 0$
implies $S_{\Lambda,0}=S_{0,0}$, and then identify
what appear to be very few remaining exceptions which include
those arising from conformal embeddings. The \ADEs\ invariants
include all the automorphism invariants,
for which $Z_{0,\Lambda}=\del 0\Lambda$. Such an invariant
is basically an automorphism of the fusion ring.
There is a permutation $\sigma$ of \ddAgx m\ such that
$Z_{\Lambda,\Lambda'}=\del \Lambda{\sigma(\Lambda')}$.
The \ADEs\ invariants also include the simple current
invariants for which $Z_{\Lambda,\Lambda'}\neq 0$ implies
$\Lambda=J\cdot\Lambda'$ for a simple current $J$.
Automorphism and simple current invariants constitute
the $\mathcal{A}$ and $\mathcal{D}$ type modular invariants.
Note that there are two kinds of modular invariants:
\[ \bearll
\sum |\chi_i|^2 \,, \qquad & \mbox{type I} \\[.4em]
\sum \chi_i \chib_{\sigma(i)}\,, \qquad & \mbox{type II}
\eear \]
where $\chi_i$ are (possibly extended) characters and
$\sigma$ is a permutation of the (extended) fusion rules.
The type II invariants where the characters are properly
extended (i.e.\ at least one $\chi_i$ is a proper sum over two
or more $\chi_\Lambda$'s) finally constitute the $\mathcal{E}_7$
modular invariants. The type I modular invariants where
the characters $\chi_i$ are proper extensions
are also called block-diagonal.
In fact any $\SUn$ block-diagonal
modular invariant can be interpreted as a
completely diagonal invariant of a larger theory
embedding the $\SUn$ level $k$ WZW theory.
In the case of a conformal inclusion the larger theory is given
in terms of a $G$ (necessarily level 1) WZW theory with $G$
a simple Lie group, and in the orbifold inclusion case the larger
theory is given in terms of a simple current extension of the
$\SUn$ theory, and the $\SUn$ theory itself  can the be thought of
as the $\bbZ_n$ orbifold of the extended object. For $\SUz$ and
$\SUd$ both cases actually exhaust all the block-diagonal modular
invariants.\footnote{However, for $\SUn$ with larger $n$ there
appear also block-diagonal modular invariants which arise from
level-rank duality but neither from conformal nor orbifold inclusions,
e.g.\ for $\mathit{SU}(10)$ at level 2. This kind of invariants
will not be treated in this paper.}

The modular invariants appear to be labelled naturally,
in the case of $\SUz$ and $\SUd$, by graphs. The $\SUz$
modular are labelled by A-D-E Dynkin diagrams in the sense that
the non-vanishing diagonal entries of the
modular invariant are given by the conformal characters labelled
by the Coxeter exponents of the labelling ADE graph. Recall the
eigenvalues of the (adjacency matrix of the) D and E graphs
constitute subsets of the vertices of the A graph with
the same Coxeter number,
and their labels are called Coxeter exponents.  For example in
case of $\SUz$ at level $16$ there are three modular
invariants. In each case the diagonal part of the invariant
is described by a certain subset $I=\{j\}$ of the
vertices of \ddA {17}. The (adjacency matrix of the) graph
of \ddA {17} has eigenvalues $\{2\cos((j+1)\pi/h)\}$ where
$j=0,1,2,...,16$ labels the vertices of \ddA {17} and $h=18$ is the
Coxeter number of \ddA {17}. Then $I$ is the set of the Coxeter
exponents, i.e.\ the set
$\{2\cos((j+1)\pi/h),\,\,j\in I\}$ gives all the eigenvalues
of the Dynkin diagram, \ddA {17}, \ddD {10} or \ddE 7.
The completely diagonal invariant then
corresponds to the graph \ddA {17} itself. In this
way all $\SUz$ modular invariants are described by
A-D-E graphs.

In the subfactor theory only A-D-E Dynkin diagrams with
\ddA {\ell+1}, \ddD {2\ell+2} ($\ell=1,2,...$), \ddE 6 \ and \ddE 8,
appear as the (dual) principal graphs (or fusion graphs) of
subfactors with index less than four. In the
rational conformal field theory of $\SUz$ models described
by A-D-E Dynkin diagrams, one may argue that there is a degeneracy
so that only \Deven, \ddE 6 and \ddE 8, namely the type I cases
need be counted. For example in the case of $k=16$,
the modular invariant for \ddE 7
reduces to that of \ddD {10} under the simple interchange of
blocks $\chi_8$ and $\chi_2+\chi_{14}$.

A classification of $\SUd$ modular invariants was completed
by \cite{gann2}. In analogy with the A-D-E classification for $\SUz$,
we label these \ddAg\ (the completely diagonal invariants), \ddDg\
(the simple current invariants) and the exceptional \ddEg\ invariants.
We should also throw in their conjugates $Z^\rmc$,
$(Z^\rmc)_{\Lambda,\Lambda'}=Z_{\bar{\Lambda},\Lambda'}$
(here $\bar{\Lambda}$ labels the conjugate representation)
--- although
$\mathcal{D}^{(6)}=\mathcal{D}^{(6)\rmc}$,
$\mathcal{D}^{(9)}=\mathcal{D}^{(9)\rmc}$,
$\mathcal{E}^{(12)}=\mathcal{E}^{(12)\rmc}$,
$\mathcal{E}^{(24)}=\mathcal{E}^{(24)\rmc}$.
For $\SUz$, the automorphism invariants are the A-series
and the \Dodd -series. For $\SUd$
they are \ddAgx m and $\mathcal{A}^{(m)\rmc}$ for all
$m$ and $\mathcal{D}^{(m)}$ and $\mathcal{D}^{(m)\rmc}$
for $m\neq 0$ mod $3$. The \ADEs\ invariants for $\SUz$ are
the A-series, D-series, and the \ddE 7 exceptional (hence
the name \ADEs). In the $\SUd$ case the \ddAg\ invariants
are \ddAgx m and $\mathcal{A}^{(m)\rmc}$, the \ddDg\
invariants are $\mathcal{D}^{(m)}$ and $\mathcal{D}^{(m)\rmc}$,
and the \ddEgx 7 invariants are the two Moore and Seiberg
invariants \ddEgms and $\mathcal{E}^{(12)\rmc}_\mathrm{MS}$.
The other invariants $\mathcal{E}^{(8)}$, $\mathcal{E}^{(12)}$,
$\mathcal{E}^{(24)}$ correspond to conformal embeddings
$\SUd_5\subset\mathit{SU}(6)_1$,
$\SUd_9\subset(\mathrm{E}_6)_1$,
$\SUd_{21}\subset(\mathrm{E}_7)_1$,
respectively (cf.\ \ddE 6 and \ddE 8 for $\SUz$).
The simple current invariants for $\SUz$ are the A and D
series, and for $\SUd$ are again the \ddAg\ and \ddDg\
series (but not their conjugations).

Di Francesco and Zuber initiated a program
to associate graphs to these invariants \cite{frzu1,frzu2}.
These graphs are three colourable and such that their
eigenvalues (``exponents''), constituting again a subset
of the set of eigenvalues of the $\cA$ graph and thus being
labelled by its vertices, match the
non-vanishing diagonal entries of the modular invariant.
They also associated graphs to several $\SUn$ modular
invariants with higher $n$ and their concept is quite
general, however, there may be difficulties associating
a graph to some invariants --- unlike the $\SUz$ case.

We also consider in Section \ref{othappl} the
modular invariants of the extended $U(1)$ current algebras
as treated in \cite{bumt} and the minimal model
modular invariants which arise from coset theories
$(\SUz_{m-2}\otimes\SUz_1)/\SUz_{m-1}$ and are labelled by
pairs $(\cG_1,\cG_2)$ of A-D-E graphs, associated to
levels $(m-2,m-1)$.

\subsection{Outline of this paper}

A conformal inclusion directly provides a net of subfactors in
terms of the von Neumann algebras of local loop groups in the
vacuum representation of the larger theory. For the orbifold
case we start with the level $k$ vacuum representation of the
loop group $\SUn$ and construct a net of subfactors by a DHR
construction of fields implementing automorphisms, constituting
a simple current extension in terms of bounded operators.
The construction is possible and yields moreover a local extended
net exactly at the levels where the orbifold modular invariants
occur. In both cases we arrive at a net of subfactors
$\cN\subset\cM$ satisfying the necessary conditions to apply
the procedure of $\alpha$-induction elaborated in our previous
paper.

As a consequence of Wassermann's work, to the level $k$
\per s of the loop group $\LSUn$ correspond (DHR superselection)
sectors of local algebras $\NIo$, where $\Io\subset S^1$ is some
proper interval. These sectors are labelled by admissible weights
$\Lambda$ and it is proven that their sector products obey the
well-known $\SUn_k$ fusion rules \cite{wass3}, giving rise to a
fusion algebra $W=W(n,k)$. The irreducible subsectors obtained
by $\alpha$-induction of these sectors generate
sector algebra $V$. The results of our previous paper, in particular
the homomorphism property of $\alpha$-induction, allows to read off
partially the structure of $V$ in terms of the fusion rules in $W$,
but it does in general not determine the multiplication table
completely. However, in many examples it provides enough
information to resolve the puzzle.

The homomorphism property of $\alpha$-induction also implies
that $V$ carries a representation of $W$ which therefore
decomposes into a direct sum over the characters $\gamma_\Lambda$
of the fusion algebra $W$ which are labelled by admissible weights
$\Lambda$ as well. The representation matrix associated to the first
fundamental weight $\Lambda_{(1)}$ can be interpreted as
the adjacency matrix of a graph (which is in fact the fusion
graph of $\alpha_{\Lambda_{(1)}}$), and its eigenvalues
are the evaluation of the characters $\gamma_\Lambda$
in the decomposition  of this representation of $W$.
The weights $\Lambda$ labelling the characters which in fact
appear this way are called exponents as can be recognized as a
generalization of the Coxeter exponents in the $\SUz$ case.

As a consequence of $\alpha\sigma$-reciprocity, proved in our
previous paper, there is a fusion subalgebra $T\subset V$
generated by the (localized) sectors of the larger theory which
correspond to the blocks in the modular invariant. It is widely
believed in general but proven only for several cases that their
sector products coincide with the Verlinde fusion rules known
in conformal field theory. Provided that this is true
for the embedding theory at hand we show
that the interplay of S-matrices diagonalizing the fusion rules
and implementing modular transformations at the same time forces
a conformal character $\chi_\Lambda$ to appear in the modular
invariant if and only if $\Lambda$ is an exponent.

\subsection{Preliminaries}
\lablsec{prelim2}

Here we briefly review our basic notation and results, however, for
precise definitions and statements we refer the reader to our previous
paper \cite{boev1}. There we considered certain nets of subfactors
$\cN\subset\cM$ on the punctured circle, i.e.\ we were dealing with
a family of subfactors $N(I)\subset M(I)$ on a Hilbert space $\cH$,
indexed by the set $\Jz$ of open intervals $I$ on the unit circle
$S^1$ that do neither contain nor touch a distinguished point
``at infinity'' $z\in S^1$. The defining representation of $\cN$
possesses a subrepresentation $\pio$ on a distinguished
subspace $\cH_0$ giving rise to another net
$\cA=\{A(I)=\pio(N(I))\,,\,\,I\in\Jz\}$. We assumed this net
to be strongly additive (which is equivalent to strong
additivity of the net $\cN$) and
to satisfy Haag duality, $A(I)=\CA(I')'$, where $\CA(I')$ denotes
the $C^*$-algebra generated by all $A(J)$, with intervals
$J\in\Jz$ and $J\subset I'$, the (interior of the)
complement of $I$, and also
locality of the net $\cM$. Fixing an interval $\Io\in\Jz$ we
used the crucial observation in \cite{lore} that there is an
endomorphism $\can$ of the $C^*$-algebras $\cM$ into $\cN$ (the
$C^*$-algebras associated to the nets are denoted by the same
symbols as the nets itself, as usual) such that it restricts to
a canonical endomorphism of $\MIo$ into $\NIo$. By $\canr$ we
denote its restriction to $\cN$. We defined a map
$\DelNIo\rightarrow\End(\cM)$, $\lambda\mapsto\ala$,
called $\alpha$-induction, where $\DelNIo$ is the set of
transportable endomorphisms localized in $\Io$. Explicitly,
\[ \ala = \cani \circ \Ad (\eps \lambda\canr)
\circ \lambda \circ \can \,, \]
with statistics operators $\eps \lambda\canr$. As endomorphisms
in $\DelNIo$ leave $\NIo$ invariant one can consider elements of
$\DelNIo$ as elements of $\End(\NIo)$, and therefore it makes
sense to define the quotient $\LTSN$ by inner equivalence in $\NIo$.
Similarly, the endomorphisms $\ala$ leave $\MIo$ invariant,
hence we can consider them also as elements of $\End(\MIo)$
and form their inner equivalence classes $[\ala]$ in $\MIo$.
We derived that in terms of these equivalence classes,
called sectors, $\alpha$-induction $[\lambda]\mapsto[\ala]$
preserves the natural additive and multiplicative structures.
Crucial for our analysis is also the formula
\[ \la\ala,\amu\ra_\MIo = \la \canr\circ\lambda,\mu\ra_\NIo\,,
\qquad \lambda,\mu\in\DelNIo\,, \]
where for endomorphisms $\rho,\sigma$ of an infinite factor $M$
we denote
\[ \la \rho,\sigma \ra_M = \dim\,\Hom_M(\rho,\sigma)=
\dim\, \{ t\in M : t\,\rho(m)=\sigma(m)\,t\,,\,\,\, m\in M\}\,.\]
We also have a map $\End(\cM)\rightarrow\End(\cN)$,
$\beta\mapsto\sib$, called $\sigma$-restric\-tion.
Let $\DelMIo\subset\End(\cM)$ denote the set of transportable
endomorphisms localized in $\Io$, and $\DelMIoO\subset\DelMIo$
the subset of endomorphisms leaving $M(I)$ for any $I\in\Jz$
with $\Io\subset I$ invariant. (If the net $\cM$ is Haag dual
then $\DelMIoO=\DelMIo$.) If $\beta\in\DelMIoO$ then $\sib$
leaves $\NIo$ invariant and hence we can consider $\beta$
and $\sib$ as elements of $\End(\MIo)$ and $\End(\NIo)$,
respectively, and we derived $\alpha\sigma$-reciprocity,
\[ \la \ala,\beta \ra_\MIo = \la \lambda,\sib \ra_\NIo\,,
\qquad \lambda\in\DelNIo\,,\quad \beta\in\DelMIoO \,. \]
If one starts with a certain set $\cW$ of sectors in
$\LTSN$ one obtains a set $\cV$ of sectors of $\MIo$ by
$\alpha$-induction, and the above results provide close
connections between the algebraic structures of $\cW$ and
$\cV$, conveniently formulated in the language of sector
algebras.

\section{Application of $\alpha$-induction to conformal inclusions}

In this section we develop our first main application of
$\alpha$-induction. We consider nets of subfactors which arise
from conformal inclusions of $\SUn$.

\subsection{The general method}

We first explain that conformal inclusions of $\SUn$ give rise
to quantum field theoretical nets of subfactors so that we can
apply the machinery of $\alpha$-induction developed in our
previous paper.
Let $H_k\subset G_1$ be a conformal inclusion of $H=\SUn$
at level $k$ with $G$ a connected compact simple Lie group..
Then there is an associated
block-diagonal modular invariant of $\SUn$,
\be
Z=\sum_{t\in\cT} |\chi^\mathrm{ext}_t|^2 \,,\qquad
\chi^\mathrm{ext}_t = \sum_{\Lambda\in\ASU nk}
b_{t,\Lambda} \,\chi_\Lambda \,.
\labl{bldmi}
Here $\cT$ denotes the labelling set of \per s $(\pi^t,\cH^t)$
of $\LG$ at level 1, $\chi^\mathrm{ext}_t$ the characters of $\cH^t$,
and $\chi_\Lambda$ the characters of the level $k$
\per\ spaces $\cH_\Lambda$ of $\LSUn$, and
$(\pi_\Lambda,\cH_\Lambda)$ appears in the decomposition
of $(\pi^t,\cH^t)$ with
multiplicity $b_{t,\Lambda}$. Thus we have in terms of
the \per s
\be
\pi^t|_{\LH} = \bigoplus_{\Lambda\in\ASU nk}
b_{t,\Lambda} \,\pi_\Lambda \,.
\labl{branchHG}
Now let us define a net of subfactors $\cN\subset\cM$
on the Hilbert space $\cH\equiv\cH^0$ by
\be
N(I)=\pi^0(\LIH)''\,, \qquad M(I)=\pi^0(\LIG)''\,,
\ee
and also the net $\cA$ by
\be
A(I) = \pio(\LIH)''
\ee
for intervals $I\subset S^1$. For conformal embeddings
the index of the subfactors $N(I)\subset M(I)$ is finite
(see e.g.\ \cite{wass?}, \cite{reh4}), and as the nets
$\cM$ and $\cA$ constitute M\"obius covariant precosheaves
on the circle they satisfy Haag duality on the closed
circle \cite{brgl} and hence in particular locality.
Moreover, we have a vector $\Omega\in\cH$ which is cyclic
and separating for each $M(I)$ ($I\neq S^1$ any interval
such that $\bar{I}\neq S^1$) on $\cH$
and $N(I)$ on $\cH_0\subset\cH$. The modular group
of $M(I)$ associated to the state
$\omega(\cdot)=\la \Omega,\cdot\,\Omega\ra$ is geometric,
i.e.\ its action restricts to a geometric action on the
loop group elements, see \cite{wass3} for the case
$G=\mathit{SU}(m)$ and \cite{wass?} for the general case
that $G$ is any compact simple Lie group. Hence the
modular group leaves the subalgebra $N(I)$ invariant for
$\SUn\subset G$ is a subgroup. Therefore
there is a normal conditional expectation $E_I$ from $M(I)$
onto $N(I)$ and preserves the state $\omega$ by Takesaki's
theorem \cite{take}. Furthermore, $E_I$ is unique
and faithful as the inclusion is also irreducible.
The net $\cN\subset\cM$ is standard (by the
Reeh-Schlieder theorem) and hence the Jones
projection $e_N$ from $\cH$ onto $\cH_0$ does not depend
on the interval $I$. Therefore we have
$E_I(m)\Omega=E_I(m)e_N\Omega=e_Nme_N\Omega=e_Nm\Omega$
for any $I$ and $m\in M(I)$. Hence $E_I(m)\Omega=E_J(m)\Omega$
for any pair $I\subset J$ since $\Omega$ is separating
for $M(J)$. We conclude that we have a faithful normal
conditional expectation $E$ from $\cM$ onto $\cN$ and it
obviously preserves the vector state $\omega$.

Now let us remove a ``point at infinity'' $z\in S^1$ and
take the set $\Jz$ as the index set of our
nets $\cA$, $\cN$, $\cM$. Then we are clearly dealing with
directed nets. For $H=\SUn$, Haag duality on the closed circle,
$A(I)=A(I')'$, has been proven directly by Wassermann
\cite{wass3}, as has strong additivity or ``irrelevance of points''
i.e.\ $A(I)=A(I_1)\vee A(I_2)$ if the intervals $I_1$ and
$I_2$ are obtained by removing one single point from the
interval $I$. Moreover, $A(I)=\bigvee_n A(J_n)$ for any
sequence of increasing intervals $J_n$ tending to $I$ \cite{wass?}.
Both arguments imply that we have Haag duality even on
the punctured circle, $A(I)=\CA(I')'$. In fact, as the proofs
in \cite{wass?} are formulated for any compact connected simple
Lie group we similarly have Haag duality on the punctured circle
for $\cM$, $M(I)=\CM(I')'$. (For $G=\mathit{SO}(m)$ (level 1)
this has also been proven directly in \cite{bock2}.) As $\pio$
appears (precisely once) in $\pi^0|_\LH$ we conclude that the net
$\cN$ has a Haag dual subrepresentation, and the corresponding
net is given by $\cA=\{A(I)=\pio(N(I))\,,\,\,I\in\Jz\}$
(note that we take, by abuse
of notation, the same symbol $\pio$ for the subrepresentation
of the net $\cN$ and for the vacuum representation of
$\LH$). Let us summarize the discussion in the following

\begin{proposition}
Starting from a conformal inclusion $\SUn_k\subset G_1$
with $G$ a compact connected simple Lie group the net
$\cN\subset\cM$ (over the index set $\Jz$) defined as above
is a quantum field theoretical net of subfactors of finite
index where $\cM$ is Haag dual (hence local) and $\cN$ is
strongly additive and has a Haag dual subrepresentation.
\end{proposition}

As the \per s of $\LH=\LSUn$
satisfy local equivalence \cite{wass3},
\[ \pi_\Lambda (\LIH) \simeq \pio (\LIH) \,, \]
we have by the standard arguments endomorphisms
$\lambda_{0;\Lambda}\in\DelAIo$ that correspond to $\pi_\Lambda$
for some interval $\Io\in\Jz$. Wassermann has
related the $\LSUn$ fusion rules to the (relative tensor)
product of bimodules, and this is equivalent
to the composition of endomorphisms. Hence we have complete
information about the sector products
$[\lambda_{0;\Lambda}]\times[\lambda_{0;\Lambda'}]$.
Equivalently, we can also take the lifted
endomorphisms
\[ \lambda_\Lambda = \pioi \circ \lambda_{0;\Lambda} \circ \pio
\in \DelNIo \,, \]
and then we clearly have the same sector product rules.

By Eq.\ (27) of \cite{boev1} we have
\[ [\canr] = \bigoplus_{\Lambda\in\ASU nk}
b_{0,\Lambda} \, [\lambda_\Lambda] \]
where this decomposition corresponds to the
the vacuum block in the modular invariant,
$\chi^\mathrm{ext}_0 =
\sum_{\Lambda\in\ASU nk} b_{0,\Lambda} \chi_\Lambda$, see \erf{bldmi}.
Our procedure is then as follows.
Recall that a sector basis is a finite set of irreducible sectors
with finite statistical dimension which contains the identity sector
and is closed under sector products and conjugation. A sector
basis canonically defines an algebra called sector algebra.
(We refer again to \cite{boev1} for precise definitions.)
We take the sector basis
$\cW\equiv\cW(n,k)=\{ [\lambda_\Lambda]\,,\,\,\Lambda\in\ASU nk \}
\subset\LTSN$ and we denote by $W\equiv W(n,k)$ the associated
fusion algebra. By $\alpha$-induction (see Theorem 4.2 of \cite{boev1})
we obtain a sector algebra $V$ with sector basis
$\cV\subset\Sect(\MIo)$, consisting of the distinct irreducible
subsectors of the $[\alpha_\Lambda]$. (We write $\alpha_\Lambda$ for
$\alpha_{\lambda_{\Lambda}}$.) Picking endomorphisms
$\lambda_{\Lambda_{(p)}}$,
associated to the $p$-th fundamental representation,
$p=1,2,...,n-1$ ($\Lambda_{(p)}$ denotes the $p$-th fundamental
weight) and forming $\alf p$, we can compute the sector products
$[\alf p]\times\as\Lambda$ for all
$\Lambda\in\ASU nk$. In many cases, the homomorphism $[\alpha]$ is
surjective and therefore all the fusion rules in $V$
can be read off from the fusion rules in $W$. But even for those
of our examples where the homomorphism $[\alpha]$ is not surjective
we can at least determine the fusion rules $[\alf p]\times[\beta]$ for
all $[\beta]\in\cV$, and thus we can draw the associated fusion graphs.

Since the \per s of a loop group of any connected compact simple
satisfy local equivalence \cite{wass?} we have endomrphisms
$\beta_t\in\DelMIoO=\DelMIo$, $t\in\cT$, corresponding to the level
$1$ \per s of $\LG$. As we know the branching rules of the
decomposition of $\pi^t|_\LH$, \erf{branchHG}, and as
$\sigma$-restriction corresponds to the restriction of
representations it follows
$[\sigma_{\beta_t}]=\bigoplus_{\Lambda\in\ASU nk} b_{t,\Lambda}
[\lambda_\Lambda]$. As a consequence of $\alpha\sigma$-reciprocity,
$\la\alpha_\Lambda,\beta_t\ra_\MIo=\la\lambda_\Lambda,
\sigma_{\beta_t}\ra_\NIo=b_{t,\Lambda}$, $\Lambda\in\ASU nk$,
$t\in\cT$, we conclude (cf.\ Theorem 4.3 of \cite{boev1})
that $\cT\subset\cV$ and that the associated fusion algebra $T$ must
be a sector subalgebra $T\subset V$. (We identify the labelling set
$\cT$ itself with the associated sector basis:
$\cT\equiv\{[\beta_t]\}\subset\LTSM$, $t\equiv[\beta_t]$.)
It is widely believed
but in general not known whether the endomorphisms associated to
the \per s of a loop group $\LG$ obey the Verlinde fusion rules
of the corresponding WZW theory. However, for the level $1$ theories
which are relevant here, this is proven for many cases including
$G=\mathit{SU}(m)$ as a special (and most trivial) case of
Wassermann's analysis \cite{wass3} and $G=\mathit{SO}(m)$ as
done in \cite{bock2}, moreover, for $G=\Gtwo$ it follows from our
treatment of the conformal embedding $\SUz_{28}\subset(\Gtwo)_1$.

Now recall that Di Francesco, Petkova and Zuber
(see \cite{frzu2,pezu2} or \cite{franc,frms}) associated
certain graphs to modular invariants by some empirical
procedure. For these graphs they constructed fusion algebras
(which are possibly not uniquely determined), and they
discovered for the block-diagonal modular invariants
some subalgebras spanned by a subset of the vertices,
called {\em marked vertices}, which obey the fusion rules
of the extended theory. Indeed, in our examples we rediscover
their graphs by drawing the fusion graphs of $\alf p$. The
elements of $\cT$ turn out to represent exactly the marked
vertices, and our theory provides an explanation why the
graph algebras (which are in fact the fusion algebras $V$)
possess subalgebras corresponding to the fusion rules of
the extended theory.

\subsection{Example: $\SUz_{10}\subset\SOf_1$}

We consider the conformal inclusion
$\SUz_{10}\subset\SOf_1$. The corresponding $\SUz$
modular invariant is the \ddE 6 one,
\be
Z_{\mathrm{E}_6}= |\chi_0 + \chi_6|^2 + |\chi_4 + \chi_{10}|^2
+ |\chi_3 + \chi_7|^2 \,.
\labl{ZEs}
The three blocks come from the basic ($0$), the vector (v)
and the spinor (s) representation of $\mathit{LSO}(5)$ at level 1.
For $\LSUz$ at level $10$ there are $11$ \per s $\pi_j$,
labelled by the (doubled, thus integer valued) spin
$j=0,1,2,...,10$. Let
$\lambda_j\in\DelNIo$ be corresponding endomorphisms.
The fusion algebra $W\equiv W(2,k)$ is characterized by the
fusion rules
\be
\ls {j_1} \times \ls {j_2} =
\bigoplus_{j=|j_1-j_2|\,,\,\,j+j_1+j_2\,\,\mathrm{even}}
^{\mathrm{min} (j_1+j_2,2k-(j_1+j_2))} \ls j \,,
\labl{fusuz}
and here $k=10$. From the vacuum block in \erf{ZEs} we read off
$[\canr]=\ls 0\oplus \ls 6$. By Theorem 3.9 of \cite{boev1}
we obtain (writing $\alpha_j$ for $\alpha_{\lambda_j}$)
\[ \la \alpha_{j_1},\alpha_{j_2} \ra_\MIo =
\la \lambda_{j_1},\canr\circ\lambda_{j_2} \ra_\NIo =
\la \lambda_{j_1},\lambda_{j_2} \ra_\NIo +
\la \lambda_{j_1},\lambda_6\circ\lambda_{j_2} \ra_\NIo \,. \]
We find this way
\[ \la \alpha_j , \alpha_j \ra_\MIo = \left\{ \bearll
1 & \for j=0,1,2,8,9,10 \\
2 & \for j=3,4,5,6,7 \eear \right. . \]
We further compute $\la \alpha_3,\alpha_9\ra_\MIo=1$, hence
$\as 3=\as 9 \oplus \asx 3 1$ with $\asx 3 1$ irreducible.
As $\la \alpha_3,\alpha_j\ra_\MIo=0$ for $j=0,1,2,8,10$,
there is no irreducible $\as j$ that equals $\asx 3 1$.
Checking all other $\la \alpha_{j_1},\alpha_{j_2} \ra_\MIo$ we finally
find that there are six different irreducible sectors, i.e.\
elements of $\cV$, namely
$\as 0$, $\as 1$, $\as 2$, $\as 9$, $\as {10}$, and
$\asx 3 1$, and we have the identity $\as 8 = \as 2$.
The reducible $\as j$'s decompose into the elements of $\cV$
as follows,
\[ \bearll
\as 3 = \asx 3 1 \oplus \as 9 \,, \quad &
\as 4 = \as 2 \oplus \as {10} \,, \qquad
\as 5 = \as 1 \oplus \as 9 \\[.5em]
\as 6 = \as 0 \oplus \as 2 \,, &
\as 7 = \as 1 \oplus \asx 3 1 \,.  \eear \]
We are in the fortunate situation that we can write all
elements of $\cV$ as (integral) linear combinations of
$\as j$'s, i.e.\ the homomorphism $[\alpha]$ is surjective.
Thus we can determine their fusion rules from those of
$\LSUz$. For instance, we compute
\[ \bearll \asx 3 1 \times \as 1 &= (\as 3 \times \as 1)
\ominus (\as 9 \times \as 1 ) \\[.5em]
&=(\as 2 \oplus \as 4) \ominus (\as 8 \oplus \as {10})=\as 2
\,. \eear \]
In particular, we can draw the fusion graph for
$[\alf 1]\equiv \as 1$. It is straightforward to check that this
is \ddE 6, Fig.\ \ref{E6}.
\thinlines
\setlength{\unitlength}{3pt}
\begin{figure}[tb]
\begin{center}
\begin{picture}(60,30)
\put(10,10){\line(1,0){40}}
\put(30,10){\line(0,1){10}}
\put(10,10){\circle*{1}}
\put(20,10){\circle*{1}}
\put(30,10){\circle*{1}}
\put(40,10){\circle*{1}}
\put(50,10){\circle*{1}}
\put(30,20){\circle*{1}}
\put(10,10){\circle{2.5}}
\put(50,10){\circle{2.5}}
\put(30,20){\circle{2.5}}
\put(7,5){$\as 0$}
\put(17,5){$\as 1$}
\put(27,5){$\as 2$}
\put(37,5){$\as 9$}
\put(47,5){$\as {10}$}
\put(26,23){$\asx 3 1$}
\end{picture}
\end{center}
\caption{\ddE 6}
\label{E6}
\end{figure}

The homomorphism
$[\alpha]:W\rightarrow V$ induces an induction-restriction graph
connecting \ddA {11} and \ddE 6. We just draw an edge from each
spin $j$ vertex of \ddA {11} to the vertices of \ddE 6 that
represent the irreducible subsectors in the decomposition of
$\as j$. For example, we draw from the spin $j=4$ vertex
one line to the vertex $\as 2$ and one to $\as {10}$.
Completing the picture we obtain a graph with two connected
components $\Gamma_1$ and $\Gamma_2$ corresponding to the even
and odd spins, respectively, see Figs.\ \ref{G1}, \ref{G2}.
\begin{figure}[tb]
\unitlength 0.3mm
\thinlines
\begin{center}
\begin{picture}(300,100)
\multiput(50,80)(40,0){6}{\circle*{5}}
\multiput(70,20)(80,0){3}{\circle*{5}}
\put(50,80){\line(1,-3){20}}
\put(90,80){\line(-1,-3){20}}
\put(90,80){\line(1,-1){60}}
\put(130,80){\line(1,-3){20}}
\put(170,80){\line(-1,-3){20}}
\put(210,80){\line(-1,-1){60}}
\put(210,80){\line(1,-3){20}}
\put(250,80){\line(-1,-3){20}}
\put(40,90){$[\lambda_0]$}
\put(80,90){$[\lambda_6]$}
\put(120,90){$[\lambda_2]$}
\put(160,90){$[\lambda_8]$}
\put(200,90){$[\lambda_4]$}
\put(240,90){$[\lambda_{10}]$}
\put(60,5){$\as 0$}
\put(140,5){$\as 2$}
\put(220,5){$\as {10}$}
\end{picture}
\end{center}
\caption{$\Gamma_1$}
\label{G1}
\end{figure}
\begin{figure}[tb]
\unitlength 0.6mm
\thinlines
\begin{center}
\begin{picture}(110,50)
\multiput(10,30)(20,0){5}{\circle*{1}}
\multiput(30,10)(20,0){3}{\circle*{1}}
\put(10,30){\line(1,-1){20}}
\put(30,30){\line(1,-1){20}}
\put(50,30){\line(1,-1){20}}
\put(50,30){\line(-1,-1){20}}
\put(70,30){\line(-1,-1){20}}
\put(90,30){\line(-1,-1){20}}
\put(30,10){\line(0,1){20}}
\put(70,10){\line(0,1){20}}
\put(3,34){$\ls 1$}
\put(23,34){$\ls 7$}
\put(43,34){$\ls 5$}
\put(63,34){$\ls 3$}
\put(83,34){$\ls 9$}
\put(23,1){$\as 1$}
\put(42,1){$\asx 3 1$}
\put(63,1){$\as 9$}
\end{picture}
\caption{$\Gamma_2$}
\label{G2}
\end{center}
\end{figure}
These graphs are actually well known as graphs connecting
\ddA {11} and \ddE 6, cf.\ \cite{pas1,evgo,kaw1,ghj}.
Indeed, one can show that $\Gamma_1$ is the principal graph for
the inclusion $\NIo\subset\MIo$, and $\Gamma_1$ and $\Gamma_2$.
We plan to come back to this fact in
a separate publication.

Now we turn to the discussion of the marked vertices. Let
$\beta_0,\beta_\rmv,\beta_\rms\in\DelMIo$ be endomorphisms
corresponding to the level 1 basic, vector and spinor representation
of $\mathit{LSO}(5)$ (as constructed in \cite{bock2}).
>From the blocks in \erf{ZEs} we can read off the decomposition
of the $\sigma$-restricted endomorphisms,
\[ [\sigma_{\beta_0}]=\ls 0 \oplus \ls 6 \,, \quad
[\sigma_{\beta_\rmv}]=\ls 4 \oplus \ls {10} \,, \quad
[\sigma_{\beta_\rms}]=\ls 3 \oplus \ls 7 \,. \]
By $\alpha\sigma$-reciprocity, we conclude that
$[\beta_0]=[\id_\MIo]$ must appear in $\as 0$ and $\as 6$,
$[\beta_\rmv]$ in $\as 4$ and $\as {10}$ and
$[\beta_\rms]$ in $\as 3$ and $\as 7$ with
multiplicity one. Hence we conclude
\[ \as 0 = [\beta_0]\,, \qquad \as {10} = [\beta_\rmv] \,,
\qquad \asx 3 1 = [\beta_\rms] \,. \]
In Fig.\ \ref{E6} we encircled the marked vertices (and we
will do it also in the following examples).
It is easy to check that $\as 0$, $\as {10}$ and $\asx 3 1$
indeed obey the Ising fusion rules, e.g.
\[  \asx 3 1 \times \asx 3 1
= (\as 3 \ominus \as 9) \times
(\as 3 \ominus \as 9) = \as 0 \oplus \as {10} \]
as it is well known for the end vertices in the graph algebra of
\ddE 6. This finds now an explanation by the machinery of
$\alpha$-induction and $\sigma$-restriction. Put differently,
our theory proves again the result of \cite{bock2}, namely that
the endomorphisms associated to the level 1 \per s of
$\mathit{LSO}(5)$ obey the Ising fusion rules.

\subsection{More examples}

(i) {\it Example:} $\SUz_{28}\subset (\Gtwo)_1$. The corresponding modular
invariant is the \ddE 8 one,
\[ Z_{\mathrm{E}_8}= |\chi_0 + \chi_{10} + \chi_{18} + \chi_{28}|^2
+ |\chi_6 + \chi_{12} + \chi_{16} + \chi_{22}|^2 \,. \]
The two blocks come from the \per s $\pi_0$ and $\pi_\phi$ of
$\mathit{L}(\Gtwo)$ at level $1$. With
\[ [\canr] = \ls 0 \oplus \ls {10} \oplus \ls {18} \oplus \ls {28} \]
we can determine the structure of the induced sector algebra $V$.
We omit the straightforward calculations and just present the result
here. We find that the sector basis $\cV$ has elements,
given by $\as 0$, $\as 1$, $\as 2$, $\as 3$, $\as 4$, $\asx 5 1$,
$\asx 5 2$ and $\asx 6 1$. The decompositions of the reducible
$\as j$'s read
\[ \bearll
\as 5 = \asx 5 1 \oplus \asx 5 2\,, &
\as 6 = \as 4 \oplus \asx 6 1 \,,\\[.4em]
\as 7 = \as 3 \oplus \asx 5 1\,, &
\as 8 = \as 2 \oplus \as 4 \,,\\[.4em]
\as 9 = \as 1 \oplus \as 3 \oplus \asx 5 2 \,,\qquad &
\as {10} = \as 0 \oplus \as 2 \oplus \as 4 \,,\\[.4em]
\as {11} = \as 1 \oplus \as 3 \oplus \asx 5 1 \,, &
\as {12} = \as 2 \oplus \as 4 \oplus \asx 6 1 \,,\\[.4em]
\as {13} = \as 3 \oplus \asx 5 1 \oplus \asx 5 2 \,, &
\as {14} = 2 \, \as 4 \,,
\eear \]
and we have $\as {28-j} = \as j$. The fusion graph of $\as 1$
is in fact \ddE 8, given in Fig.\ \ref{E8}.
\thinlines
\setlength{\unitlength}{3pt}
\begin{figure}[tb]
\begin{center}
\begin{picture}(90,30)
\put(10,10){\line(1,0){60}}
\put(50,10){\line(0,1){10}}
\put(10,10){\circle*{1}}
\put(20,10){\circle*{1}}
\put(30,10){\circle*{1}}
\put(40,10){\circle*{1}}
\put(50,10){\circle*{1}}
\put(60,10){\circle*{1}}
\put(70,10){\circle*{1}}
\put(50,20){\circle*{1}}
\put(10,10){\circle{2.5}}
\put(70,10){\circle{2.5}}
\put(7,4){$\as 0$}
\put(17,4){$\as 1$}
\put(27,4){$\as 2$}
\put(37,4){$\as 3$}
\put(47,4){$\as 4$}
\put(57,4){$\asx 5 1$}
\put(67,4){$\asx 6 1$}
\put(46,23){$\asx 5 2$}
\end{picture}
\end{center}
\caption{\ddE 8}
\label{E8}
\end{figure}
The marked vertices are given by
\[ \as 0 = [\beta_0] \,, \qquad \asx 6 1 = [\beta_\phi] \,,\]
and it is easy to check that they indeed obey the Lee-Yang
fusion rules
\[ \asx 6 1 \times \asx 6 1 = \as 0 \oplus \asx 6 1 \]
of $(\Gtwo)_1$, i.e.\ here our theory {\it proves} that the
endomorphisms associated to the $(\Gtwo)_1$ \per s obey these
fusion rules.

(ii) {\it Example:} $\SUz_4\subset\SUd_1$. The
corresponding modular invariant is the \ddD 4 one,
\[ Z_{\mathrm{D}_4} = |\chi_0+\chi_4|^2 + 2\,|\chi_2|^2 \,.\]
The first block comes from the vacuum representation
$\pi_{(0,0)}$ and the second one from the \per s
$\pi_{(1,0)}$ and $\pi_{(1,1)}$ of $\LSUd$ at level $1$
which both restrict to the spin $2$ representation
of $\LSUz$ at level $4$. With $[\canr]=\ls 0 \oplus \ls 4$ we find
that $\cV$ has four elements, namely $\as 0$, $\as 1$, $\asx 2 1$,
$\asx 2 2$ where we have the decomposition
$\as 2=\asx 2 1 \oplus \asx 2 2$, and also $\as {4-j}=\as j$.
Note that we cannot isolate $\asx 2 1$ and $\asx 2 2$. Thus
in this case the homomorphism $[\alpha]$ is not surjective!
However, since
\[ \as 1 \times \as 2 = \as 1 \oplus \as 3 = 2\, \as 1 \]
it follows that
\[ \as 1 \times \asx 2 i = \as 1 \,, \qquad i=1,2\,,\]
and we find
\[ \as 1 \times \as 1 = \as 0 \oplus \as 2 =
\as 0 \oplus \asx 2 1 \oplus \asx 2 2 \,, \]
hence the fusion graph of $\as 1$ is uniquely determined
to be \ddD 4, see Fig.\ \ref{D4}.
\thinlines
\setlength{\unitlength}{3pt}
\begin{figure}[tb]
\begin{center}
\begin{picture}(40,30)
\put(6,10){\line(1,0){14}}
\put(20,10){\line(1,1){10}}
\put(20,10){\line(1,-1){10}}
\put(6,10){\circle*{1}}
\put(20,10){\circle*{1}}
\put(30,20){\circle*{1}}
\put(30,0){\circle*{1}}
\put(6,10){\circle{2.5}}
\put(30,20){\circle{2.5}}
\put(30,0){\circle{2.5}}
\put(3,5){$\as 0$}
\put(15,5){$\as 1$}
\put(33,19){$\asx 2 1$}
\put(33,-1){$\asx 2 2$}
\end{picture}
\end{center}
\caption{\ddD 4}
\label{D4}
\end{figure}

The $\SUd_1$ \per s obey $\bbZ_3$ fusion rules, and from
$\alpha\sigma$-reciprocity we conclude that the
marked vertices are given by
\[ \as 0 = [\beta_{(0,0)}]\,,\qquad
\asx 2 1 = [\beta_{(1,0)}]\,,\qquad
\asx 2 2 = [\beta_{(1,1)}]\,. \]
(Clearly we have the freedom to define which is
$\asx 2 1$ and which $\asx 2 2$.)

(iii) {\it Example:} $\SUd_3 \subset \mathit{SO}(8)_1$.
We now turn to the treatment of the $\SUd$ conformal embeddings.
We shall label the $\LSUd$ level $k$
\per s by pairs of integers $(p,q)$, $k\ge p\ge q\ge 0$, that give
the lengths of the rows of the associated Young tableaux. Thus the
(first) fundamental representation has the label $(1,0)$. We denote
the endomorphism that corresponds to the \per\ labelled by $(p,q)$
by $\lad pq$. Thus the sectors $\lasd pq$ constitute the sector
basis of the fusion algebra $W(3,k)$. Recall that the the fusion of the
sector $\lasd 10$ that corresponds to the fundamental representation is
\[ \lasd pq \times \lasd 10 = \lasd {p+1}q \oplus
\lasd p{q+1} \oplus \lasd {p-1}{q-1} \,, \]
where it is understood that on the r.h.s.\ only sectors inside
\ddAgx {k+3}\ contribute.
Now for the conformal embedding $\SUd_3 \subset \mathit{SO}(8)_1$,
the corresponding modular invariant reads
\[ Z_{\mathcal{D}^{(6)}}=
|\chi_{(0,0)} + \chi_{(3,0)} + \chi_{(3,3)}|^2 +
3\,|\chi_{(2,1)}|^2 \,, \]
thus we have $[\canr]=\lasd 00 \oplus \lasd 30 \oplus \lasd 33$.
We find that $\cV$ has six elements,
$\asd 00$, $\asd 10$, $\asd 11$, $\asdx 21 1$, $\asdx 21 2$
and $\asdx 21 3$. Here the only non-trivial decomposition is
$\asd 21 = \asdx 21 1 \oplus \asdx 21 2 \oplus \asdx 21 3$, and
we have $\asd pq=\asd {3-q}{p-q}$. The fusion graph of $\asd 10$
is indeed \ddDgx 6, see Fig.\ \ref{ddDD6}.
\thinlines
\begin{figure}[tb]
\unitlength 0.6mm
\begin{center}
\begin{picture}(80,60)
\put(0,30){\line(4,2){40}}
\put(0,30){\line(4,-2){40}}
\put(40,10){\line(4,1){40}}
\put(40,10){\line(4,2){40}}
\put(40,10){\line(4,3){40}}
\put(39.55,10){\line(0,1){40}}
\put(40.45,10){\line(0,1){40}}
\put(40,50){\line(4,-1){40}}
\put(40,50){\line(4,-2){40}}
\put(40,50){\line(4,-3){40}}
\put(0,30){\circle*{2}}
\put(40,10){\circle*{2}}
\put(40,50){\circle*{2}}
\put(80,20){\circle*{2}}
\put(80,30){\circle*{2}}
\put(80,40){\circle*{2}}
\put(0,30){\circle{5}}
\put(80,20){\circle{5}}
\put(80,30){\circle{5}}
\put(80,40){\circle{5}}
\put(20,40){\vector(-2,-1){0}}
\put(20,20){\vector(2,-1){0}}
\put(40,30){\vector(0,1){0}}
\put(60,45){\vector(4,-1){0}}
\put(60,40){\vector(2,-1){0}}
\put(60,35){\vector(4,-3){0}}
\put(60,25){\vector(-4,-3){0}}
\put(60,20){\vector(-2,-1){0}}
\put(60,15){\vector(-4,-1){0}}
\put(-24,28){$\asd 00$}
\put(33,3){$\asd 10$}
\put(33,55){$\asd 11$}
\put(85,40){$\asdx 21 1$}
\put(85,28){$\asdx 21 2$}
\put(85,16){$\asdx 21 3$}
\end{picture}
\end{center}
\caption{\ddDgx 6}
\label{ddDD6}
\end{figure}
The marked vertices are $\asd 00$, $\asdx 21 1$, $\asdx 21 2$
and $\asdx 21 3$ and hence obey the $\bbZ_2\times\bbZ_2$ fusion
rules of $\mathit{SO}(8)_1$.

The other D-type block-diagonal modular invariants,
namely \ddD {2\varrho+2} for $\SUz$ and \ddDgx {3\varrho+3}
for $\SUd$, $\varrho=2,3,4,...$, do not come from conformal
inclusions. This will be discussed in the following
section.

\begin{figure}[tb]
\unitlength 0.8mm
\begin{center}
\begin{picture}(100,90)
\path(50,90)(38.45,70)
\put(44.225,80){\vector(-1,-2){0}}
\path(61.55,70)(50,90)
\put(55.775,80){\vector(-1,2){0}}
\path(38.45,70)(61.55,70)
\put(50,70){\vector(1,0){0}}
\path(38.45,70)(15.36,70)
\put(26.905,70){\vector(-1,0){0}}
\path(15.36,70)(26.91,50)
\put(21.135,60){\vector(1,-2){0}}
\path(26.91,50)(38.45,70)
\put(32.68,60){\vector(1,2){0}}
\path(26.91,50)(15.36,30)
\put(21.135,40){\vector(-1,-2){0}}
\path(15.36,30)(38.45,30)
\put(26.905,30){\vector(1,0){0}}
\path(38.45,30)(26.91,50)
\put(32.68,40){\vector(-1,2){0}}
\path(38.45,30)(50,10)
\put(44.225,20){\vector(1,-2){0}}
\path(50,10)(61.55,30)
\put(55.775,20){\vector(1,2){0}}
\path(61.55,30)(38.45,30)
\put(50,30){\vector(-1,0){0}}
\path(61.55,30)(84.64,30)
\put(73.095,30){\vector(1,0){0}}
\path(84.64,30)(73.09,50)
\put(78.865,40){\vector(-1,2){0}}
\path(73.09,50)(61.55,30)
\put(67.32,40){\vector(-1,-2){0}}
\path(73.09,50)(84.64,70)
\put(78.865,60){\vector(1,2){0}}
\path(84.64,70)(61.55,70)
\put(73.095,70){\vector(-1,0){0}}
\path(61.55,70)(73.09,50)
\put(67.32,60){\vector(1,-2){0}}
\path(61.55,70)(26.91,50)
\put(44.23,60){\vector(-2,-1){0}}
\path(26.91,50)(61.55,30)
\put(44.23,40){\vector(2,-1){0}}
\path(61.55,30)(61.55,70)
\put(61.55,50){\vector(0,1){0}}
\path(38.45,70)(38.45,30)
\put(38.45,50){\vector(0,-1){0}}
\path(38.45,30)(73.09,50)
\put(55.77,40){\vector(2,1){0}}
\path(73.09,50)(38.45,70)
\put(55.77,60){\vector(-2,1){0}}
\put(50,90){\circle*{1.5}}
\put(15.36,70){\circle*{1.5}}
\put(38.45,70){\circle*{1.5}}
\put(61.55,70){\circle*{1.5}}
\put(84.64,70){\circle*{1.5}}
\put(26.91,50){\circle*{1.5}}
\put(73.09,50){\circle*{1.5}}
\put(15.36,30){\circle*{1.5}}
\put(38.45,30){\circle*{1.5}}
\put(61.55,30){\circle*{1.5}}
\put(84.64,30){\circle*{1.5}}
\put(50,10){\circle*{1.5}}
\put(50,90){\circle{3.75}}
\put(15.36,70){\circle{3.75}}
\put(15.36,30){\circle{3.75}}
\put(84.64,70){\circle{3.75}}
\put(84.64,30){\circle{3.75}}
\put(50,10){\circle{3.75}}
\put(43,96){$\asd 50$}
\put(43,1){$\asdx 20 1$}
\put(25,23){$\asd 10$}
\put(62,23){$\asd 54$}
\put(62,74){$\asd 40$}
\put(25,74){$\asd 51$}
\put(-2,28){$\asd 00$}
\put(89,28){$\asd 55$}
\put(-2,69){$\asdx 22 1$}
\put(89,69){$\asdx 30 1$}
\put(11,49){$\asd 11$}
\put(75.7,49){$\asd 44$}
\end{picture}
\end{center}
\caption{$\mathcal{E}^{(8)}$}
\label{Ed8}
\end{figure}
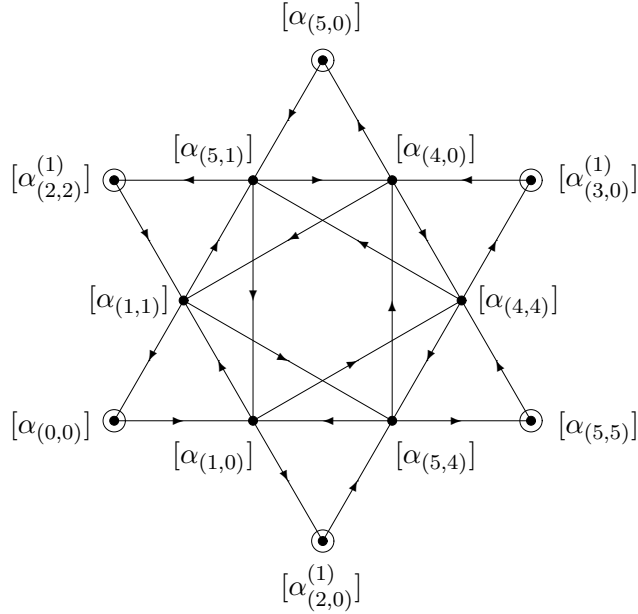

(iv) {\it Example:} $\SUd_5\subset \mathit{SU}(6)_1$.
The corresponding modular invariant reads
\[ \bearll
Z_{\mathcal{E}^{(8)}} &= |\chid 00 + \chid 42|^2 +
|\chid 20 + \chid 53|^2 + |\chid 22 + \chid 52|^2 \\[.5em]
&\qquad\qquad + \,|\chid 30 + \chid 33|^2 +
|\chid 31 + \chid 55|^2 + |\chid 32 + \chid 50|^2
\,,
\eear \]
hence
\[ [\canr] = \lasd 00 \oplus \lasd 42 \,. \]
By computing all the numbers
\[ \la \alpha_{(p,q)}, \alpha_{(r,s)} \ra_\MIo
= \la \canr\circ \lad pq, \lad rs \ra_\NIo \]
(where we denote $\alpha_{(p,q)}=\alpha_{\lad pq}$) we find that
$\cV$ has 12 elements, and the
reducible $\asd pq$'s decompose into these irreducibles as
\[ \bearll
\asd 20 & = \asd 44 \oplus \asdx 20 1 \,,\\[.5em]
\asd 21 & = \asd 51 \oplus \asd 54 \,,\\[.5em]
\asd 22 & = \asd 40 \oplus \asdx 22 1 \,,\\[1.0em]
\asd 30 & = \asd 54 \oplus \asdx 30 1 \,,\\[.5em]
\asd 31 & = \asd 10 \oplus \asd 40 \oplus \asd 55 \,,\\[.5em]
\asd 32 & = \asd 11 \oplus \asd 44 \oplus \asd 50 \,,\\[.5em]
\asd 33 & = \asd 51 \oplus \asdx 30 1 \,,\\[1.0em]
\asd 41 & = \asd 11 \oplus \asd 44 \,,\\[.5em]
\asd 42 & = \asd 00 \oplus \asd 51 \oplus \asd 54 \,,\\[.5em]
\asd 43 & = \asd 10 \oplus \asd 40 \,,\\[1.0em]
\asd 52 & = \asd 10 \oplus \asdx 22 1 \,,\\[0.5em]
\asd 53 & = \asd 11 \oplus \asdx 20 1 \,.
\eear \]
We find that the homomorphism $[\alpha]$ is surjective as we can
invert these formula, namely we obtain
\[ \bearll
\asdx 20 1 &= \asd 20 \ominus \asd 44 \,,\\[.5em]
\asdx 22 1 &= \asd 22 \ominus \asd 40 \,,\\[.5em]
\asdx 30 1 &= \frac12 \left( \asd 30 \oplus \asd 33
\ominus \asd 51 \ominus \asd 54 \right) \,.
\eear \]
It is then a straightforward calculation to determine the
fusion rules of $V$, and the fusion graph of $\asd 10$ is
given by Figure \ref{Ed8}.
\thinlines

The marked vertices are given by $\asd 00$, $\asdx 22 1$,
$\asd 50$, $\asdx 30 1$, $\asd 55$ and $\asdx 20 1$, and one
may check that they in fact obey the $\bbZ_6$ fusion rules
of $\mathit{SU}(6)_1$.

\subsection{A non-commutative sector algebra}

{\it Example:} $\SUf_4\subset \mathit{SO}(15)_1$.
labelling the \per s of $\SUf_4$ with partitions
$(p_1,p_2,p_3)\in\bbZ^3$, $4\ge p_1\ge p_2 \ge p_3 \ge 0$,
the corresponding modular invariant reads
\[ \bearll Z &= |\chif 000 + \chif 310 + \chif 332 + \chif 440 |^2
\\[.4em] & \qquad + |\chif 211 + \chif 400 + \chif 431 + \chif 444 |^2
+ 4\, |\chif 321|^2 \,,
\eear \]
where the blocks correspond to the basic, the vector and the
spinor representation of $\mathit{SO}(15)_1$. With
\[ [\canr] = \lasf 000 + \lasf 310 + \lasf 332 + \lasf 440 \]
we can compute the table
($\aef {p_1}{p_2}{p_3}\equiv\alpha_{\lambda_{(p_1,p_2,p_3)}}$)
\[ \la \aef {p_1}{p_2}{p_3}, \aef {q_1}{q_2}{q_3} \ra_\MIo
= \la \canr\circ\lambda_{(p_1,p_2,p_3)},\lambda_{(q_1,q_2,q_3)}
\ra_\NIo \,. \]
Evaluating this table we first obtain that $\asf 000$, $\asf 100$,
$\asf 110$, $\asf 111$ and $\asf 400$ are distinct irreducible
sectors and we find identities
\[ \bearll
\asf 000 &= \asf 440 \,, \\[.4em]
\asf 100 &= \asf 333 = \asf 410 = \asf 441 \,, \\[.4em]
\asf 110 &= \asf 200 = \asf 222 = \asf 330 = \asf 411 \\[.4em]
         & \quad = \asf 420 = \asf 433 = \asf 442 \,, \\[.4em]
\asf 111 &= \asf 300 = \asf 430 = \asf 443 \,, \\[.4em]
\asf 400 &= \asf 444 \,.
\eear \]
We find seven further elements in $\cV$, namely
$\asfx 210i$, $\asfx 220i$, $\asfx 221i$, $i=1,2$, and
$\asfx 3211$, and the reducible $\asf {p_1}{p_2}{p_3}$ decompose
in this basis as follows,
\[ \bearll
\asf 210 &= \asf 111 \oplus \asfx 210 1 \oplus \asfx 210 2 \,,\\[.4em]
\asf 211 &= \asf 400 \oplus \asfx 220 1 \oplus \asfx 220 2 \,,\\[.4em]
\asf 220 &= \asfx 220 1 \oplus \asfx 220 2 \,,\\[.4em]
\asf 221 &= \asf 100 \oplus \asfx 221 1 \oplus \asfx 221 2 \,,
\eear \]
\[ \bearll
\asf 310 &= \asf 000 \oplus \asfx 220 1 \oplus \asfx 220 2 \,,\\[.4em]
\asf 311 &= \asf 100 \oplus \asfx 221 1 \oplus \asfx 221 2 \,,\\[.4em]
\asf 320 &= \asf 100 \oplus \asfx 221 1 \oplus \asfx 221 2 \,,\\[.4em]
\asf 321 &= 2 \, \asf 110 \oplus 2 \, \asfx 321 1 \,,\\[.4em]
\asf 322 &= \asf 111 \oplus \asfx 210 1 \oplus \asfx 210 2 \,,\\[.4em]
\asf 331 &= \asf 111 \oplus \asfx 210 1 \oplus \asfx 210 2 \,,\\[.4em]
\asf 332 &= \asf 000 \oplus \asfx 220 1 \oplus \asfx 220 2 \,,
\eear \]
\[ \bearll
\asf 421 &= \asf 111 \oplus \asfx 210 1 \oplus \asfx 210 2 \,,\\[.4em]
\asf 422 &= \asfx 220 1 \oplus \asfx 220 2 \,,\\[.4em]
\asf 431 &= \asf 400 \oplus \asfx 220 1 \oplus \asfx 220 2 \,,\\[.4em]
\asf 432 &= \asf 100 \oplus \asfx 221 1 \oplus \asfx 221 2 \,.
\eear \]
The marked vertices are $\asf 000$, $\asf 400$ and $\asfx 321 1$,
corresponding to the basic, vector and spinor representation of
$\mathit{SO}(15)_1$, respectively.  That the spinor representation
$\pi_\rms$ of $\mathit{SO}(15)_1$ restricts to two copies of
$\pi_{(3,2,1)}$, i.e.\ $b_{\rms,(3,2,1)}=2$, implies in particular
that $\asfx 321 1$ appears in the decomposition of $\asf 321$ with
multiplicity $2$ by $\alpha\sigma$-reciprocity.

Using the $\SUf_4$ fusion rules, i.e.\ of
$W(4,4)$, we obtain the following sector products by the
homomorphism property of $\alpha$-induction,
\bea
\asf 100 \times \asf 100 &=& 2\, \asf 110 \,, \label{fu100}\\
\asf 100 \times \asf 110 &=& 2\, \asf 111 \oplus \asfx 210 1
  \oplus \asfx 210 2 \,, \label{fu110}\\
\asf 100 \times \asf 111 &=& \asf 000 \oplus \asf 400
  \oplus \asfx 220 1   \oplus \asfx 220 2 \,, \label{fu111}\\
\asf 100 \times \asf 400 &=& \asf 100 \,,\label{fu400}\\
\asf 100 \times \asfx 321 1 &=& \asfx 210 1 \oplus \asfx 210 2
\,.\label{fu321}
\eea
However, as the homomorphism $[\alpha]:W\rightarrow V$ is not surjective
we cannot isolate $\asfx 210 i$, $\asfx 220 i$ and $\asfx 221 i$, $i=1,2$.
First we can only compute
\bea
\asf 100 \times \left( \asfx 210 1 \oplus \asfx 210 2 \right)
&=& 2 \, \asfx 220 1 \oplus 2 \, \asfx 220 2  \,,\label{fu210} \\
\asf 100 \times \left( \asfx 220 1 \oplus \asfx 220 2 \right)
&=& 2 \, \asf 100 \oplus 2 \, \asfx 221 1 \oplus
   2 \, \asfx 221 2  \,, \label{fu220} \\
\asf 100 \times \left( \asfx 221 1 \oplus \asfx 221 2 \right)
&=& 2 \, \asf 110 \oplus 2 \, \asfx 321 1 \,. \label{fu221}
\eea
Now recall that the statistical dimension of the \per\ of
$\SUn_k$ labelled by a partition $(p_1,p_2,...,p_{n-1})$,
$k\ge p_1 \ge p_2 \ge ... \ge p_{n-1} \ge p_n \equiv 0$,
is given by
\[ d_{(p_1,p_2,...,p_{n-1})} = \prod_{1\le i<j \le n}
\frac{\sin\left(\frac{(p_i-p_j+j-i)\pi}{n+k}\right)}
{\sin\left(\frac{(j-i)\pi}{n+k}\right)} \,. \]
With $n=k=4$ we obtain $d_{(1,0,0)}=\sin(\pi/8)^{-1}$. Since
the marked vertex $\asfx 321 1$ has statistical dimension
(we write $\dafx {p_1}{p_2}{p_3}i \equiv d_{\aefx {p_1}{p_2}{p_3}i}$)
$\dafx 321 1 =\sqrt2 \equiv 4 \sin(\pi/8)\cos(\pi/8)$
we obtain from \erf{fu321}
$\dafx 210 1 + \dafx 210 2 = d_{(1,0,0)} \dafx 321 1
= 4 \cos (\pi/8)$. So we may and do assume without loss of
generality that $\dafx 210 1 \le 2\cos(\pi/8)$. As
$4\cos^2(\pi/8)=2+\sqrt2<4$ it follows that
$\asfx 210 1 \times \asfxb 210 1$ decomposes into at most
three irreducible sectors. Therefore we conclude by
\erf{fu111} that
\[ \bearll\la \aef 100 \circ \aefx 210 1 , \aef 100 \circ \aefx 210 1
\ra_\MIo = \\[.4em]
\qquad\qquad\qquad = \la \aef 111 \circ \aef 100 , \aefx 210 1 \circ
\aefxb 210 1 \ra_\MIo \le 3 \,, \eear\]
and thus $\asf 100 \times \asfx 210 1$ cannot contain an irreducible
sector with multiplicity larger than one. But we also have
\[ \la \aef 100 \circ \aefx 210 1, \aef 220 \ra_\MIo =
\la \aefx 210 1, \aef 111 \circ \aef 220 \ra_\MIo =2 \]
since one checks $\asf 111 \times \asf 220 = 2\asf 210$. It follows
by comparison with \erf{fu210}
\be
\asf 100 \times \asfx 210 i = \asfx 220 1 \oplus \asfx 220 2\,,
\qquad i=1,2\,,
\labl{fu210i}
and $\dafx 210 i=2\cos(\pi/8)$, $i=1,2$.

We have $\asfb 210 = \asf 221$, and hence with a
suitable choice of notation
$\asfxb 210 i = \asfx 221 i$ for $i=1,2$. One checks
\[ \left(  \asfx 210 1 \oplus \asfx 210 2 \right) \times
\asf 111 = 2 \, \asf 110 \oplus 2\, \asfx 321 1 \,, \]
and since $2+\sqrt2=d_{(1,1,0)} \neq \dafx 321 1=\sqrt2$
we find
\be \asfx 210 i \times \asf 111 = \asf 110 \oplus \asfx 321 1\,,
\qquad i=1,2\,,
\labl{fuc210i}
and conjugation yields
\be
\asf 100 \times \asfx 221 i = \asf 110 \oplus \asfx 321 1\,,
\qquad i=1,2\,,
\labl{fu221i}

We have $\asfb 220 = \asf 220$, and this is
\[ \asfxb 220 1 \oplus \asfxb 220 2 = \asfx 220 1 \oplus \asfx 220 1 \,,\]
hence conjugation of \erf{fu210i} yields
\[ \asf 111 \times \asfx 221 i = \asfx 220 1 \oplus \asfx 220 2 \,,
\qquad i=1,2 \,. \]
Thus we find for $i,j=1,2$,
\[ \la \aef 100 \circ \aefx 220 i, \aefx 221 j \ra_\MIo
= \la \aefx 220 i, \aef 111 \circ \aefx 221 j \ra_\MIo = 1 \,, \]
and similarly we obtain (by use of \erf{fu111})
\[ \la \aef 100 \circ \aefx 220 i, \aef 100 \ra_\MIo
= \la \aefx 220 i, \aef 111 \circ \aef 100 \ra_\MIo = 1  \]
for $j=1,2$. It follows now from \erf{fu220}
\be
\asf 100 \times \asfx 220 i = \asf 100 \oplus \asfx 221 1
\oplus \asfx 221 2 \,, \qquad i=1,2 \,.
\labl{fu220i}
We have succeeded to compute $\asf 100 \times [\beta]$ for
each $[\beta]\in\cV$, and thus we can draw the fusion graph
given in Fig.\ \ref{noncom1}.
\thinlines
\begin{figure}[tb]
\unitlength 0.6mm
\begin{center}
\begin{picture}(200,100)
\put(0,40){\line(2,1){80}}
\put(0,40){\line(2,-1){80}}
\put(40,40){\line(1,1){40}}
\put(40,40){\line(1,-1){40}}
\put(79.6,40){\line(0,1){40}}
\put(80.4,40){\line(0,1){40}}
\put(79.6,40){\line(0,-1){40}}
\put(80.4,40){\line(0,-1){40}}
\put(80,40){\line(1,1){40}}
\put(80,40){\line(2,1){80}}
\put(80,40){\line(1,-1){40}}
\put(80,40){\line(2,-1){80}}
\put(120,40){\line(-1,1){40}}
\put(120,40){\line(-1,-1){40}}
\put(120,40){\line(0,1){40}}
\put(120,40){\line(0,-1){40}}
\put(120,40){\line(1,1){40}}
\put(120,40){\line(1,-1){40}}
\put(160,40){\line(-2,1){80}}
\put(160,40){\line(-2,-1){80}}
\put(160,40){\line(-1,1){40}}
\put(160,40){\line(-1,-1){40}}
\put(160,40){\line(0,1){40}}
\put(160,40){\line(0,-1){40}}
\put(200,40){\line(-2,1){80}}
\put(200,40){\line(-2,-1){80}}
\put(200,40){\line(-1,1){40}}
\put(200,40){\line(-1,-1){40}}
\put(80,0){\circle*{2}}
\put(120,0){\circle*{2}}
\put(160,0){\circle*{2}}
\put(0,40){\circle*{2}}
\put(40,40){\circle*{2}}
\put(80,40){\circle*{2}}
\put(120,40){\circle*{2}}
\put(160,40){\circle*{2}}
\put(200,40){\circle*{2}}
\put(80,80){\circle*{2}}
\put(120,80){\circle*{2}}
\put(160,80){\circle*{2}}
\put(0,40){\circle{5}}
\put(40,40){\circle{5}}
\put(200,40){\circle{5}}
\put(40,60){\vector(-2,-1){0}}
\put(40,20){\vector(2,-1){0}}
\put(60,60){\vector(-1,-1){0}}
\put(60,20){\vector(1,-1){0}}
\put(80,20){\vector(0,1){0}}
\put(80,60){\vector(0,1){0}}
\put(96,64){\vector(1,-1){0}}
\put(96,56){\vector(1,1){0}}
\put(104,24){\vector(-1,-1){0}}
\put(104,16){\vector(-1,1){0}}
\put(112,64){\vector(2,-1){0}}
\put(112,56){\vector(2,1){0}}
\put(120,64){\vector(0,-1){0}}
\put(120,24){\vector(0,-1){0}}
\put(128,24){\vector(-2,-1){0}}
\put(128,16){\vector(-2,1){0}}
\put(136,64){\vector(1,-1){0}}
\put(136,24){\vector(1,-1){0}}
\put(144,64){\vector(-1,-1){0}}
\put(144,24){\vector(-1,-1){0}}
\put(160,64){\vector(0,-1){0}}
\put(160,24){\vector(0,-1){0}}
\put(168,56){\vector(-2,1){0}}
\put(168,24){\vector(2,1){0}}
\put(180,60){\vector(-1,1){0}}
\put(180,20){\vector(1,1){0}}
\put(-14,29){$\asf 000$}
\put(14,38.3){$\asf 400$}
\put(56,38.3){$\asf 110$}
\put(124,38.3){$\asfx 220 1$}
\put(161.5,38.3){$\asfx 220 2$}
\put(195.5,29){$\asfx 321 1$}
\put(69,-9){$\asf 100$}
\put(109,-9){$\asfx 221 1$}
\put(149,-9){$\asfx 221 2$}
\put(109,85){$\asfx 210 1$}
\put(69,85){$\asf 111$}
\put(149,85){$\asfx 210 2$}
\end{picture}
\end{center}
\caption{Fusion graph of $\asf 100$}
\label{noncom1}
\end{figure}

Similarly one finds for the sector products of $\asf 110$
\[ \bearll
\asf 110 \times \asf 110 &= \asf 000 \oplus \asf 400
  \oplus 2\, \asfx 220 1 \oplus 2\, \asfx 220 2 \,, \\[.4em]
\asf 110 \times \asf 111 &= 2\, \asf 100 \oplus \asfx 221 1
  \oplus \asfx 221 2 \,, \\[.4em]
\asf 110 \times \asf 400 &= \asf 110 \,, \\[.4em]
\asf 110 \times \asfx 321 1 &= \asfx 220 1 \oplus \asfx 220 2 \,,
\eear \]
and also (we omit some details)
\[ \bearlll
\asf 110 \times \asfx 210  i &= \asf 100 \oplus \asfx 221 1
  \oplus \asfx 221 2 \,,  \qquad & i=1,2\,,\\[.4em]
\asf 110 \times \asfx 220  i &= 2\, \asf 110 \oplus \asfx 321 1
 & i=1,2\,,\\[.4em]
\asf 110 \times \asfx 221  i &= \asf 111 \oplus \asfx 210 1
  \oplus \asfx 210 2 \,,  \qquad & i=1,2\,.
\eear \]
These equations are visualized in in the (disconnected) fusion graph
of $\asf 110$, see Fig.\ \ref{noncom2}.
\thinlines
\begin{figure}[tb]
\unitlength 0.6mm
\begin{center}
\begin{picture}(200,100)
\put(119.6,40){\line(0,1){40}}
\put(120.4,40){\line(0,1){40}}
\put(119.6,40){\line(0,-1){40}}
\put(120.4,40){\line(0,-1){40}}
\put(160,0){\line(0,1){80}}
\put(200,0){\line(0,1){80}}
\put(120,0){\line(1,1){80}}
\put(120,80){\line(1,-1){80}}
\put(120,0){\line(1,2){40}}
\put(120,80){\line(1,-2){40}}
\put(0,20){\line(2,1){40}}
\put(0,60){\line(2,-1){40}}
\put(80,20){\line(0,1){40}}
\put(40,40.5){\line(2,1){40}}
\put(40,39.5){\line(2,1){40}}
\put(40,40.5){\line(2,-1){40}}
\put(40,39.5){\line(2,-1){40}}
\put(0,20){\circle*{2}}
\put(0,60){\circle*{2}}
\put(40,40){\circle*{2}}
\put(80,60){\circle*{2}}
\put(80,40){\circle*{2}}
\put(80,20){\circle*{2}}
\put(120,80){\circle*{2}}
\put(120,0){\circle*{2}}
\put(160,80){\circle*{2}}
\put(160,0){\circle*{2}}
\put(200,80){\circle*{2}}
\put(200,0){\circle*{2}}
\put(0,20){\circle{5}}
\put(0,60){\circle{5}}
\put(80,40){\circle{5}}
\put(-11,11){$\asf 000$}
\put(-11,67){$\asf 400$}
\put(69,11){$\asfx 220 2$}
\put(69,67){$\asfx 220 1$}
\put(84,38.3){$\asfx 321 1$}
\put(10,38.3){$\asf 110$}
\put(109,-9){$\asf 100$}
\put(149,-9){$\asfx 221 1$}
\put(189,-9){$\asfx 221 2$}
\put(149,85){$\asfx 210 1$}
\put(109,85){$\asf 111$}
\put(189,85){$\asfx 210 2$}
\end{picture}
\end{center}
\caption{Fusion graph of $\asf 110$}
\label{noncom2}
\end{figure}

We now want to show that for this example the $\alpha$-induced
sector algebra is in fact non-commutative! (The appearance of
a non-commutative sector structure associated to the conformal
embedding $\SUf_4\subset\mathit{SO}(15)_1$ was first observed
by Xu \cite{xu1} in his framework.) From Eqs.\
(\ref{fuc210i}) and (\ref{fu221i}) we obtain
(recall $\asfxb 210 i = \asfx 221 i$)
\[ \bearl
\la \aefx 210 i \circ \aefx 210 j ,
\aef 100 \circ \aef 100 \ra_\MIo =\\[.4em]
\qquad\qquad = \la \aefx 210 j \circ \aef 111 ,
\aefx 221 i \circ \aef 100 \ra_\MIo
= 2 \,, \quad i,j=1,2 \,,
\eear \]
but from $\asf 100 \times \asf  100 = 2 \asf 110$ we conclude
that $\asf 110$ is a subsector of $\asfx 210 i \times \asfx 210 j$,
and by matching the statistical dimensions we find indeed
\[ \asfx 210 i \times \asfx 210 j = \asf 110 \,, \qquad i,j=1,2 \,, \]
and hence
\be
\bearl
\la \aefx 210 i \circ \aefx 221 i , \aefx 221 i \circ
\aefx 210 i \ra_\MIo =\\[.4em]
\qquad\qquad = \la \aefx 210 i \circ \aefx 210 i ,
\aefx 210 i \circ \aefx 210 i \ra_\MIo
= 1 \,, \quad i=1,2 \,.
\eear
\labl{oneic}
But $\asfx 210 i \times \asfx 221 i$ as well as
$\asfx 221 i \times \asfx 210 i$ must contain the identity
sector $\asf 000$, and also other sectors since
$\dafx 210 i = \dafx 221 i = 2 \cos (\pi/8) >1$, $i=1,2$.
Because \erf{oneic} tells us that these products have only
the identity sector in common we have shown
\[ \asfx 210 i \times \asfx 221 i \neq \asfx 221 i
\times \asfx 210 i \,, \qquad i=1,2\,. \]
Indeed one can compute these products as follows. Since
\[ \asfx 210 i \times \asf 100 = \asf 100 \times \asfx 210 i
= \asfx 220 1 \oplus \asfx 220 2 \,, \quad i=1,2\,, \]
it follows
\[ \bearl
\la \aefx 210 i \circ \aefx 221 i , \aef 111 \circ
\aef 100 \ra_\MIo =\\[.4em]
\qquad\qquad = \la \aef 100 \circ \aefx 210 i ,
\aef 100 \circ \aefx 210 i \ra_\MIo
= 2 \,, \quad i=1,2 \,,
\eear \]
and
\[ \bearl
\la \aefx 221 i \circ \aefx 210 i , \aef 100 \circ
\aef 111 \ra_\MIo =\\[.4em]
\qquad\qquad = \la \aefx 210 i \circ \aef 100 ,
\aefx 210 i \circ \aef 100 \ra_\MIo
= 2 \,, \quad i=1,2 \,,
\eear \]
and from \erf{fu111} we conclude that both
$\asfx 210 i \times \asfx 221 i$ and $\asfx 221 i \times \asfx 210 i$
must contain, besides the identity sector, one of the sectors
$\asf 400$, $\asfx 220 1$ and $\asfx 220 2$. Let us assume that
$\asfx 210 1 \times \asfx 221 1$ contains $\asf 400$. Then,
because of a mismatch of quantum dimensions of $\sqrt2$, it
contains necessarily a third sector (which is determined to
be $\asfx 321 1$). Since now $\asfx 221 1 \times \asfx 210 1$
cannot contain $\asf 400$ it contains either $\asfx 220 1$ or
$\asfx 220 2$, and as then the quantum dimensions match this
means that $\asfx 221 1 \times \asfx 210 1$ decomposes into two
irreducible sectors whereas $\asfx 210 1 \times \asfx 221 1$
decomposes into three. However, this contradicts
\[ \bearl \la \aefx 210 1 \circ \aefx 221 1 , \aefx 210 1 \circ \aefx 221 1
\ra_\MIo = \\[.4em]
\qquad\qquad\qquad\qquad  =\la \aefx 221 1 \circ \aefx 210 1 ,
\aefx 221 1 \circ \aefx 210 1 \ra_\MIo \,.
\eear \]
It follows, with a suitable choice of notation,
\[ \bearll
\asfx 210 1 \times \asfx 221 1 &= \asf 000 \oplus \asfx 220 1 \,,\\[.4em]
\asfx 221 1 \times \asfx 210 1 &= \asf 000 \oplus \asfx 220 2 \,.
\eear \]

Petkova and Zuber obtained the fusion graphs of Figs.\
\ref{noncom1} and \ref{noncom2} in a completely different and
more empirical way (Fig.\ A.2.\ in \cite{pezu2}). The
non-commutativity of $V$ nicely explains why they could
not find non-negative structure constants associated to these
graphs: They were searching for a (commutative) fusion
algebra.

\section{The treatment of orbifold inclusions}

We have seen that conformal inclusions can be described in
terms of nets of subfactors. For orbifold inclusions the
extended net is not a priori given. However, it is argued
in \cite{mose} that orbifold type modular invariants arise
from extensions of current algebras by some simple currents.
The conformal dimensions of these simple currents are
necessarily integers. In the following we will describe this
idea in our ``bounded operator framework''. The techniques we
use are not essentially new. Similar and often more general
statements can be found in particular in \cite{reh1,reh2}.
However, we prefer to give a self-contained presentation and to
avoid unnecessary generality if this simplifies our arguments.

Starting from
the vacuum net $\cA$ where $A(I)=\pio(\LISUn)''$ as usual we
will construct a net of subfactors $\cN\subset\cM$ describing
the analogue of the conformal inclusions now for the
orbifold type modular invariants, and we will see that the
extended net obtained by this construction is local exactly
for the levels where the orbifold modular invariants appear.

\subsection{Construction of the extended net}

We call an automorphism $\sioh\in\DelAIo$ a {\em simple current of order}
$n$, if $n=2,3,4,...$ is the smallest positive integer such that
$\sioh^n$ is equivalent to the identity, i.e.\ $\sioh^n=\Ad(Y)$
for a unitary $Y\in\fB(\cH_0)$, and then $Y\in\AIo$ by
Haag duality. For our construction we need an equivalent automorphism
$\sio$ which is periodic i.e.\ $\sio^n$ is exactly the identity. We call
$\rho_0\in\DelAIo$ a {\em fixed point of the simple current}
$\sioh$ if $[\sioh\circ\rho_0]=[\rho_0]$. The following lemma
gives a sufficient criterion for the possibility of a choice of
a periodic automorphism (cf.\ \cite{izu1}, Prop.\ 3.3, or
\cite{reh1}, Lemmata 4.4 and 4.5).

\begin{lemma}
Let $\sioh\in\DelAIo$ be a simple current of order $n$. If
there is an irreducible fixed point $\rho_0\in\DelAIo$ of
$\sioh$ then there is a simple current $\sio\in\DelAIo$
such that $[\sio]=[\sioh]$ and $\sio^n=\id$.
\lablth{simfix}
\end{lemma}

\bproof
Since $[\sioh\circ\rho_0]=[\rho_0]$ there is a unitary
$U\in\AIo$ such that $\sioh\circ\rho_0=\Ad(U)\circ\rho_0$.
We set $\sio=\Ad(U^*)\circ\sioh$. Then $\sio^n$ is clearly
inner, namely
\[ \sio^n=(\Ad(U^*)\circ\sioh)^n=
\Ad(U^*\sioh(U^*)\sioh^2(U^*)\cdots\sioh^{n-1}(U^*))\circ\sioh^n
=\Ad(Z) \,, \]
where $Z=U^*\sioh(U^*)\sioh^2(U^*)\cdots\sioh^{n-1}(U^*)Y$.
(Recall $\sioh^n=\Ad(Y)$.) Now we have
$\rho_0=\sio^n\circ\rho_0=\Ad(Z)\circ\rho_0$ and thus
$Z\in\rho_0(\AIo)'\cap\AIo$. Since we assumed
$\rho_0$ to be irreducible it follows $Z\in\bbC\bfe$ and
hence $\sio^n=\Ad(Z)=\id$.
\eproof

>From now on we assume that there is an irreducible fixed point
$\rho_0\in\DelAIo$ for the simple current $\sioh\in\DelAIo$
of order $n$, and hence we have an equivalent periodic
automorphism $\sio\in\DelAIo$, i.e.\ $\sio^n=\id$, and also
$\sio\circ\rho_0=\rho_0$. The following construction
is basically the construction of the field group and algebra
as in \cite{dhr1b}. Recall that $\cH_0$ is the vacuum
Hilbert space where $\cA$  lives on. We set
\[ \cH = \bigoplus_{p=0}^{n-1} \cH_0 \,. \]
For a vector $\Psi\in\cH$ we denote by $\Psi_p\in\cH_0$ its $p$-th
component with respect to this decomposition. We define a
representation $\pi$ of $\cA$ on $\cH$ by
$\pi(a)=\bigoplus_{p=0}^{n-1} \sio^p(a)$, i.e.
\[ (\pi(a) \Psi)_p = \sio^p(a) \Psi_p \,, \qquad
a\in\cA\,, \quad \Psi\in\cH \,, \quad p=0,1,2,\ldots,n-1\,. \]
Then the net $\cN$ is defined in terms of local algebras by
\[ N(I) = \pi(A(I)) \,, \qquad I\in\Jz \,. \]
Pick a unitary $U_I$ such that
$\sioI=\Ad(U_I)\circ\sio\in\DelAI$ for some $I\in\Jz$.
We define unitary {\em field operators} $\fUI\in\fB(\cH)$ by
\[ (\fUI \Psi)_p = \sio^{p-1}(U_I^*) \Psi_{p-1} \,,
\qquad \Psi \in\cH\,,\quad p=0,1,2,\ldots,n-1\,,\,\,\,
(\mbox{mod}\,\, n)\,. \]
It is easy to check that
\[ (\fUI^* \Psi)_p = \sio^p(U_I) \Psi_{p+1} \,,
\qquad \Psi \in\cH\,,\quad p=0,1,2,\ldots,n-1\,,\,\,\,
(\mbox{mod}\,\, n)\,. \]
Then we find
\[ \bearll
(\fUI^* \pi(a) \fUI \Psi)_p
&= \sio^p (U_I) \, (\pi(a) \fUI \Psi)_{p+1} \\[.4em]
&= \sio^p (U_I) \, \sio^{p+1} (a) \, (\fUI \Psi)_{p+1} \\[.4em]
&= \sio^p (U_I) \, \sio^{p+1} (a) \, \sio^p (U_I^*) \Psi_p \\[.4em]
&= \sio^p \circ \sioI (a) \Psi_p \,,
\eear \]
hence
\be
\fUI^* \, \pi(a) \, \fUI = \pi \circ \sioI (a) \,, \qquad a\in\cA\,.
\labl{fields}
We use the following notation: For $\lambda_0\in\End(\cA)$
we define $\lambda\in\End(\cN)$ by
$\lambda(\pi(a))=\pi(\lambda_0(a))$, $a\in\cA$.
One checks easily that
$\lambda\in\DelNIo$ if $\lambda_0\in\DelAIo$.
Then \erf{fields} reads $\fUI^* x \fUI = \sigma_I (x)$
for $x\in\cN$, and in particular $\fUI\in N(I')'$,
i.e.\ fields are relatively local to observables.
Now we define the extended net $\cM$ in terms of local
algebras $M(I)$ being generated by $N(I)$ and $\fUI$,
\[ M(I) = \la N(I) \,,\, \fUI \ra \,, \qquad I\in\Jz \,. \]
Note that we have
\[ \fUI = f_\bfe \pi(U_I^*) \]
since
\[ (\fUI \Psi)_p = \sio^{p-1}(U_I^*) \Psi_{p-1} =
\pi(U_I^*) \Psi_{p-1} = (f_\bfe \pi(U_I^*) \Psi)_p \,. \]
Therefore the definition of $M(I)$ is independent on the special
choice of $U_I$ because if $\Ad(\hat{U}_I)\circ\sio$ is also
localized in $I$ then $U_I\hat{U}_I^*\in A(I)$ by Haag duality
and hence $\fUI$ and $f_{\hat{U}_I}$ differ only by an element in
$N(I)$.

Note that our construction is such that (obviously by taking
$U_{\Io}=\bfe$) we have $\MIo\cong\AIo\rtimes_{\sio}\bbZ_n$.
We want to show that this is similar for any $I\in\Jz$.

\begin{lemma}
For any $I\in\Jz$ there is a unitary $W\in A(I)$ such
that $\tilde{\sigma}_{0;I}=\Ad(W^*)\circ\sioI\in\DelAI$
fulfills $\tilde{\sigma}_{0;I}^n=\id$.
\end{lemma}

\bproof
Since the irreducible fixed point $\rho_0$ is transportable
there is a unitary $U_{\rho_0;\Io,I}$ such that
$\rho_{0;I}=\Ad(U_{\rho_0;\Io,I})\circ\rho_0\in\DelAI$,
and hence
\[ \bearll
\sioI\circ\rho_{0;I} &= \Ad(U_I)\circ\sio\circ\Ad(U_{\rho_0;\Io,I})
\circ\rho_0 \\[.4em]
&= \Ad(U_I\sio(U_{\rho_0;\Io,I}))\circ\sio\circ\rho_0 \\[.4em]
&= \Ad(U_I\sio(U_{\rho_0;\Io,I}))\circ\rho_0 \\[.4em]
&= \Ad(U_I\sio(U_{\rho_0;\Io,I})U_I^*)\circ\rho_{0;I} \,.
\eear \]
Now $W=U_I\sio(U_{\rho_0;\Io,I})U_I^*\in A(I)$ by Haag duality,
hence $\rho_{0;I}$ is an irreducible fixed point for $\sioI$.
Then, by the same argument as in Lemma \ref{simfix} we find
that $\tilde{\sigma}_{0;I}=\Ad(W^*)\circ\sioI$
fulfills $\tilde{\sigma}_{0;I}^n=\id$.
\eproof

Now we take this unitary $W$ such that
$\tilde{\sigma}_{0;I}=\Ad(W^*)\circ\sioI$ and
$\tilde{\sigma}_{0;I}^n=\id$, and then we define
$\tilde{U}_I=W^*U_I$ and set
\[ \fUIt = \fUI \pi(W) \in M(I) \,. \]
Then it follows from \erf{fields} that
\be
\fUIt^* \, \pi(a) \, \fUIt = \pi \circ
\tilde{\sigma}_{0;I} (a) \,, \qquad a\in\cA \,.
\labl{gields}

\begin{lemma}
With a suitable choice of the phase of $W$ we have $\fUIt^n=\bfe$.
\end{lemma}

\bproof
We have $\sio=\Ad(\tilde{U}_I)\circ\tilde{\sigma}_{0;I}$. Choose
$J\in\Jz$ such that $J\supset I\cup\Io$. Then for any $a\in A(J)$
we find
\[ a = \sio^n(a) = (\Ad(\tilde{U}_I)\circ\tilde{\sigma}_{0;I})^n(a)=
\Ad(X)\circ\tilde{\sigma}_{0;I}^n=XaX^* \,, \]
where $X=\tilde{U}_I\tilde{\sigma}_{0;I}(\tilde{U}_I)
\tilde{\sigma}_{0;I}^2 (\tilde{U}_I) \cdots
\tilde{\sigma}_{0;I}^{n-1}(\tilde{U}_I)$, and therefore
$X\in A(J)' \cap A(J) =\bbC\bfe$. If $X=\xi\bfe$, $\xi\in\bbC$,
the we can replace $W$ by $\xi^{1/n}W$ i.e.\
$\tilde{U}_I$ by $\xi^{-1/n}\tilde{U}_I$
to achieve $X=\bfe$. Now we compute
\[ \bearll
\fUIt^n &= f_\bfe \, \pi (\tilde{U}_I) \, \fUIt^{n-1} \\[.4em]
&= f_\bfe \, \fUIt^{n-1} \,
   \pi(\tilde{\sigma}_{0;I}^{n-1} (\tilde{U}_I)) \\[.4em]
&= f_\bfe f_\bfe \, \pi (\tilde{U}_I) \, \fUIt^{n-2} \,
   \pi(\tilde{\sigma}_{0;I}^{n-1} (\tilde{U}_I)) \\[.4em]
&= f_\bfe^2 \, \fUIt^{n-2} \, \pi(\tilde{\sigma}_{0;I}^{n-2}
   (\tilde{U}_I) \tilde{\sigma}_{0;I}^{n-1} (\tilde{U}_I)) \\[.4em]
&= \ldots = f_\bfe^n \, \pi(X) = \bfe \,,
\eear \]
where we used \erf{gields}.
\eproof

\erf{gields} holds in particular for $a\in A(I)$, moreover,
$\tilde{\sigma}_{0;I}^p$ is outer for $p\neq 0$ (mod $n$),
and $\fUIt^n=\bfe$. By the uniqueness of the crossed
product we find

\begin{corollary}
We have $M(I)\cong A(I) \rtimes_{\tilde{\sigma}_{0;I}}\bbZ_n$
for any $I\in\Jz$. In particular, each $M(I)$ is a factor.
\end{corollary}

Let $\Omega_0\in\cH_0$ denote the vacuum vector. Then $\Omega_0$
is cyclic and separating for each local algebra $A(I)$. Let
$\Omega\in\cH$ denote the vector given by $\Omega_p=\del p0 \Omega_0$.
It is clear from our construction that $\Omega$ is cyclic and
separating for each $M(I)$, that is, our net $\cN\subset\cM$
is standard.

Fixing $U_I$ for any $I\in\Jz$ it is clear that each $m\in M(I)$
can be uniquely written as
\[ m= \sum_{p=0}^{n-1} x_p \, \fUIt^p \,, \qquad x_p\in N(I) \,. \]
Then the map
\[ E^{M(I)}_{N(I)}: M(I) \rightarrow N(I)\,, \qquad
m \mapsto E^{M(I)}_{N(I)}(m) = x_0 \,, \]
is a faithful normal conditional expectation. It also
satisfies $E^{M(J)}_{N(J)} |_{M(I)} = E^{M(I)}_{N(I)}$ for
$I\subset J$ and preserves the vector state
$\omega=\la\Omega,\cdot\,\Omega\ra$. We summarize the
discussion in the following

\begin{proposition}
The net $\cN\subset\cM$ is a standard net of subfactors
with a standard conditional expectation.
\end{proposition}

Note that $\cN\subset\cM$ is even a quantum field theoretical
net of subfactors by relative locality $M(I)\subset N(I')'$.
However, $\cM$ will in general not be local itself. The
requirement of locality of $\cM$ imposes restrictions on
our simple current $\sio$.

For $\lambda_0,\mu_0\in\DelAIo$ we denoted by $\lambda$ and $\mu$
the corresponding endomorphisms in $\DelNIo$. For the statistics
operators we use the notation
$\epspm \lambda\mu = \pi( \epsopm {\lambda_0}{\mu_0})$ (and
$\eps \lambda\mu= \epsp \lambda\mu$) as in the previous paper
\cite{boev1}. Recall that for disjoint intervals $I_1,I_2\in\Jz$
we write $I_2>I_1$ (respectively $I_2<I_1$) if $I_1$ lies
clockwise (respectively counter-clockwise) to $I_2$ relative
to the point ``at infinity'' $z$.

\begin{lemma}
For $I_1\cap I_2=\emptyset$ we have
$f_{U_{I_2}} f_{U_{I_1}} = \epspm \sigma\sigma
f_{U_{I_1}} f_{U_{I_2}}$ with the $+$-sign if $I_2>I_1$ and
the $-$-sign if $I_2<I_1$.
\end{lemma}

\bproof
We compute
\[ \bearll
f_{U_{I_2}} f_{U_{I_1}} &= f_\bfe \pi (U_{I_2}^*)
f_\bfe \pi (U_{I_2}^*) \\[.4em]
&= \pi (\sio^{-1} (U_{I_2}^*)\sio^{-2} (U_{I_1}^*)) \, f_\bfe^2 \\[.4em]
&= \pi (\sio^{-1} (U_{I_2}^*)\sio^{-2} (U_{I_1}^*)
\sio^{-2} (U_{I_2})\sio^{-1} (U_{I_1}))
\, f_\bfe \pi(U_{I_1})^* f_\bfe \pi(U_{I_2})^* \\[.4em]
&= \sigma^{-2} \circ\pi (\sio (U_{I_2}^*) U_{I_1}^*
U_{I_2} \sio (U_{I_1}))
\, f_{U_{I_1}} f_{U_{I_2}} \\[.4em]
&= \sigma^{-2} \circ\pi (\epsopm \sio\sio)
\, f_{U_{I_1}} f_{U_{I_2}} \\[.4em]
&= \sigma^{-2} (\epspm \sigma\sigma)
\, f_{U_{I_1}} f_{U_{I_2}} \,,
\eear \]
where we recognized the definition of the statistics operator
in Subsection 2.3 of \cite{boev1}, and $\epspm \sigma\sigma$
are just scalars since $\epsopm \sio\sio\in\sio^2(\AIo)'\cap\AIo$,
hence we can omit the symbol $\sigma^{-2}$.
\eproof

This leads us immediately to the following

\begin{corollary}
The net $\cM$ is local if and only if $\epso \sio\sio = \bfe$.
\lablth{localM}
\end{corollary}

In Subsection \ref{spst} we will use Corollary \ref{localM}
to analyze for which levels we have a local extended net if
we take for $\sio$ the simple current corresponding to the
weight $k\Lambda_{(1)}$ of the $\LSUn$ level $k$ theory.

For completeness we also add the following

\begin{proposition}
If the net $\cM$ is local then it is in fact Haag dual.
\end{proposition}

\bproof
Let $I\in\Jz$ be arbitrary. We have to show that
$\CM(I')'=M(I)$. As $\CM(I')'\supset M(I)$ follows
from locality we only have to show the reverse inclusion.
Thus assume $X\in\CM(I')'$, and we have to show that
$X\in M(I)$. Choose an interval $J\in\Jz$
such that $\Io\cup I\subset J$. Then in particular
$X\in\CM(J')'$ and therefore
$X\pi(a)=\pi(a)X$ for all $a\in\CA(J')$. This
reads in matrix components (corresponding to the
decomposition of $\cH$ into $n$ copies of $\cH_0$)
$X_{p,q}\sio^q(a)=\sio^p(a)X_{p,q}$, $p,q\in\bbZ_n$, but
$\sio$ acts trivially on $\CA(J')$ as $\Io\subset J$. Hence
$X_{p,q}\in\CA(J')'=A(J)$ by Haag duality of $\cA$.
Now choose $K\in\Jz$ such that $K\subset J'$. Then
we have in particular $X f_{U_K}= f_{U_K} X$. From
this we obtain for the matrix components
$X_{p,q+1} \sio^q(U_K^*)=\sio^{p-1}(U_K^*)X_{p-1,q}$,
$p,q\in\bbZ_n$, and hence
\[ \bearll
X_{p+1,q+1} &= \sio^p(U_K^*) X_{p,q} \sio^q (U_K) \\[.4em]
&= \sio^p (U_K^* \sio^{-p} (X_{p,q}) \sio^{q-p}(U_K)) \\[.4em]
&= \sio^p (U_K^* \cdot \sigma_{0;K} \circ \sio^{-p} (X_{p,q})
   \cdot \sio^{q-p}(U_K)) \\[.4em]
&= \sio^p (\sio^{1-p}(X_{p,q})U_K^*\sio^{q-p}(U_K)) \\[.4em]
&= \sio (X_{p,q}) \sio^p (\epso {\sio^{q-p}}\sio) \\[.4em]
&= \sio (X_{p,q}) \,
\eear \]
where we used that $\sio^{-p}(X_{p,q})\in A(J)$ since
$\Io\subset J$ and $\sigma_{0;K}$ acts trivially on $A(J)$
since $K\subset J'$, and also that
\[ \epso {\sio^{q-p}}\sio)=\epso\sio\sio \sio(\epso\sio\sio)
\cdots \sio^{q-p-1}(\epso\sio\sio)=\bfe \]
since $\epso\sio\sio=\bfe$ as $\cM$ is local. We conclude that
$X_{p+k,q+k}=\sio^k(X_{p,q})$, $p,q,k\in\bbZ_n$, and by
setting $\tilde{a}_p=X_{0,p}\in A(J)$ this means that $X$ can be
written as $X=\sum_{p\in\bbZ_n} \pi(\tilde{a}_p) f_\bfe^p$,
i.e.\ $X\in M(J)$, but then we can also alternatively write
$X=\sum_{p\in\bbZ_n} \pi(a_p) \fUI ^p$ with $a_p\in A(J)$
since also $I\subset J$. Because we assumed that $X\in\CM(I')'$
we must have in particular that $X\pi(b)=\pi(b)X$ whenever
$b\in\CA(I')$. Now
\[ X \, \pi(b) = \sum_{p\in\bbZ_n} \pi(a_p) \, \fUI^p \, \pi(b)
= \sum_{p\in\bbZ_n} \pi(a_p) \, \pi(b) \, \fUI^p \]
by relative locality of fields and observables, and
\[ \pi(b) \, X= \sum_{p\in\bbZ_n} \pi(b) \, \pi(a_p) \, \fUI^p \,, \]
hence
\[ \sum_{p\in\bbZ_n} \pi(a_p) \, \pi(b) \, \fUI^p  =
\sum_{p\in\bbZ_n} \pi(b) \, \pi(a_p) \, \fUI^p \,. \]
Multiplication by $\fUI^{-q}$ from the right and
application of the conditional expectation yields
$\pi(a_q)\pi(b)=\pi(b)\pi(a_q)$ for all $b\in\CA(I')$,
$q\in\bbZ_n$. It follows $a_q\in\CA(I')'=A(I)$, $q\in\bbZ_n$,
and therefore $X=\sum_{p\in\bbZ_n} \pi(a_p) \fUI^p\in M(I)$.
\eproof

\subsection{Endomorphisms of the extended net}

We have lifted endomorphisms $\lambda_0$ of $\cA$
to endomorphisms $\lambda$ of $\cN$ by
$\lambda\circ\pi=\pi\circ\lambda_0$. Next we consider the
$\alpha$-induced endomorphisms $\ala\in\End(\cM)$.
In the following we assume that $\cM$ is Haag dual,
i.e.\ that $\epso \sio\sio=\bfe$. For notation we refer
again to our previous paper \cite{boev1} and to
Subsection \ref{prelim2}.

\begin{lemma}
For $\lambda_0\in\DelAIo$ we have
$\ala(f_\bfe) = f_\bfe \eps \lambda\sigma$.
\lablth{dendinv}
\end{lemma}

\bproof
By applying $\can$ to $f_\bfe^* x f_\bfe = \sigma(x)$,
$x\in\cN$, we find
$\can(f_\bfe)\in\Hom_\NIo(\canr\circ\sigma,\canr)$.
Hence by the BFE, Eq.\ (22) of \cite{boev1}, we obtain
\[ \can(f_\bfe) \, \canr(\eps \lambda\sigma) \eps \lambda\canr
= \eps \lambda\canr \cdot \lambda\circ\can (f_\bfe) \,, \]
and therefore
\[ \ala (f_\bfe) \equiv \cani \circ \Ad (\eps \lambda\canr) \circ
\lambda \circ \can (f_\bfe) = f_\bfe \eps \lambda\sigma \,, \]
proving the lemma.
\eproof

Now we ask when $\ala$ is localized. For the sake of simplicity,
we restrict the discussion to irreducible $\lambda_0$.
Define the {\em monodromy} by
$Y(\lambda_0,\sio)=\epso {\lambda_0}\sio \epso \sio{\lambda_0}$.
Note that for irreducible $\lambda_0$ the monodromy
is a scalar as
$Y(\lambda_0,\sio)\in\sio\circ\lambda_0(\AIo)'\cap\AIo=\bbC\bfe$,
i.e.\ $Y(\lambda_0,\sio)=\omega \bfe$, $\omega\in\bbC$.
Therefore we have $\epso{\lambda_0}\sio =
Y(\lambda_0,\sio) \epso \sio{\lambda_0} ^*
= \omega \epsom {\lambda_0}\sio$.

\begin{lemma}
For $\lambda_0\in\DelAIo$ irreducible
$\ala$ is localized in $\Io$ if and only if the monodromy
$Y(\lambda_0,\sio)$ is trivial, i.e.\ $\omega=1$.
\lablth{lloc}
\end{lemma}

\bproof
It is clear that $\ala(x)\equiv\lambda(x)=x$ for any $x\in N(I)$
with $I\cap\Io=\emptyset$ since $\lambda_0\in\DelAIo$. Thus
we have to check whether $\ala(\fUI)=\fUI$ whenever
$I\cap\Io=\emptyset$. By definition
\[ \bearll
\ala(\fUI) &= \ala(f_\bfe)\lambda(\pi(U_I^*)) =
f_\bfe \pi(\epso {\lambda_0}\sio \lambda_0(U_I^*)) \\[.4em]
&= \fUI \pi (U_I\epso {\lambda_0}\sio \lambda_0(U_I^*)) \,.
\eear \]
For $I\in\Jz$ such that $I\cap\Io=\emptyset$ we distinguish
two cases.

Case 1: $I>\Io$. We can choose an interval $I_+\in\Jz$ such
that $I_+>\Io$ and $I_+>I$. Since $I>\Io$ we can choose
some $J_+\in\Jz$ such that $J_+\supset I\cup I_+$ but
$J_+\cap\Io=\emptyset$. For any unitary $\Usiop$ such that
$\sigma_{0,+}=\Ad(\Usiop)\circ\sio\in\Delta_\cA(I_+)$
the statistics operator can be written as
$\epso {\lambda_0}\sio = \Usiop^* \lambda_0(\Usiop)$.
Since $\sioI=\Ad(U_I)\circ\sio$ we have
$\sigma_{0,+}=\Ad(V_+)\circ\sioI$ with $V_+=\Usiop U_I^*$,
and hence $V_+\in A(J_+)$ by Haag duality. Then
\[ U_I \epso {\lambda_0}\sio \lambda_0 (U_I^*)
= V_+^* \lambda_0 (V_+) = V_+^* V_+ = \bfe \]
since $J_+\cap\Io=\emptyset$. Hence
$\ala(\fUI)=\fUI$ for $I>\Io$.

Case 2: $I<\Io$. Recall
$\epso {\lambda_0}\sio = \omega \epsom {\lambda_0}\sio$,
hence
\[ U_I \epso {\lambda_0}\sio \lambda_0 (U_I^*) =
\omega U_I \epsom {\lambda_0}\sio \lambda_0 (U_I^*) \,. \]
We can choose an interval $I_-\in\Jz$ such
that $I_-<\Io$ and $I_-<I$. Since $I<\Io$ we can choose
some $J_-\in\Jz$ such that $J_-\supset I\cup I_-$ but
$J_-\cap\Io=\emptyset$. For any unitary $\Usiom$ such that
$\sigma_{0,-}=\Ad(\Usiom)\circ\sio\in\Delta_\cA(I_-)$
the statistics operator can be written as
$\epsom {\lambda_0}\sio = \Usiom^* \lambda_0(\Usiom)$.
Then
$\sigma_{0,-}=\Ad(V_-)\circ\sioI$ with
$V_-=\Usiom U_I^*\in A(J_-)$, and
\[ U_I \epsom {\lambda_0}\sio \lambda_0 (U_I^*)
= V_-^* \lambda_0 (V_-) = V_-^* V_- = \bfe\,, \]
and hence
$\ala(\fUI)= \omega\fUI$ for $I<\Io$. The
statement follows.
\eproof

The next step is the transportability. Note that
$\DelMIoO=\DelMIo$ since $\cM$ is Haag dual.
We have the following (cf.\ Prop.\ 5.2 in \cite{reh2})

\begin{lemma}
For $\lambda_0\in\DelAIo$ irreducible we have
$\ala\in\DelMIo$ if and only if the monodromy
$Y(\lambda_0,\sio)$ is trivial, i.e.\ $\omega=1$.
\lablth{blaMIo}
\end{lemma}

\bproof
After Lemma \ref{lloc} all we have to
show is that $\ala$ is transportable if $\omega=1$. Since
$\lambda_0\in\DelAIo$ there is for any $J\in\Jz$
a unitary $U\equiv U_{\lambda_0;\Io,J}$ such that
$\tilde{\lambda}_0\in\Delta_\cA(J)$. Define
$\tilde{\alpha}_\lambda=\Ad(\pi(U))\circ\ala$. It is clear
that $\tilde{\alpha}_\lambda(x)=x$ whenever $x\in N(I)$ with
$I\cap J=\emptyset$. We show that also
$\tilde{\alpha}_\lambda(\fUI)=\fUI$ in that case.
We again distinguish two cases.

Case 1: $I>J$. We choose $I_+\in\Jz$ such that
$I_+>\Io$ and $I_+>J$. Since $I>J$ there is a
$K_+\in\Jz$ such that $K_+\supset I_+\cup I$ but
$K_+\cap J=\emptyset$. As before, we choose a unitary
$\Usiop$ such that $\sigma_{0,+}=\Ad(\Usiop)\circ\sio\in\Delta_\cA(I_+)$.
Then $\sigma_{0,+}=\Ad(V_+)\circ\sioI$ and therefore
$V_+=\Usiop U_I^*\in A(K_+)$,
hence $\tilde{\lambda}_0(V_+)=V_+$.
Since $I_+>\Io$ and $I_+>J$ we also have $\sigma_{0,+}(U)=U$.
Now we compute
\[ \bearll
\tilde{\alpha}_\lambda(\fUI)
&= \Ad (\pi(U)) \circ \ala(\fUI) \\[.4em]
&= \pi(U) \, \fUI \, \pi(U_I \epso {\lambda_0}\sio
   \lambda_0 (U_I^*) U^* ) \\[.4em]
&= \fUI \, \pi( \sioI(U) U_I \epso {\lambda_0}\sio
   \lambda_0 (U_I^*) U^* ) \\[.4em]
&= \fUI \, \pi( U_I \sio(U) \epso {\lambda_0}\sio
   U^* \tilde{\lambda}_0 (U_I^*)) \\[.4em]
&= \fUI \, \pi( U_I \sio(U) \Usiop^* \lambda_0(\Usiop)
   U^* \tilde{\lambda}_0 (U_I^*)) \\[.4em]
&= \fUI \, \pi( U_I \Usiop^* \sigma_{0,+} (U) U^*
   \tilde{\lambda}_0 (\Usiop) \tilde{\lambda}_0 (U_I^*)) \\[.4em]
&= \fUI \, \pi(V_+^* \tilde{\lambda}_0 (V_+))
   = \fUI \,.
\eear \]

Case 2: $I<J$. We choose $I_-\in\Jz$ such that
$I_-<\Io$ and $I_-<J$. Since $I<J$ there is a
$K_-\in\Jz$ such that $K_-\supset I_-\cup I$ but
$K_-\cap J=\emptyset$. Let
$\sigma_{0,-}=\Ad(\Usiom)\circ\sio\in\Delta_\cA(I_-)$.
Then $V_-=\Usiop U_I^*\in A(K_-)$,
hence $\tilde{\lambda}_0(V_-)=V_-$;
since $I_-<\Io$ and $I_-<J$ we also have $\sigma_{0,-}(U)=U$.
If $\omega=1$ then $\epso {\lambda_0}\sio = \epsom {\lambda_0}\sio$,
so we can compute analogously
\[ \bearll
\tilde{\alpha}_\lambda(\fUI)
&= \fUI \, \pi( U_I \sio(U) \epsom {\lambda_0}\sio
   U^* \tilde{\lambda}_0 (U_I^*)) \\[.4em]
&= \fUI \, \pi( U_I \sio(U) \Usiom^* \lambda_0(\Usiom)
   U^* \tilde{\lambda}_0 (U_I^*)) \\[.4em]
&= \fUI \, \pi( U_I \Usiom^* \sigma_{0,-} (U) U^*
   \tilde{\lambda}_0 (\Usiom) \tilde{\lambda}_0 (U_I^*)) \\[.4em]
&= \fUI \, \pi(V_-^* \tilde{\lambda}_0 (V_-))
   = \fUI \,.
\eear \]
We have shown that $\tilde{\alpha}_\lambda$ is localized in $J$. Since
$J\in\Jz$ was arbitrary it follows $\ala\in\DelMIo$.
\eproof

Our construction of the net $\cM$ is such that (Proposition 2.10
and the discussion in Subsection 2.4 in \cite{boev1})
\be
[\canr] = \bigoplus_{p=0}^{n-1} \,\, [\sigma^p] \,.
\labl{canrorb}

\begin{lemma}
For any $\lambda_0\in\DelAIo$ we have
\be
\Hom_\MIo(\ala,\ala) = \left\{ t=\sum_{p=0}^{n-1} \pi(T_p) f_\bfe^p
\,,\,\,\, T_p \in \Hom_\AIo(\sio^{-p}\circ\lambda_0,\lambda_0) \right\} \,.
\labl{Hombla}
\lablth{bladec}
\end{lemma}

\bproof
Suppose $t\in\Hom_\MIo(\ala,\ala)$. We can write
$t=\sum_{p=0}^{n-1} \pi(T_p) f_\bfe^p$ with
$T_p\in\AIo$. Now from
$t\cdot\ala\circ\pi(a)=\ala\circ\pi(a)\cdot t$ for all
$a\in\AIo$ we obtain
\[ \sum_{p=0}^{n-1} \pi(T_p) f_\bfe^p \cdot \pi\circ\lambda_0(a)\equiv
\sum_{p=0}^{n-1} \pi(T_p \cdot \sio^{-p} \circ\lambda_0(a) )f_\bfe^p
= \sum_{p=0}^{n-1} \pi( \lambda_0(a) T_p )f_\bfe^p \,. \]
It follows $T_p\in\Hom_\AIo(\sio^{-p}\circ\lambda_0,\lambda_0)$.
It remains to be shown that then
$t \ala (f_\bfe) = \ala (f_\bfe) t$. From the
BFE, Eq.\ (11) in \cite{boev1}, we obtain
\[ \sio(T_p) \, \epso {\sio^{-p}}\sio \sio^{-p} ( \epso {\lambda_0}\sio )
= \epso {\lambda_0}\sio \, T_p \]
But
\[ \epso {\sio^{-p}}\sio \equiv \epso {\sio^{n-p}}\sio =
\epso \sio\sio \sio( \epso \sio\sio ) \cdots \sio^{n-p-1}
( \epso \sio\sio ) = \bfe \]
as $\epso \sio\sio=\bfe$. Hence we find
\[ \sio(T_p) \, \sio^{-p} (\epso {\lambda_0}\sio)
= \epso {\lambda_0}\sio \, T_p \,. \]
Now we compute
\[ \bearll
t\, \ala (f_\bfe) &= \sum_{p=0}^{n-1} \pi(T_p) \, f_\bfe^{p+1}
\eps \lambda\sigma \\[.4em]
&= \sum_{p=0}^{n-1} f_\bfe \cdot \pi\circ\sio(T_p) \cdot f_\bfe^p
\eps \lambda\sigma \\[.4em]
&= \sum_{p=0}^{n-1} f_\bfe \cdot \pi\circ\sio(T_p) \cdot \sigma^{-p}
(\eps \lambda\sigma) f_\bfe^p \\[.4em]
&= \sum_{p=0}^{n-1} f_\bfe \, \pi (\sio (T_p) \sio^{-p}
(\epso {\lambda_0}\sio)) \, f_\bfe^p \\[.4em]
&= \sum_{p=0}^{n-1} f_\bfe \, \pi (\epso {\lambda_0}\sio T_p) \,
f_\bfe^p \\[.4em]
&= \sum_{p=0}^{n-1} f_\bfe \eps \lambda\sigma \, \pi(T_p)
f_\bfe^p \\[.4em]
&= \ala (f_\bfe) \, t \,,
\eear \]
thus we have indeed \erf{Hombla}.
\eproof

For the fixed point $\rho_0$ of $\sio$ we have
$\Hom_\AIo(\sio\circ\rho_0,\rho_0)=\bbC\bfe$.
By Lemma \ref{bladec} it is not hard to see that then
$\Hom_\MIo(\aro,\aro)$ is an $n$-di\-men\-sio\-nal commutative
algebra, i.e.\
$\Hom_\MIo(\aro,\aro)\cong\bbC\oplus\bbC\oplus\ldots\oplus\bbC$,
and therefore
$[\aro]$ decomposes in $n$ distinct irreducible sectors.
Since $\sigma_\aro=\can\circ\aro|_\cN=\canr\circ\rho$
and thus $[\sigma_\aro]=[\canr\circ\rho]=n\,[\rho]$
we arrive at the following

\begin{corollary}
We have
$[\aro]=\bigoplus_{p=0}^{n-1}[\delta_p]$ with $[\delta_p]$
distinct and irreducible.  Moreover, $[\sigma_{\delta_p}]=[\rho]$
for all $p=0,1,\ldots,n-1$.
\end{corollary}

\subsection{Spin and statistics}
\lablsec{spst}

We found that the extended net is local (and even Haag dual)
if and only if $\epso \sio\sio=\bfe$. In fact
$\epso \sio\sio$ can be computed by the spin and statistics
connection. In the following the conformal dimensions
$h_\Lambda$, which are by definition the lowest eigenvalues
of the rotation generator $L_0$ in the \per s
$(\pi_\Lambda,\cH_\Lambda)$, $\Lambda\in\ASU nk$, will
play an important role. They are given by
\[ h_\Lambda = \frac{(\Lambda|\Lambda + 2 \rho)}{2(k+n)} \,, \]
where $\rho=\sum_{i=1}^{n-1} \Lambda_{(i)}$ and
$(\cdot|\cdot)$ is the symmetric bilinear form. Recalling that
$(\Lambda_{(i)}|\Lambda_{(j)})=i(n-j)/n$ for $1\le i\le j\le n-1$
one may obtain for $\Lambda=\sum_{i=1}^{n-1} m_i \Lambda_{(i)}$
\be
h_\Lambda = \sum_{1\le i\le j\le n-1} m_i m_j \frac{i(n-j)}{n(k+n)}
\,\,- \,\,\sum_{i=1}^{n-1} m_i^2 \frac{i(n-i)}{2n(k+n)}
\,\,+ \,\,\sum_{i=1}^{n-1} m_i \frac{i(n-i)}{2(k+n)} \,,
\labl{cdim}
where we used the Dynkin labelling, i.e.\ $m_i\in\bbN_0$ and
$\sum_{i=1}^{n-1} m_i \le k$. Now let
$\lambda_{0;\Lambda}\in\DelAIo$ denote the endomorphisms
corresponding to the \per s $(\pi_\Lambda,\cH_\Lambda)$,
$\Lambda\in\ASU nk$. Then $\sio=\lambda_{0;k\Lambda_{(1)}}$
is a simple current of order $n$, and its fusion rules correspond
to the $\bbZ_n$-rotation of $\ASU nk$. It has a fixed point
if $k$ is a multiple of $n$, namely
$\rho_0=\lambda_{0;\Lambda_R}$, where $\Lambda_R=\frac kn \Lambda_{(1)}
+ \frac kn \Lambda_{(2)} + \ldots + \frac kn \Lambda_{(n-1)}$.
Therefore we first require
$k\in n\bbN$ so that we can construct the extended net $\cM$
by means of $\sio$ as explained in the previous subsections.
Then we ask when $\cM$ is local.

\begin{proposition}
The net $\cM$ is local if and only if $k\in 2n\bbN$
if $n$ is even and $k\in n\bbN$ if $n$ is odd.
\lablth{locSUn}
\end{proposition}

\bproof
By Corollary \ref{localM} the net $\cM$ is local if and
only if $\epso \sio\sio=\bfe$. Since $\sio$ is an
automorphism we have $\epso \sio\sio=\kappa_{\sio}\bfe$,
where $\kappa_{\sio}\in\bbC$ is the statistical phase.
By the conformal spin and statistics theorem \cite{gulo2}
we have $\kappa_{\sio}=\E^{2\pi\I h_{\sio}}$ where $h_{\sio}$
is the infimum of the spectrum of the rotation generator $L_0$
in the representation $\pio\circ\sio$. But this is the conformal
dimension, $h_{\sio}=h_{k\Lambda_{(1)}}$, and by \erf{cdim}
\[ h_{k\Lambda_{(1)}}= k  \, \frac{n-1}{2n} \,. \]
Therefore $\epso \sio\sio=\bfe$ if and only if
$k(n-1)/2n\in\bbN$, the statement follows.
\eproof

The next step is to ask for which $\Lambda\in\ASU nk$ we have
$\alpha_\Lambda\equiv\alpha_{\lambda_\Lambda}\in\DelMIo$. For
$\Lambda=m_1\Lambda_{(1)}+m_2\Lambda_{(2)}+...+m_{n-1}\Lambda_{(n-1)}$
we denote $|\Lambda|=\sum_{i=1}^{n-1} i\,m_i$. Recall that the
$\bbZ_n$-rotation $\sigma$ on $\ASU nk$ is defined by
\[ \sigma(\Lambda) = (k-m_1-\ldots - m_{n-1}) \Lambda_{(1)}
+ m_1 \Lambda_{(2)} + m_2 \Lambda_{(3)} + \ldots +
m_{n-2} \Lambda_{(n-1)} \,.\]

\begin{proposition}
We have $\alpha_\Lambda\in\DelMIo$ if and only if
$|\Lambda|\in n\bbZ$.
\lablth{mave}
\end{proposition}

\bproof
By Lemma \ref{blaMIo} we have $\alpha_\Lambda\in\DelMIo$ if and only if
$Y(\lambda_{0;\Lambda},\sio)=\bfe$. By Lemma 3.3 of \cite{frs2}
we have for any $T\in\Hom_\AIo(\lambda_{0;\sigma(\Lambda)},\sio
\circ\lambda_{0;\Lambda})$
\[ Y(\lambda_{0;\Lambda},\sio) T =
\frac{\kappa_{\lambda_{0;\sigma(\Lambda)}}}{\kappa_{\sio} \,
\kappa_{\lambda_{0;\Lambda}}} T \,, \]
where the $\kappa$'s are statistical phases.
Since $[\lambda_{0;\sigma(\Lambda)}]=[\sio\circ\lambda_{0,\Lambda}]$
and since $\lambda_{0;\Lambda}$ is irreducible we can take
$T$ unitary, hence
\[  Y(\lambda_{0;\Lambda},\sio) =
\frac{\kappa_{\lambda_{0;\sigma(\Lambda)}}}{\kappa_{\sio} \,
\kappa_{\lambda_{0;\Lambda}}} \bfe \,. \]
Using again the conformal spin and statistics theorem we find
\[ \frac{\kappa_{\lambda_{0;\sigma(\Lambda)}}}{\kappa_{\sio} \,
\kappa_{\lambda_{0;\Lambda}}} =
\E^{2\pi\I (h_{\lambda_{0;\sigma(\Lambda)}} -
h_{\sio} - h_{\lambda_{0;\Lambda}})} \equiv
\E^{2\pi\I (h_{\sigma(\Lambda)} - h_{k\Lambda_{(1)}} -
h_\Lambda )} \,. \]
Now by Lemma 2.7 of \cite{kota} we have
\[ h_{\sigma(\Lambda)} - h_\Lambda = \frac 1n \left(
\frac{(n-1)k}2 - |\Lambda| \right) \,, \]
hence $h_{\sigma(\Lambda)} - h_{k\Lambda_{(1)}} -
h_\Lambda=-|\Lambda|/n$. Therefore
$Y(\lambda_{0;\Lambda},\sio) = \bfe$ if and only if
$|\Lambda|\in n\bbZ$.
\eproof

{\it Remark.} If we label the \per s of $\LSUn$ at level $k$
by partitions (or Young tableaux) $(p_1,p_2,...,p_{n-1})$
with $p_i=\sum_{j=i}^{n-1} m_j$ then Proposition \ref{mave}
reads $\alpha_{(p_1,\ldots,p_{n-1})}\in\DelMIo$ if and only
if $\sum_{i=1}^{n-1} p_i\in n\bbZ$.

By Proposition \ref{mave} it should be clear that for
the orbifold modular invariants the sectors corresponding to
the marked vertices are (the irreducible subsectors of)
$\alpha_{(p_1,\ldots,p_{n-1})}$ with
$\sum_{i=1}^{n-1} p_i\in n\bbZ$, as these $\alpha$-induced
endomorphisms are localized and transportable endomorphisms
of the extended net $\cM$. Moreover,
as we will see by the treatment of the examples,
their $\sigma$-restriction corresponds to the block structure
of the corresponding orbifold modular invariants.
The $\SUn_k$ sectors that do not appear in the blocks of
the modular invariant can be identified as ``twisted sectors''
if we consider the $\SUn_k$ theory as the $\bbZ_n$ orbifold
of the extended theory. In fact, $\alpha$-induction of these
sectors does not provide localized sectors; here we only
obtain ``solitonic'' localization of the $\alpha$-induced
endomorphisms.

For $\SUz$ the \per s are labelled by the spin
$j\equiv m_1=0,1,2,...,k$. Then \erf{cdim} reduces to
\[ h_j = \frac{j(j+2)}{4k+8} \,. \]
First we find by Proposition \ref{locSUn} that we can construct
the local extended net for $k=4\varrho$, $\varrho\in\bbN$, since
then $h_k=\varrho\in\bbZ$. The rotation $\sigma$ is now the flip
$\sigma(j)=k-j$. Hence
\[ h_{\sigma(j)} - h_j = \frac{(k-j)(k-j+2)}{4k+8} - \frac{j(j+2)}{4k+8}
= \frac{k-2j}4 = \varrho - \frac j2 \,, \]
i.e.\ $\alpha_j\in\DelMIo$ if and only if $j\in2\bbZ$.

\subsection{Examples}
\lablsec{Dex}

We now consider some examples for the application of
$\alpha$-induction to the extended net coming from an
orbifold block-diagonal modular invariant. The simplest case is
the $\SUz$ \ddD 4 modular invariant but we have already discussed
this case as the extended net here coincides with the net
associated to the $\SUd_1$ theory. The \ddD {2\varrho+2}
modular invariants with $\varrho>1$ do not come from conformal
inclusions. They appear at level $k=4\varrho$ and can be
written as
\[ Z_{\mathrm{D}_{2\varrho+2}} = \frac12
\sum_{k \ge j \ge 0 \atop j\in 2\bbZ}
| \chi_j + \chi_{k-j} |^2 \,. \]
Let us first illustrate this at the next case in the
D-series, namely \ddD 6. The \ddD 6 invariant appears at level 8,
thus we start with the fusion algebra $W(2,8)$. The simple current
is given by $\ls 8$ and indeed $h_8=2$. \erf{canrorb} now reads
$[\canr]=\ls 0 \oplus \ls 8$ and from this we get immediately
that $\as 4$ decomposes into two irreducible sectors, say
$\as 4 = \asx 4 1 \oplus \asx 4 2$, all other $\as j$
are irreducible and $\as {8-j}=\as j$. The fusion rules
involving $\as j$, $j=0,1,2,3$, can be read off from those
of $\ls j$ by the homomorphism property of $\alpha$-induction,
so one only has to find the fusion rules involving $\asx 4i$,
$i=1,2$. One checks that the following fusion rules,
\[ \bearll
\as 1 \times \asx 4 i &=  \as 3 \,,\qquad i=1,2\,,\\[.4em]
\as 2 \times \asx 4 i &= \as 2 \oplus \asx 4 {i+1} \,,
\qquad i=1,2 \,\,\,(\mod \,\,2)\,,\\[.4em]
\as 3 \times \asx 4 i &= \as 1 \oplus \as 3 \,,\qquad i=1,2\,,\\[.4em]
\asx 4 i \times \asx 4 i &= \as 0 \oplus \asx 4 i \,,
\qquad i=1,2\,,\\[.4em]
\asx 4 1 \times \asx 4 2 &= \as 2 \,,
\eear \]
determine a well-defined fusion algebra with unit $\as 0$.
The fusion graph of $\as 1$ is easily seen to be
\ddD 6, see Fig.\ \ref{D6}.
\thinlines
\setlength{\unitlength}{3pt}
\begin{figure}[tb]
\begin{center}
\begin{picture}(70,30)
\put(8,10){\line(1,0){42}}
\put(50,10){\line(1,1){10}}
\put(50,10){\line(1,-1){10}}
\put(8,10){\circle*{1}}
\put(22,10){\circle*{1}}
\put(36,10){\circle*{1}}
\put(50,10){\circle*{1}}
\put(60,20){\circle*{1}}
\put(60,0){\circle*{1}}
\put(8,10){\circle{2.5}}
\put(36,10){\circle{2.5}}
\put(60,20){\circle{2.5}}
\put(60,0){\circle{2.5}}
\put(5,5){$\as 0$}
\put(19,5){$\as 1$}
\put(33,5){$\as 2$}
\put(46,5){$\as 3$}
\put(63,19){$\asx 4 1$}
\put(63,-1){$\asx 4 2$}
\end{picture}
\end{center}
\caption{\ddD 6}
\label{D6}
\end{figure}

For arbitrary $\varrho=1,2,3,...$, $k=4\varrho$, the fusion
algebra can be characterized as follows. We have $2\varrho+2$
irreducible sectors $\as j$, $j=0,1,2,...,2\varrho-1$, and
$\asx {2\varrho} 1$ and $\asx {2\varrho} 2$. The fusion rules
are given from those in $W(2,4\varrho)$, see \erf{fusuz},
i.e.
\[ \as {j_1} \times \as {j_2} =
\bigoplus_{j=|j_1-j_2|\,,\,\,j+j_1+j_2\,\,\mathrm{even}}
^{\mathrm{min} (j_1+j_2,2k-(j_1+j_2))} \as j \,,\]
for $j=0,1,2,...,2\varrho$, where we identify $\as {k-j}=\as j$
and $\as {2\varrho} = \asx {2\varrho} 1 \oplus \asx {2\varrho} 2$
on the r.h.s. Thus associativity, the homomorphism
property of $[\alpha]$ and compatibility with
$\la\alpha_j,\alpha_{j'}\ra_\MIo
=\la\canr\circ\lambda_j,\lambda_{j'}\ra_\NIo$
where $[\canr]=\ls 0 \oplus \ls k$ are
automatically guaranteed, and the fusion graph of $\as 1$ is
already determined to be \ddD {2\varrho+2}. We only have to
specify the fusion rules involving the isolated
$\asx {2\varrho} i$, $i=1,2$. But it is shown in \cite{izu1} that
the fusion graph \ddD {2\varrho+2} of $\as 1$ already determines
all the (endomorphism) fusion rules; they are given by
\[
\as j \times \asx {2\varrho} i = \left\{ \bearll
\as {2\varrho-j} \oplus \as {2\varrho-j+2} \oplus \ldots
\oplus \as {2\varrho-3} \oplus \as {2\varrho-1}\,,
& j\in 2\bbZ+1  \\[.4em]
\as {2\varrho-j} \oplus \as {2\varrho-j+2} \oplus \ldots
\oplus \as {2\varrho-2} \oplus \asx {2\varrho} i \,,
& j\in 4\bbZ  \\[.4em]
\as {2\varrho-j} \oplus \as {2\varrho-j+2} \oplus \ldots
\oplus \as {2\varrho-2} \oplus \asx {2\varrho} {i+1} \,,
& j\in 4\bbZ+2  \eear \right.
\]
for $0<j<2\varrho$ and $i=1,2\,\,(\mod\,\,2)$. Of course
$\as 0 \times \as {2\varrho} _\pm = \as {2\varrho} _\pm$, and
\[ \bearll
\asx {2\varrho} i \times \asx {2\varrho} i \!\!\!\!
&= \left\{ \bearll
\as 0 \oplus \as 4 \oplus \ldots \oplus
\as {2\varrho-4} \oplus \asx {2\varrho} i \,,\,
& \varrho = 2,4,6,...  \\[.4em]
\as 2 \oplus \as 6 \oplus \ldots \oplus
\as {2\varrho-4} \oplus \asx {2\varrho} {i+1} \,,
& \varrho = 1,3,5,...  \eear \right.
\\[1.4em]
\asx {2\varrho} i \times \asx {2\varrho} {i+1} \!\!\!\!
&= \left\{ \bearll
\as 2 \oplus \as 6 \oplus \ldots \oplus
\as {2\varrho-6} \oplus \as {2\varrho-2} \,,
& \varrho = 2,4,6,...  \\[.4em]
\as 0 \oplus \as 4 \oplus \ldots \oplus
\as {2\varrho-6} \oplus \as {2\varrho-2} \,,
& \varrho = 1,3,5,...  \eear \right.
\eear \]
for $i=1,2\,\,(\mod\,\,2)$.

Next we consider the \ddDgx {3\varrho+3},
$\varrho\in\bbN$, (block-diagonal) modular invariant that
appears at level $k=3\varrho$,
\[ Z_{\mathcal{D}^{(3\varrho+3)}} = \frac13
\sum_{k\ge p \ge q \ge 0 \atop p+q\in 3\bbZ} | \chi_{(p,q)}
+ \chi_{\sigma(p,q)} + \chi_{\sigma^2(p,q)} |^2 \,, \]
where $\sigma$ is the $\bbZ_3$ rotation of the \ddAgx {k+3}
graph,
\[ \sigma(p,q) = (k-q,p-q)\,,\qquad \sigma^2 (p,q) =
(k-p+q,k-p) \,. \]
This is an orbifold invariant and it can be treated completely
analogously to the \Deven\ invariants of $\SUz$. The vacuum block
gives us the $[\canr]$,
\[ [\canr] = \lasd 00 \oplus \lasd k0 \oplus \lasd kk \,.\]
Using the $\LSUd$ fusion rules at level $k$, in particular those
for the simple currents
\[ \lasd pq \times \lasd k0 = [\lambda_{\sigma(p,q)}] \,, \qquad
\lasd pq \times \lasd kk = [\lambda_{\sigma^2(p,q)}] \,, \]
we find
\[ \bearll \la \asd pq , \asd rs \ra &= \del {(p,q)}{(r,s)} +
\del {\sigma(p,q)}{(r,s)} + \del {\sigma^2(p,q)}{(r,s)} \\[.5em]
&= \del pr \del qs + \del {k-q}r \del {p-q}s
+ \del {k-p+q}r \del {k-p}s \eear \,.\]
Hence we have identifications
\[ \asd pq = [\alpha_{\sigma(p,q)}]=[\alpha_{\sigma^2(p,q)}] \]
and all $\asd pq$ are irreducible apart from the fixed point
$(p,q)=(2\varrho,\varrho)$, where
$\la \asd {2\varrho}\varrho,\asd {2\varrho}\varrho \ra= 3$,
so that it decomposes into three irreducible sectors as follows,
\[ \asd {2\varrho}\varrho = \asdx {2\varrho}\varrho 1 \oplus
\asdx {2\varrho}\varrho 2 \oplus \asdx {2\varrho}\varrho 3 \,.\]
One easily checks that the fusion graphs of $\asd 10$ are
the orbifold graphs \ddDgx {k+3} which were first discovered
in \cite{kost} in the context of statistical mechanical
models and then in \cite{evka2} in the subfactor context.

\section{Graphs and intertwining matrices}

In this section we define several fusion matrices and study
their properties. Using some ideas of Xu \cite{xu1}, we
establish identities between these matrices
which allow to identify them with certain matrices considered
by Di Francesco and Zuber.

\subsection{Some matrices and their properties}

Let again $\cW\equiv\cW(n,k)=\{[\lambda_\Lambda], \Lambda\in\ASU nk\}$
be the canonical sector basis for $\SUn_k$. Recall that the
structure constants of $W$ can be written as
\[ \N \Lambda{\Lambda'}{\Lambda''} =
\la \lambda_\Lambda\circ\lambda_{\Lambda'},\lambda_{\Lambda''}\ra_\NIo
\,, \qquad \Lambda,\Lambda',\Lambda''\in\ASU nk \,,\]
and this defines matrices $N_\Lambda$ by
$(N_\Lambda)_{\Lambda',\Lambda''}=
\N {\Lambda'}\Lambda{\Lambda''}$. We set
$A_p=N_{\Lambda_{(p)}}$,
for the fundamental weights $\Lambda_{(p)}$, $p=1,2,...,n-1$.
Note that $A_1$ is the adjacency matrix of the first fusion graph
of the fundamental representation, i.e. of $\ASU nk$ considered
as a graph. For either a conformal inclusion or
an orbifold inclusion as discussed above let $V$ denote
the sector algebra with basis $\cV$ obtained by
$\alpha$-induction. We denote
$\alpha_\Lambda\equiv\alpha_{\lambda_\Lambda}$.
First of all we claim

\begin{lemma}
For either a conformal or an orbifold inclusion,
$\alf 1$ is always irreducible.
\end{lemma}

\bproof
Irreducibility of $\alf 1$ means that
$\la \alf 1, \alf 1 \ra_\MIo=1$. We have
\[ \bearll
\la \alf 1,\alf 1 \ra_\MIo &= \la \canr\circ\lambda_{\Lambda_{(1)}},
\lambda_{\Lambda_{(1)}} \ra_\NIo = \la \canr,
\lambda_{\Lambda_{(1)}} \circ
\lambda_{\Lambda_{(n-1)}} \ra_\NIo \\[.4em]
&= 1 + \la \canr , \lambda_{\Lambda_{(1)}+\Lambda_{(n-1)}} \ra_\NIo \,,
\eear \]
as $[\overline{\lambda}_{\Lambda_{(1)}}]=[\lambda_{\Lambda_{(n-1)}}]$
and $[\lambda_{\Lambda_{(1)}}]\times[\lambda_{\Lambda_{(n-1)}}]=
[\lambda_0]\oplus[\lambda_{\Lambda_{(1)}+\Lambda_{(n-1)}}]$ and
since $[\lambda_0]=[\id]$ appears in the decomposition of $[\canr]$
precisely once. Using the formula for the conformal dimension,
\erf{cdim}, one checks that
$h_{\Lambda_{(1)}+\Lambda_{(n-1)}}=n/(k+n)\notin\bbZ$. However, all
subsectors of $[\canr]$ must have integer conformal dimension (and
this corresponds to T-invariance in the modular invariant picture):
For the conformal inclusion case, the decomposition of $[\canr]$
corresponds to the decomposition of the restricted vacuum
representation. In the orbifold inclusion case we have
$[\canr]=\bigoplus_{p=0}^{n-1}[\sigma^p]$,
$[\sigma^p]=[\lambda_{k\Lambda_{(p)}}]$, and
$h_{k\Lambda_{(p)}}=kp(n-p)/2n\in\bbZ$ as $k\in 2n\bbN$
if $n$ is even and $k\in n\bbN$ if $n$ is odd. We
conclude that
$\la \canr , \lambda_{\Lambda_{(1)}+\Lambda_{(n-1)}} \ra_\NIo=0$,
proving irreducibility of $\alf 1$.
\eproof

We define the following collection of non-negative integers,
\[ \V \Lambda ab = \la \beta_a \circ
\alpha_\Lambda,\beta_b\ra_\MIo\,,
\qquad \Lambda\in\ASU nk\,,\quad a,b\in\cV, \]
where $\beta_a$ are representative endomorphisms of $a\equiv[\beta_a]$
(and we will use the label $0$ for the identity sector of $\MIo$
as well). This defines square matrices $V_\Lambda$,
$\Lambda\in\ASU nk$, by $(V_\Lambda)_{a,b}=\V \Lambda ab$,
as well as rectangular matrices $V_{(a)}$, $a\in\cV$,
by $(V_{(a)})_{\Lambda,b}=\V \Lambda ab$. Also, we set
$G_p=V_{\Lambda_{(p)}}$, $p=1,2,...,n-1$. Hence
$G_p$ is the adjacency matrix of the fusion graph of
$\alpha_{\Lambda_{(p)}}$.

\begin{lemma}
The matrices $V_\Lambda$ and $V_{(a)}$ have the following
properties.
\begin{enumerate}
\item $V_0=\bfe_{d}$,
\item $A_p \, V_{(a)}=V_{(a)} \, G_p$, $a\in\cV$, $p=1,2,\ldots,n-1$,
\item $V_\Lambda \, V_{\Lambda'} = \sum_{\Lambda''}
\N \Lambda{\Lambda'}{\Lambda''} \cdot V_{\Lambda''}$.
\end{enumerate}
\lablth{123}
\end{lemma}

\bproof
Ad 1. We obviously have $\V 0ab= \del ab$ as $\cV$ is a sector basis.

Ad 2. We compute
\[ \bearll
(A_p \, V_{(a)})_{\Lambda,b} &= \sum_{\Lambda'\in\ASU nk}
(A_p)_{\Lambda,\Lambda'} \, \V {\Lambda'}ab \\[.4em]
&= \sum_{\Lambda'\in\ASU nk} \la \lambda_\Lambda
\circ \lambda_{\Lambda_{(p)}},
\lambda_{\Lambda'}\ra_\NIo \la \beta_a \circ
\alpha_{\Lambda'},\beta_b\ra_\MIo \\[.4em]
&= \la \beta_a \circ \alpha_{\lambda_\Lambda\circ\lambda_{\Lambda_{(p)}}},
\beta_b\ra_\MIo \\[.4em]
&= \sum_{c\in\cV} \la \beta_a \circ \alpha_\Lambda,\beta_c\ra_\MIo
\la \beta_c \circ \alpha_{\Lambda_{(p)}},\beta_b\ra_\MIo \\[.4em]
&=  \sum_{c\in\cV} \V \Lambda ac \, (G_p)_{c,b} \\[.4em]
&= (V_{(a)}\,G_p)_{\Lambda,b}\,,
\eear \]
where we used the fact that the $[\lambda_\Lambda]$'s and
$[\beta_a]$'s constitute sector bases, and in the third
equality we used the additive homomorphism property of
$\alpha$-induction.

Ad 3. We compute
\[ \bearll
(V_\Lambda \, V_{\Lambda'})_{a,b} &= \sum_{c\in\cV}
\V \Lambda ac \, \V {\Lambda'} cb \\[.4em]
&= \sum_{c\in\cV} \la \beta_a \circ \alpha_\Lambda,
\beta_c \ra_\MIo \la \beta_c \circ \alpha_{\Lambda'},
\beta_b\ra_\MIo \\[.4em]
&= \la \beta_a \circ \alpha_\Lambda \circ
\alpha_{\Lambda'}, \beta_b\ra_\MIo \\[.4em]
&= \sum_{\Lambda''\in\ASU nk} \N \Lambda{\Lambda'}{\Lambda''}
\la \beta_a \circ \alpha_{\Lambda''},
\beta_b\ra_\MIo \\[.4em]
&= \sum_{\Lambda''\in\ASU nk}
\N \Lambda{\Lambda'}{\Lambda''} \, (V_{\Lambda''})_{a,b} \,,
\eear \]
where we again used the homomorphism property of
$\alpha$-induction.
\eproof

By some abuse of notation we also denote the
sector product matrices associated to $V$ by $N_a$, i.e.\
$(N_a)_{b,c}=\N bac$ with
\[ \N bac = \la \beta_b\circ\beta_a,\beta_c \ra_\MIo\,,
\qquad a,b,c\in\cV \,. \]
Analogous to the commutative case \cite{kwai}, these matrices realize
the ``regular'' representation of the sector algebra
$V$; we have
\[ N_a \, N_b = \sum_{c\in\cV} \N abc \cdot N_c \,,
\qquad a,b\in\cV\,, \]
since
\[ \bearll
(N_a\,N_b)_{d,e} = \sum_{f\in\cV} \N daf \, \N fbe
&= \sum_{f\in\cV} \la \beta_d\circ\beta_a,\beta_f \ra_\MIo
\la \beta_f\circ\beta_b,\beta_e \ra_\MIo \\[.4em]
&= \la \beta_d \circ \beta_a \circ \beta_b,
   \beta_e \ra_\MIo \\[.4em]
&= \sum_{c\in\cV} \N abc \, \la \beta_d \circ \beta_c,
   \beta_e \ra_\MIo \\[.4em]
&= \sum_{c\in\cV} \N abc \, (N_c)_{d,e} \,,
\eear \]
where we used that $\cV$ is a sector basis.

Note that Lemma \ref{123} (3.) reflects basically the
homomorphism property of $\alpha$-induction.  Using the
decomposition of $[\alpha_\Lambda]$ one can similarly
derive $V_\Lambda=\sum_{a\in\cV} \V \Lambda 0a\cdot N_a$
for $\Lambda\in\ASU nk$. The following lemma reflects
the commutativity of $[\alpha_\Lambda]$ with each $[\beta_a]$,
proven in Proposition 3.16 of \cite{boev1}.

\begin{lemma}
We have $V_\Lambda\, N_a = N_a \, V_\Lambda$ for any
$\Lambda\in\ASU nk$ and $a\in\cV$.
\lablth{VN=NV}
\end{lemma}

\bproof
We compute
\[ \bearll
(V_\Lambda\,N_a)_{b,c} = \sum_{d\in\cV} \V \Lambda bd \N dac
&= \sum_{d\in\cV} \la \beta_b \circ \alpha_\Lambda,\beta_d \ra_\MIo
\la \beta_d\circ\beta_a,\beta_c\ra_\MIo \\[.4em]
&= \la \beta_b \circ \alpha_\Lambda \circ \beta_a,
\beta_c\ra_\MIo \\[.4em]
&= \la \beta_b \circ\beta_a \circ \alpha_\Lambda,
\beta_c\ra_\MIo \\[.4em]
&= \sum_{d\in\cV} \la \beta_b \circ \beta_a, \beta_d\ra_\MIo
\la \beta_d \circ \alpha_\Lambda,\beta_c\ra_\MIo \\[.4em]
&= \sum_{d\in\cV} \N bad \cdot \V \Lambda dc
= (N_a \, V_\Lambda )_{b,c} \,,
\eear \]
where we used Proposition 3.16 of \cite{boev1}.
\eproof

\subsection{Modular invariants and exponents of graphs}

Let us briefly recall some facts about fusion algebras
(see e.g.\ \cite{kwai}). If $W$ is a fusion algebra with
sector basis $\cW=\{w_0,w_1,...,w_{d-1}\}$ and structure
constants $\N ijk$ then the matrices $N_i$ defined by
$(N_i)_{j,k}=\N ijk$ form the regular representation
of $W$, and since they constitute a family of normal,
commuting matrices they can be simultaneously diagonalized
by a unitary matrix $S$. Then the diagonal matrices
$S^*N_iS$ form a direct sum over all the irreducible
(one-dimensional) representations of $W$ i.e.\ over
its characters. These
representations $\rho_j$ are labelled by
$j=0,1,...,d-1$ and are given by
\[ \rho_j(w_i) = \frac{\Sm ij}{\Sm 0j} \,,
\qquad i=0,1,2,\ldots,d-1, \]
where $\Sm ij$ are the matrix elements of $S$.

Now let us start with a conformal or orbifold inclusion of
$\SUn$ at level $k$, and let $\cV$ again denote the sector
basis obtained by $\alpha$-induction from the sector basis
$\cW=\cW(n,k)$ corresponding to the \per s of $\SUn_k$.
Recall that $\cT\subset\cV$ are the sectors corresponding
to the marked vertices, generating a commutative
sector subalgebra $T\subset V$ by Theorem 4.3 of
\cite{boev1}; for details see also Subsection 2.1.
Note that $\N bac = \N c{\co a}b$, thus $N_{\co a}$ is the
transpose matrix of $N_a$.
Since $\cT$ is closed under conjugation and by
Lemma \ref{VN=NV}, the matrices
$N_t$, $t\in\cT$, and $V_\Lambda$, $\Lambda\in\ASU nk$,
form a family of normal, commuting matrices and hence
can be simultaneously diagonalized in a suitable orthonormal
basis that we denote by $\{\psi^i\,,\,\,\,i=1,2,...,D\}$;
here $D=|\cV|$. As the matrices $V_\Lambda$ constitute
a representation of the fusion algebra $W\equiv W(n,k)$ by
Lemma \ref{123} they decompose in the one-dimensional
irreducible representations $\gamma_\Phi$ of $W$,
which are labelled by weights $\Phi\in\ASU nk$ and are
given by
\[ \gamma_\Phi (\Lambda) = \frac{\Sm \Lambda \Phi}{\Sm 0 \Phi} \,,
\qquad \Lambda\in \ASU nk \,,\]
where $\Sm \Lambda \Phi$ denote the entries of the matrix $S$
that diagonalizes the fusion rules of the endomorphisms associated
to the $\LSUn$ level $k$ theory. Due to Wassermann's result
\cite{wass3} these endomorphisms obey the fusion rules given
by the Verlinde formula in terms of the modular S-matrix
$\cS$, therefore the modular S-matrix $\cS$ diagonalizes the
endomorphism fusion rules, i.e.\ we have indeed $S=\cS$.

We conclude that we have a map $\Phi:\{1,2,...,D\}\rightarrow\ASU nk$,
$i\mapsto\Phi(i)$, such that
\[ V_\Lambda= \sum_{i=1}^D \gamma_{\Phi(i)}(\Lambda)
|\psi^i\ra\la\psi^i| \,, \qquad\Lambda\in\ASU nk\,, \]
i.e.\ in components
\[ \V \Lambda ab = \sum_{i=1}^D \frac{\Sm \Lambda{\Phi(i)}}{\Sm 0{\Phi(i)}}
\psiv ia \psivs ib \,, \qquad\Lambda\in\ASU nk\,,\quad a,b\in\cV \]
The image of $\Phi$ is the set of weights $\Omega\in\ASU nk$ such
that $\gamma_\Omega$ appears in the $V_\Lambda$'s. Since in particular
$G_p=V_{\Lambda_{(p)}}$, $p=1,2,...,n-1$, we call these weights
$\Omega$ {\em exponents} and denote the set of exponents
$\Exp=\Bild \Phi$. In other words, $\Exp$ labels the joint
spectrum of the matrices $V_\Lambda$.
Similarly, as the $N_t$'s with $t\in\cT$ give
a representation of the fusion algebra $T$ of the extended
theory by Theorem 4.3 of \cite{boev1} we have a map
$s:\{1,2,...,D\}\rightarrow\cT$, $i\mapsto s(i)$, such that
\[ N_t = \sum_{i=1}^D \eta_{s(i)}(t)
|\psi^i\ra\la\psi^i| \,, \qquad t\in\cT\,, \]
where
\[ \eta_s (t) = \frac{\Sex ts}{\Sex 0s}\,,\qquad t\in\cT\,,\]
are the one-dimensional representations of $T$ and $\Sex ts$
denote the entries of a matrix $S^\mathrm{ext}$ that diagonalizes
the (endomorphism!) fusion rules of $T$. It is widely believed for
general conformal field theories (and even conjectured e.g.\ in
\cite{frga}, Conjecture 4.48) that endomorphisms representing the
sectors of a conformal field theory obey the Verlinde fusion rules
given in terms of the modular S-matrix, i.e.\ that we can choose
$S^\mathrm{ext}=\cS^\mathrm{ext}$, where $\cS^\mathrm{ext}$ is
the S-matrix coming from the modular transformation of the
extended characters. However, a proof exists only for several
particular cases, see below.

Thus we have in components
\[ \N atc = \sum_{i=1}^D \frac{\Sex t{s(i)}}{\Sex 0{s(i)}}
\psiv ia \psivs ic \,, \qquad t\in\cT\,,\quad a,c\in\cV \,. \]
For $\Lambda\in\ASU nk$ and $t\in\cT$ we define
$\Eig(\Lambda,t)$ to be the space spanned by those
$\psi^i$ which correspond simultaneous eigenvalues
$\gamma_\Lambda(\Lambda')$ of $V_{\Lambda'}$ and $\eta_t(t')$
of $N_{t'}$ for all $t'\in\cT$, $\Lambda'\in\ASU nk$, i.e.
\[ \Eig(\Lambda,t) = \mathrm{span} \{ \psi^i \,\,:\,\, i\in
\Phi^{-1}(\Lambda) \cap s^{-1}(t) \} \,, \]
so in particular $\Eig (\Lambda,t)=0$ iff $\Lambda\notin\Exp$.
So far the vectors $\psi^i$ are fixed up to unitary transformations
in each $\Eig(\Lambda,t)$.

\begin{lemma}
We have $\psiv it = \frac{\Sex t{s(i)}}{\Sex 0{s(i)}}\psiv i0$ for
any $t\in\cT$ and $i=1,2,...,D$.
\lablth{pstiSS}
\end{lemma}

\bproof
Clearly we have $\N 0tc = \del ct$. Hence
\[ \psiv it = \sum_{c\in\cV} \N 0tc \psiv ic =
\sum_{c\in\cV} \sum_{j=1}^D \frac{\Sex t{s(j)}}{\Sex 0{s(j)}}
\psiv j0 \psivs jc \psiv ic = \sum_{j=1}^D
\frac{\Sex t{s(j)}}{\Sex 0{s(j)}} \psiv j0 \del ij
= \frac{\Sex t{s(i)}}{\Sex 0{s(i)}} \psiv i0 \, \]
by orthonormality of the $\psi^i$'s.
\eproof

Let $\psi_0=(\psiv i0)_{i=1}^D$ denote the dual vector
of $0$-components, and we set
\[ \|\psi_0\|_{\Lambda,t}=
\sqrt{\sum_{i\in\Phi^{-1}(\Lambda)\cap s^{-1}(t)}
|\psiv i0|^2 }\,. \]

\begin{lemma}
If $\Eig(\Lambda,t)\neq 0$ for some $\Lambda\in\ASU nk$ and
$t\in\cT$ then $\|\psi_0\|_{\Lambda,t} \neq 0$.
\lablth{nonv}
\end{lemma}

\bproof
Since $N_a\,N_b=\sum_{c\in\cV} \N abc \cdot N_c$ and
$\N atc=\N tac$ for $t\in\cT$ by Theorem 4.3
of \cite{boev1} ($a,b,c\in\cV$)
we have  $N_a N_t = N_t N_a$ for any $a\in\cV$.
Hence we find for $i\in\Phi^{-1}(\Lambda)\cap s^{-1}(t)$
\[ V_\Omega N_a \psi^i = \gamma_\Lambda(\Omega) N_a \psi^i \,,
\quad N_u N_a \psi^i = \eta_t (u) N_a \psi^i \,, \quad
\Omega\in\ASU nk\,,\,\,\, u\in\cT \,, \]
i.e.\ $N_a \psi^i \in\Eig(\Lambda,t)$. In other words,
the matrices $N_a$ are block-diagonal in the basis $\psi^i$
corresponding to the decomposition in $\Eig(\Lambda,t)$.
It follows that there are matrices, the ``blocks''
$B_a\equiv B_a(\Lambda,t)$, $(B_a)_{i,j}=B_{a;j}^i\in\bbC$,
$i,j\in\Phi^{-1}(\Lambda)\cap s^{-1}(t)$ such that
$N_a \psi^i = \sum_{j\in\Phi^{-1}(\Lambda)\cap s^{-1}(t)}
B_{a;j}^i \psi^j$, hence in particular for the $0$-components
\[ (N_a \psi^i)_0 = \sum_{j\in\Phi^{-1}(\Lambda)\cap s^{-1}(t)}
B_{a;j}^i \psiv j0 \,. \]
Since $(N_a \psi^i)_0=\sum_{c\in\cV} \N a0c \psiv ic = \psiv ia$ we
have for any $i\in\Phi^{-1}(\Lambda)\cap s^{-1}(t)$ and any $a\in\cV$
\[ \psiv ia = \sum_{j\in\Phi^{-1}(\Lambda)\cap s^{-1}(t)}
B_{a;j}^i \psiv j0 \,. \]
It follows if $\psiv j0=0$ for all
$j\in\Phi^{-1}(\Lambda)\cap s^{-1}(t)$ then $\psiv ia=0$
for all $i\in\Phi^{-1}(\Lambda)\cap s^{-1}(t)$ and $a\in\cV$,
i.e.\ $\Eig(\Lambda,t)=0$.
\eproof

We set $D_{\Lambda,t}=\dim\,\Eig (\Lambda,t)
\equiv |\Phi^{-1}(\Lambda)\cap s^{-1}(t)|$. Our vectors $\psi^i$
are fixed up to unitary transformations (rotations) in each
$\Eig(\Lambda,t)$,
$\psi^i\mapsto\sum_{j\in\Phi^{-1}(\Lambda)\cap s^{-1}(t)} u_{i,j}
\psi^j$, with unitary matrices
$u=(u_{i,j})_{i,j\in\Phi^{-1}(\Lambda)\cap s^{-1}(t)}$. Thus we have
in particular
$\psiv i0 \mapsto\sum_{j\in\Phi^{-1}(\Lambda)\cap s^{-1}(t)} u_{i,j}
\psiv j0$. This is a rotation in $\bbC^{D_{\Lambda,t}}$ on the
sphere of radius $\|\psi_0\|_{\Lambda,t}$. As we have
shown that  $\|\psi_0\|_{\Lambda,t}\neq 0$ if
$\Eig(\Lambda,t)\neq 0$ we arrive at the following

\begin{corollary}
There is a choice of eigenvectors $\psi^i$ such that
$\psiv i0\neq 0$ for all $i=1,2,\ldots,D$, e.g.\
$\psiv i0 = D_{\Lambda,t}^{-1/2}
\|\psi_0\|_{\Lambda,t} > 0$ whenever
$i\in\Phi^{-1}(\Lambda)\cap s^{-1}(t)$.
\end{corollary}

As we now can divide by $\psiv i0$ we obtain immediately
from Lemma \ref{pstiSS} the following

\begin{corollary}
For such a choice we have for any $t\in\cT$ and any $b,c\in\cV$
\be
\N atc = \sum_{i=1}^D \frac{ \psiv ia \psiv it \psivs ic}
{\psiv i0} \,.
\ee
\end{corollary}

Let $\chi^\mathrm{ext}_t$, $t\in\cT$, denote the characters of the
extended theory and $\chi_\Lambda$, $\Lambda\in\ASU nk$, those of
$\SUn_k$. Further, let $b_{t,\Lambda}$ denote the
{\em branching coefficients}, defined by
$\chi^\mathrm{ext}_t = \sum_{\Lambda\in\ASU nk}
b_{t,\Lambda} \, \chi_\Lambda$. Let also $\cS^\mathrm{ext}$ be the
modular S-matrix of the extended theory, i.e.\ (in the
notation of \cite{kac})
\[ \chi^\mathrm{ext}_t \left(-\frac{1}{\tau},\frac{z}{\tau},
{\tt u}-\frac{(z|z)}{\tau}\right) = \sum_{v\in\cT} \Sexm tv \,
\chi^\mathrm{ext}_t (\tau,z,{\tt u}) \,. \]

\begin{lemma}
For any $\Lambda\in\ASU nk$ and $u\in\cT$ we have
\be
\sum_{v\in\cT} \Sexm uv \, b_{v,\Lambda} = \sum_{\Omega\in\ASU nk}
b_{u,\Omega} \, \Sm \Omega\Lambda \,.
\labl{SbbS}
\end{lemma}

\bproof
This is essentially the computation in \cite{kac}, p.\ 268,
here for the special case that the branching functions are
constants. By taking the S-transformation on both sides of
$\chi^\mathrm{ext}_u = \sum_{\Omega\in\ASU nk}
b_{u,\Omega} \, \chi_\Omega$
we obtain
\[ \sum _{v\in\cT} \Sexm uv \, \chi^\mathrm{ext}_v \equiv
\sum_{v\in\cT} \sum_{\Lambda\in\ASU nk} \Sexm uv \, b_{v,\Lambda} \,
\chi_\Lambda = \sum_{\Lambda,\Omega\in\ASU nk} b_{u,\Omega} \,
\Sm \Omega\Lambda \, \chi_\Lambda \,.\]
Since the full (not the Virasoro specialized!) characters $\chi_\Lambda$
are linearly independent functions the coefficients must coincide,
so we are done.
\eproof

Note that
$\V \Lambda 0t = \la \alpha_\Lambda,\beta_t \ra_\MIo =
\la \lambda_\Lambda,\sigma_{\beta_t} \ra_\NIo$
by $\alpha\sigma$-reciprocity, thus we find for the branching
coefficients $b_{t,\Lambda}=\V \Lambda 0t$.
Let $\Nres_t$ denote the restriction of the matrices
$N_t$ to $\cT$, i.e.\ $(\Nres_t)_{u,v}=\N utv$,
$t,u,v\in\cT$.

\begin{lemma}
Provided that $\cS^\mathrm{ext}$ diagonalizes the
fusion matrices $\Nres_t$, $t\in\cT$, i.e.
$\cS^\mathrm{ext}=S^\mathrm{ext}$,
we have for $\Lambda\in\ASU nk$ and $t\in\cT$
\be
b_{t,\Lambda} =
\frac{\| \psi_0 \|^2_{\Lambda,t}}{\Sex 0t \Sm 0\Lambda} \,.
\ee
\lablth{bran}
\end{lemma}

\bproof
Exploiting $\cS^\mathrm{ext}=S^\mathrm{ext}$,
multiplying \erf{SbbS} by $(\Sex ut)^*$ and
summing over $u\in\cT$ yields
\[ \bearll
b_{t,\Lambda} &= \sum_{u\in\cT} \sum_{\Omega\in\ASU nk}
(\Sex ut)^* \, b_{u,\Omega} \, \Sm \Omega\Lambda \\[.4em]
&= \sum_{u\in\cT} \sum_{\Omega\in\ASU nk}
(\Sex ut)^* \, \overline{V}_{\Omega;0}^u \, \Sm \Omega\Lambda \\[.4em]
&= \sum_{u\in\cT} \sum_{\Omega\in\ASU nk} \sum_{i=1}^D
(\Sex ut)^* \, \frac{(\Sm \Omega{\Phi(i)})^*}{\Sm 0{\Phi(i)}}
\psivs i0 \psiv iu \, \Sm \Omega\Lambda \\[.4em]
&= \sum_{u\in\cT} \sum_{i=1}^D
(\Sex ut)^* \, \frac{\del \Lambda{\Phi(i)}}{\Sm 0\Lambda}
\psivs i0 \psiv iu \\[.4em]
&= \sum_{u\in\cT} \sum_{i=1}^D
(\Sex ut)^* \, \frac{\del \Lambda{\Phi(i)}}{\Sm 0\Lambda}
|\psiv i0|^2 \frac{\Sex u{s(i)}}{\Sex 0{s(i)}} \\[.4em]
&= \sum_{i=1}^D \del t{s(i)} \del \Lambda{\Phi(i)}
\frac{|\psiv i0|^2 }{\Sex 0t \, \Sm 0\Lambda} \\[.4em]
&= \frac{\| \psi_0 \|^2_{\Lambda,t}}{\Sex 0t \Sm 0\Lambda} \,,
\eear \]
where we used
$b_{u,\Omega}=\V \Omega 0u = \overline{V}_{\Omega;0}^u$
and Lemma \ref{pstiSS}.
\eproof

Recall that the mass matrix of the modular invariant is given by
$Z_{\Lambda,\Lambda'}=\sum_{t\in\cT} b_{t,\Lambda} b_{t,\Lambda'}$.
We can now summarize Lemmata \ref{nonv} and \ref{bran} in the
following

\begin{theorem}
Provided that $\cS^\mathrm{ext}$ diagonalizes the
fusion matrices $\Nres_t$, $t\in\cT$, i.e.
$\cS^\mathrm{ext}=S^\mathrm{ext}$, we have
$b_{t,\Lambda}\neq 0$ if and only if $\Eig(\Lambda,t)\neq 0$.
In particular $Z_{\Lambda,\Lambda}\neq 0$ if and only if
$\Lambda\in\Exp$.
\lablth{bD}
\end{theorem}

Actually we would like to prove a stronger statement than
Theorem \ref{bD}, namely $b_{t,\Lambda}=\sqrt{D_{\Lambda,t}}$
because this equality holds in all our examples we have
investigated so far. Let us explain why it holds in our examples.
Let
\[ \tr Z = \sum_{\Lambda\in\ASU nk} Z_{\Lambda,\Lambda}
= \sum_{t\in\cT} \sum_{\Lambda\in\ASU nk} b_{t,\Lambda}^2 \,. \]
We clearly have
$\sum_{t\in\cT} \sum_{\Lambda\in\ASU nk} D_{\Lambda,t}=D\equiv|\cV|$,
and it is a simple observation that $\tr Z=D$ in all our examples,
hence
\[ \sum_{t\in\cT} \sum_{\Lambda\in\ASU nk} b_{t,\Lambda}^2 =
\sum_{t\in\cT} \sum_{\Lambda\in\ASU nk} D_{\Lambda,t} \,. \]
Thus, if all $b_{t,\Lambda}\in\{0,1\}$ then our derived
equivalence of $b_{t,\Lambda}\neq 0$ and $D_{\Lambda,t}\neq 0$
in Theorem \ref{bD} implies $b_{t,\Lambda}=\sqrt{D_{\Lambda,t}}$ for
all $t\in\cT$, $\Lambda\in\ASU nk$. The only case of our
examples where some $b_{t,\Lambda}>1$ appears is the conformal
embedding $\SUf_4\subset\mathit{SO}(15)_1$ where the spinor (s)
representation of $\mathit{SO}(15)_1$ restricts to two
copies of $\pi_{(3,2,1)}$, i.e.\ $b_{\rms,(3,2,1)}=2$.
Because of Theorem \ref{bD} we have
$D_{\Lambda,t}\ge b_{t,\Lambda}^2$ for all pairs
$(t,\Lambda) \neq (\rms,(3,2,1))$.
However, Petkova and Zuber \cite{pezu2} found a multiplicity
$4$ of the exponent $(3,2,1)$ in the graphs of Figs.\ \ref{noncom1}
and \ref{noncom2}, i.e.\ $\sum_{t\in\cT} D_{(3,2,1),t}=4$.
But since $b_{t,(3,2,1)}=0$ for $t\neq\rms$ implies
$D_{(3,2,1),t}=0$ for $t\neq\rms$ it follows
$D_{(3,2,1),\rms}=4$, and hence indeed
$b_{t,\Lambda}=\sqrt{D_{\Lambda,t}}$ for all
$t\in\cT,\Lambda\in\ASU 44$ because $\tr Z=D=12$.
Nevertheless we have not succeeded in proving
this equality for the general case.

\subsection{Discussion and consequences}

Let us now summarize some of the results of this
section. To each block-diagonal modular invariant of
$\SUn$, coming either from a conformal inclusion or being of
$\bbZ_n$-orbifold type, we have a net of subfactors
such that we can apply $\alpha$-induction. By doing this,
we obtain in particular a set of $n-1$ normal, mutually
commuting matrices $G_p$, $p=1,2,...,n-1$, which can be
interpreted as adjacency matrices of fusion graphs,
namely those of $[\alpha_{\Lambda_{(p)}}]$ in the
sector algebra $V$. Since
$[\overline{\alpha}_{\Lambda_{(p)}}]=[\alpha_{\Lambda_{(n-p)}}]$
we find $G_p^t=G_{n-p}$. The matrices $G_p$ can be simultaneously
diagonalized in an orthonormal basis
$\{\psi^i\,,\,\,i=1,2,...,|\cV|\}$, and the eigenvalues
of $G_p$ are given by $\Sm {\Lambda_{(p)}}\Phi / \Sm 0\Phi$,
$\Phi\in\Exp$, where $\Exp$ is a subset of $\ASU nk$.

Recall that one can define a $\bbZ_n$-valued colouring $\tau$
on the vertices of \ddAgx {n+k} (which we can identify with
the elements of $\cW$) by $\tau(\Lambda)=|\Lambda|\,\,\mod\,\,n$.
Then one has $\tau(0)=0$ and
$\N \Lambda{\Lambda'}{\Lambda''}=
\la\lambda_\Lambda\circ\lambda_{\Lambda'},
\lambda_{\Lambda''}\ra_\NIo=0$
if $\tau(\Lambda)+\tau(\Lambda')\neq\tau(\Lambda'')$.
If $[\canr]$ decomposes
only into sectors $[\lambda_\Lambda]$ of colour zero then
the elements of $\cV$ inherit the colouring from \ddAgx {n+k}:
The colour of $[\beta]\in\cV$ is set to be $\tau(\Lambda)$
if $[\beta]$ appears in $[\alpha_\Lambda]$. This is then well
defined because
$\la\alpha_\Lambda,\alpha_{\Lambda'}\ra_\MIo=
\la\canr\circ\lambda_\Lambda,\lambda_{\Lambda'}\ra_\NIo=0$
if $\tau(\Lambda)\neq\tau(\Lambda')$. That $[\canr]$ decomposes
only into sectors of colour zero is true for all orbifold
inclusions and also all conformal embeddings considered
here. Therefore the matrices $G_p$ satisfy all the axioms
postulated in \cite{pezu2}. However, as already mentioned there,
there are also counter examples e.g.\ the conformal embedding
$\mathit{SU}(9)_1\subset(\mathrm{E}_8)_1$ where $[\canr]$ has
also constituents of non-zero colour.

The sector algebra $V$ possesses a subalgebra
given by the fusion algebra of the sectors of the
extended net which restrict to the relevant sectors of
the net of the $\SUn$ theory. If the corresponding
fusion matrices, coming from these sector products, are
diagonalized by the modular S-matrix $\cS^\mathrm{ext}$
of the extended characters then the non-zero diagonal entries
$Z_{\Lambda,\Lambda}$ of the modular invariant are precisely
those with $\Lambda\in\Exp$. We conjecture that the modular
S-matrix $\cS^\mathrm{ext}$ always diagonalizes the
extended (endomorphism) fusion rules, but let us point out
the cases where it has already been proven.

First we consider
the modular invariants coming from conformal inclusions. It
follows from Wassermann's results \cite{wass3} that in
particular the endomorphisms of any $\mathit{LSU}(m)$ level 1
theory satisfy the ($\bbZ_m$) fusion rules of the Verlinde
formula, thus we have $S^\mathrm{ext}=\cS^\mathrm{ext}$ for
all conformal inclusions $\SUn\subset G$ with
$G=\mathit{SU}(m)$ for some $m$. This covers the infinite
series of inclusions
\[  \SUn_{n-2} \subset \mathit{SU}(n(n-1)/2)_1 \qquad \mbox{and}
\qquad  \SUn_{n+2} \subset \mathit{SU}(n(n+1)/2)_1 \,. \]
By the result of \cite{bock2}, the endomorphisms of the
$\mathit{LSO}(m)$ level 1 theories satisfy the well-known
$\mathit{SO}(m)_1$
fusion rules, hence $S^\mathrm{ext}=\cS^\mathrm{ext}$ also for
$G=\mathit{SO}(m)$. This covers the infinite series of inclusions
\[ \SUn_n \subset \mathit{SO}(n^2-1)_1 \,, \]
and also $\SUz_{10}\subset\mathit{SO}(5)_1$. Moreover, we have
seen from the treatment of the \ddE 8 modular invariant of $\SUz$
(at level $k=28$) that the endomorphisms of the $\mathit{L}(\Gtwo)$
level 1 theory obey the Lee-Yang fusion rules, thus we
have $S^\mathrm{ext}=\cS^\mathrm{ext}$ also for the conformal
inclusion $\SUz_{28}\subset (\Gtwo)_1$.

Now let us turn to the
orbifold modular invariants. Unfortunately, our results are
only complete for $\SUz$. We have seen that the fusion algebra
$V$ for the \ddD {2\varrho+2}
modular invariants is completely determined although the
homomorphism $[\alpha]$ is not surjective. The modular
S-matrices $\cS^\mathrm{ext}$ of the extended characters
are known and their Verlinde fusion rules are given in \cite{bage}.
They coincide exactly with the fusion rules of the
sectors $\as j$, $j=0,2,4,...,2\varrho-2$, and $\as {2\varrho} _\pm$,
the ``marked vertices'', which we gave in Subsection \ref{Dex}.
Thus we have $S^\mathrm{ext}=\cS^\mathrm{ext}$ also for these cases.
Summarizing we found that $S^\mathrm{ext}=\cS^\mathrm{ext}$ holds
for all block-diagonal modular invariants of $\SUz$, hence
its diagonal entries are labelled by some subset
$\Exp\subset\mathrm{A}_{k+1}$. As we have seen for the (non-trivial)
block-diagonal modular invariants that $G_1=N_{\as 1}$ is the
adjacency matrix of the Coxeter graphs \ddE 6, \ddE 8 or
\Deven\ (in fact since they are fusion graphs of norm
$d_{\as 1}^2= d_1^2=4\cos^2 (\pi/(k+2))$ they can only be
these graphs), the set $\Exp$ is necessarily given by the Coxeter
exponents of these graphs. Thus our theory explains in
particular why the spins of the diagonal entries of the
non-trivial block-diagonal modular invariants are given by
the Coxeter exponents of the graphs \ddE 6, \ddE 8 and
\ddD {2\varrho+2}, $\varrho\in\bbN$.

\section{Other applications}
\lablsec{othappl}

We shall also discuss some other examples for the application
of $\alpha$-induction which may be of some interest of their own.

\subsection{Inclusions of extended $U(1)$ theories}

Let $\cA_N$, $N=1,2,3,...$, denote the extension of the $U(1)$ current
algebra discussed in \cite{bumt}. It has $2N$ sectors constituting
$\bbZ_{2N}$ fusion rules. The characters are given by
\[ \K Na  (q) = \frac 1\eta \sum_{m\in\bbZ}
q^{(a+2mN)^2/4N} \,,\qquad a \in \bbZ_{2N} \,, \]
where $\eta$ is Dedekind's function.
The modular invariant partition functions
of these theories have been classified
\cite{frsz}. For each factorization $N=\ell^2pp'$, $\ell,p,p'\in\bbN$,
$p$ and $p'$ coprime, associate $r,r'\in\bbZ$ such that
$r'p'-rp=1$. Define $s=r'p'+rp$. Then
\[ \Z N\ell s = \sum_{a,b\in\bbZ_{2N}} \mm abN (\ell,s)
\, \K Na \, \Kb Nb \]
with
\[ \mm abN (\ell,s) = \left\{ \begin{array}{cl}
\sum_{c\in\bbZ_\ell} \del a{sb+2cN/\ell} & \qquad
\mbox{if} \,\,\, \ell|a \,\,\,\mbox{and}\,\,\,\ell|b \\[.4em]
0 & \qquad \mbox{otherwise} \eear \right. \]
exhaust all modular invariants. Note that $\ell=1$, $p=N$, $p'=1$,
implying $r=0$, $r'=1$, $s=1$, gives the diagonal modular invariant
$\Z N11$. Now choose an $N$ such that $\ell\ne 1$ so that $\cA_N$
is a non-maximal $U(1)$-extension in the terminology of \cite{bumt}.
Choose $p=N/\ell^2$, $p'=1$ implying $r=0$, $r'=1$ and $s=1$. The
corresponding partition function reads
\be
\Z N\ell 1 = \sum_{a\in\bbZ_{2p}} \left| \sum_{c\in\bbZ_\ell}
\K N {a\ell+2c N/\ell} \right|^2 \,.
\labl{block}
But
\[ \bearll
\sum_{c\in\bbZ_\ell} \K {\ell^2p} {a\ell+2cN/\ell} (q)
&= \eta^{-1} \sum_{c\in\bbZ_\ell} \sum_{m\in\bbZ}
q^{(a\ell+2c\ell p + 2m\ell^2p)^2/4\ell^2p} \\[.5em]
&= \eta^{-1} \sum_{m\in\bbZ} q^{(a+2mp)^2/4p}
= \K pa(q) \,,
\eear \]
hence
\[ \Z {\ell^2p} \ell 1 = \sum_{a\in\bbZ_{2p}} |\K pa|^2
= \Z p11 \,. \]
Indeed the inclusion $\cA_{N=\ell^2p}\subset\cA_p$ is of
$\bbZ_\ell$ type. Note that $\Z {\ell^2p} \ell 1$
is block-diagonal, and we can take the net of inclusions of
local algebras $\cA_N(I)\subset\cA_p(I)$, with $I\in\Jz$,
as our net of subfactors $\cN\subset\cM$.

Let us denote by $\laN Na$ the endomorphisms (which are in fact the
automorphisms constructed in \cite{bumt}) corresponding to the sectors
labelled by $a\in\bbZ_{2N}$. The $\bbZ_{2N}$
fusion rules then just read
\[ \lasN Na \times \lasN Nb = \lasN N{a+b}\,,
\qquad a,b\in\bbZ_{2N} \,. \]
Thus the associated fusion algebra $W^{(N)}$ is the
group algebra of $\bbZ_{2N}$ (with $\bbZ_{2N}$ as sector basis).
Now we want to apply the
machinery of $\alpha$-induction.
For a non-maximal $\cA_N$, $N=\ell^2p$,
we start with the block-diagonal
partition function in \erf{block} and read off the
$[\canr]$ from the vacuum block,
\be
[\canr] = \bigoplus_{c\in\bbZ_\ell} [\laN N{2c\ell p}] \,.
\labl{gammaN}
It is easy to see that the formula
$\la \alpha^{(N)}_a,\alpha^{(N)}_b \ra_\MIo = \la
\canr\circ \lambda^{(N)}_a,\lambda^{(N)}_a \ra_\NIo$
(we denote $\alpha^{(N)}_a\equiv\alpha_{\lambda^{(N)}_a}$)
and the homomorphism property of $[\alpha]$ determine
the induced sector algebra $V$ to be the group algebra of
$\bbZ_{2\ell p}=\bbZ_{2N}/\bbZ_\ell$. Now
$\bbZ_{2\ell p}$ has the subgroup $\bbZ_{2p}$, describing the fusion
rules of $\cA_p$, and this corresponds to the marked vertices.

As an illustration we discuss the $\bbZ_2$ inclusion
$\cA_4\subset\cA_1$. The $\cA_1$ theory has two sectors and it is
known to be precisely the $\suzh_1$ theory. The $\cA_4$ theory
has eight sectors, and their conformal weights
$\Delta_\nu$ are given by $\Delta_a=a^2/16$,
$a=0,\pm 1,\pm 2,\pm 3,4$. Indeed the series of $\bbZ_2$ orbifold
inclusions $\mathit{SO}(2n)_2\subset\mathit{SU}(2n)_1$ gives
the inclusion $\cA_4\subset\cA_1$ when $n=1$.
The sectors of $\cA_4$ labelled by $a=0,\pm 2,4$ are
the basic ($\circ$), the spinor (s,c) and the vector (v) modules,
the sectors labelled by $\pm 1$ and $\pm 3$ are the
twisted sectors $\sigma,\tau$ and and $\sigma',\tau'$,
respectively, in the terminology of $\mathit{SO}(2n)_2$.
The modular invariant $\Z 421$ of $\cA_4$ reads
\[ \Z 421 = |\K 40 + \K 44|^2 + |\K 42 + \K4{-2}|^2
= |\K 10|^2 + |\K 11|^2 = \Z 111 \,. \]
>From $[\canr]=[\laN 40] \oplus [\laN 44]$ we obtain that
$V$ is the group algebra of $\bbZ_4$.
The sectors $\lasN 40$, $\lasN 44$ and $\lasN 42$, $\lasN 4{-2}$,
obtained from $\lasN 10$ and $\lasN 11$ by
$\sigma$-restriction, yield
irreducible sectors $\asN 40 = \asN 44$ and $\asN 42=\asN 4{-2}$,
respectively, and constitute the $\bbZ_2\subset\bbZ_4$ subgroup.
In the fusion graph of $\asN 41$, being the $\bbZ_4$ graph
$\mathrm{A}^{(1)}_3$, these sectors represent the marked vertices.
The sectors $\lasN 4{\pm 1}$ and $\lasN 4{\pm 3}$ are not obtained
by $\sigma$-restriction of any $\cA_1$ sectors.
Note that these are precisely
the twisted sectors. Correspondingly, $\asN 41=\asN 4{-1}$ and
$\asN 43=\asN 4{-3}$ yield the elements of the sector basis
$\cV$ of $V$ which are not represented by marked vertices.

These observations generalize as follows to the block-diagonal
modular invariant $\Z N\ell 1$ for $N=\ell^2p$. We have seen that
we then can consider $\cA_N$ as the $\bbZ_\ell$ orbifold theory of
$\cA_p$. Reading off the $[\canr]$, \erf{gammaN}, from the
vacuum block we obtain that $V$ is the group algebra of
$\bbZ_{2\ell p}$.
The irreducible sectors $\asN Na$ with $a$ a multiple of $\ell$,
i.e.\ $a\in\bbZ_{2p}\subset\bbZ_{2\ell p}$,
are represented as marked vertices in the fusion
graph of $\asN N1$. Correspondingly, the sectors
$\lasN Na$, $a\in\bbZ_{2p}$, are obtained by
$\sigma$-restriction of the sector $\lasN pa$ of $\cA_p$.
Considering $\cA_N$ as the $\bbZ_\ell$
orbifold of $\cA_p$, we can interpret the other sectors
$\lasN Nb$, $b\notin\bbZ_{2p}$, as twisted sectors.

\subsection{Minimal models}

We shall also briefly discuss the treatment of the minimal models
here. The minimal models are described by the \per s of
the diffeomorphism group of the circle $\DiffS$, or, on the
level of Lie algebras, by the unitary highest weight modules of
the Virasoro algebra $\Vir (c)$, where the central charge
$c\equiv c(m)$ is given by
\[ c = 1 - \frac{6}{m(m+1)} \,, \qquad m=3,4,5,\ldots \]
These models arise as coset theories \cite{goko1}
\[ \frac{\SUz_{m-2} \otimes \SUz_1}{\SUz_{m-1}}\,. \]
The modules $H_{p,q}$, appearing at fixed $m$, are
labelled by pairs of integers $p=0,1,2,...,m-2$ and
$q=0,1,2,...,m-1$, the conformal grid. We have a double counting,
$H_{p,q}=H_{m-p-2,m-q-1}$.

In the setting of local
von Neumann algebras the minimal models have been
treated in \cite{loke} quite analogously to the
treatment of the loop groups $\LSUn$ by Wassermann.
Here we are dealing with a net $\cN$ of local von
Neumann algebras $N(I)=\pio(\DiffIS)''$, where $\pio$
is the vacuum representation of $\DiffS$ and $\DiffIS$
is the subgroup of diffeomorphisms concentrated on
an interval $I\subset S^1$. Analogously to the arguments
for $\LSUn$ we have Haag duality in the vacuum representation
and the \per s correspond localized, transportable
endomorphisms. The well known fusion rules are proven
in the bimodule setting in \cite{loke} and hence they
give the correct fusion rules for the corresponding sectors,
explicitly,
\[ \lasv pq \times \lasv {p'}{q'} =
\bigoplus_{r=|p-p'| \atop r+p+p'\,\,\mathrm{even}}^{\min(p+p',
2m-p-p'-4)}\,\,\,\,\,
\bigoplus_{s=|q-q'| \atop s+q+q'\,\,\mathrm{even}}^{\min(q+q',
2m-q-q'-2)} \lasv rs  \]
where $\lav pq\in\DelNIo$ denote the endomorphisms associated
to $H_{p,q}$. This determines the fusion algebra $W_{\Vir(c(m))}$.

The modular invariants of the minimal
models are classified, and are labelled by pairs $(\cG_1,\cG_2)$
of ADE-graphs (with Coxeter numbers $m-2$ and $m-1$) \cite{caiz}.
If we write the $\SUz$ modular invariants
appearing at level $k$ and labelled by ADE-graphs $\cG$ as
\[ Z_\cG = \sum_{j-0}^k \mm j{j'}k (\cG) \,\,
\chi_j \,\chib_{j'} \]
then the $(\cG_1,\cG_2)$ modular invariants of the minimal
model with $c=c(m)$ is given by
\[ Z_{\cG_1,\cG_2} = \frac12 \sum_{p,p'=0}^{m-2} \sum_{q,q'=0}^{m-1}
\mm p{p'}{m-2} (\cG_1) \,\,\mm q{q'}{m-1} (\cG_2)
\,\, \chi_{p,q} \, \chib_{p',q'} \]
where $\chi_{p,q}$ denotes the character of $H_{p,q}$.
The prefactor $1/2$ is due to the double counting.
Since either $m-2$ or $m-1$ is odd either $\cG_1$ or $\cG_2$
is necessarily an A-graph.

We would like to apply the procedure of
$\alpha$-induction. Although these are block diagonal
modular invariants we do not always know what the net $\cM$ is.
Recall that for $\SUz$ the \ddE 6 and \ddE 8 modular invariants
come from the conformal embedding $\SUz\subset G$ where
$G=\mathit{SO}(5)$ or $G_2$, respectively. So it is natural
to ask whether there is an extension of $\Vir$ for the
$(\cG_1,\cG_2)$ modular invariants where $\cG_1$ or $\cG_2$
is \ddE 6 or \ddE 8. For the (\ddE 6, \ddA {12})- and
(\ddE 8, \ddA {30})-invariants the natural candidate is
the coset
\[ \frac{G_1 \otimes \SUz_1}{\SUz_{m-1}} \]
where $G_1=\mathit{SO}(5)_1$ and $m=12$, or
$G_1=(G_2)_1$, $m=30$, respectively. However,
for the (\ddA {10},\ddE 6) and (\ddA {28}, \ddE 8) modular
invariants there is no such natural candidate.
For any block diagonal modular invariant of the minimal
models we proceed by {\em assuming} that there is a net
$\cM$ such that the net of subfactors $\cN\subset\cM$
has the correct properties, in particular, that the
blocks correspond to $\sigma$-restriction of representations
of the net $\cM$. Then we may go on as follows:
Let $\Gamma_k(\cG)$ denote the set of integers $j$ with
$[\lambda_j]$ appears in the $[\canr]$ we associated to the
$\SUz$ $\cG$ modular invariant, see Table \ref{GkG}.
\begin{table}
\begin{center}
  \begin{tabular}{|c|c|c|} \hline &&\\[-.9em]
Level & Graph $\cG$ &
$\Gamma_k(\cG)$
  \\[-.9em]&&\\ \hline\hline &&\\[-.9em]
  $k=1,2,3,\ldots$ & \ddA {k+1} &
$\{ 0 \}$ \\ &&\\[-.9em]
  $k=4\varrho\,,\,\,\,\varrho=1,2,3,\ldots$ & \ddD {2\varrho+2} &
$\{0,4\varrho\}$
  \\ &&\\[-.9em]
  $k=10$ & \ddE 6  &
$\{0,6\}$  \\ &&\\[-.9em]
  $k=28$ & \ddE 8  &
$\{0,10,18,28\}$  \\[-.8em] &&\\
\hline \multicolumn3c {} \\[.05em] \end{tabular}
\end{center}
\caption{The sets $\Gamma_k(\cG)$} \label{GkG}
\end{table}
For the $(\cG_1,\cG_2)$ modular invariant of the minimal model
with $c=c(m)$ define $[\canr]\in\LTSN$ by
\[ [\canr] = \bigoplus_{p\in \Gamma_{m-2}(\cG_1)} \,\,\,
\bigoplus_{q\in \Gamma_{m-1}(\cG_2)} \,\, [\lav pq] \,, \]
so that $[\canr]$ precisely correspond to the vacuum block
in $Z_{\cG_1,\cG_2}$.
(Note that one of the summations is always trivial as either
$\cG_1$ or $\cG_2$ is an A-graph.)
Then we determine the induced fusion algebra $V$ by
$\la \alpha_{p,q},\alpha_{p'q'}\ra_\MIo=
\la \canr\circ\lambda_{p,q},\lambda_{p'q'}\ra_\NIo$
where we denote $\alpha_{p,q}\equiv\alpha_{\lambda_{p,q}}$.

We have to choose an analogue of the fundamental representation of
$\LSUn$ for the minimal models. It is instructive to
discuss briefly the fusion graphs of $[\ala]$ for the choices
$\lambda=\lav 01, \lav 10, \lav 11$. First one checks that then
$[\ala]$ is irreducible, $\la\ala,\ala\ra_\MIo =1$,
in all cases. Now $\asv 01$ ($\asv 10$) generates the fusion
subalgebra corresponding to the first column (row) of the
conformal grid, being isomorphic to $W(2,m-1)$
($W(2,m-2$). Thus the fusion graph of $\asv 01$
($\asv 10$) is not connected; the identity component is just
$\cG_2$ ($\cG_1$). Now consider $\lasv 11$ that is
$\lasv 11 = \lasv 01 \times \lasv 10$. The fusion graph
of $\asv 11$ is somehow a combination of the graphs
$\cG_1$ and $\cG_2$. As an illustration, we give the result
for the (\ddA 4,\ddD 4) modular invariant ($m=5$, $c=4/5$)
\[ Z_{(\mathrm{A}_4,\mathrm{D}_4)} = \frac12
\left( \sum_{p=0}^3 |\chi_{p,0} + \chi_{p,4}|^2
+ 2\, |\chi_{p,2}|^2 \right) \,.\]
Then $[\canr]=[\lav 00] \oplus [\lav 04]$, and we find eight
distinct irreducible sectors, $\asv 00$, $\asv 11$, $\asv 20$,
$\asv 31$, $\asv 02 _\pm$, $\asv 22 _\pm$. The fusion graph of
$\asv 11$ is given in Fig.\ \ref{wok}.
\thinlines
\setlength{\unitlength}{3pt}
\begin{figure}[tb]
\begin{center}
\begin{picture}(40,50)
\put(10,10){\line(1,1){20}}
\put(10,30){\line(1,1){10}}
\put(10,30){\line(1,-1){20}}
\put(20,40){\line(1,-1){10}}
\put(20,10){\line(0,1){30}}
\multiput(10,10)(0,20){2}{\circle*{1}}
\multiput(20,10)(0,10){4}{\circle*{1}}
\multiput(30,10)(0,20){2}{\circle*{1}}
\put(-2,7){$\asv 02 _+$}
\put(16,5){$\asv 00$}
\put(32,7){$\asv 02 _-$}
\put(23,19){$\asv 11$}
\put(-2,29){$\asv 22 _+$}
\put(21,29){$\asv 20$}
\put(32,29){$\asv 22 _-$}
\put(16,42){$\asv 31$}
\end{picture}
\end{center}
\caption{Fusion graph of $\asv 11$ for the
(\ddA 4,\ddD 4) modular invariant.}
\label{wok}
\end{figure}
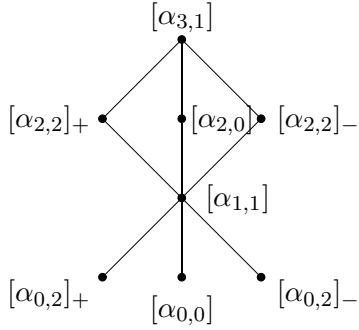

\section{Outlook}

We have applied the procedure of $\alpha$-induction and
$\sigma$-restriction of sectors to chiral conformal field
theory models, in particular to the $\SUn_k$ WZW models.
Looking at the block-diagonal modular invariants arising
from conformal or orbifold inclusions of $\SUn$, we have seen
that their classification by certain fusion graphs --- in
particular the A-D-E classification in the $\SUz$ case ---
can be understood by the induction-restriction machinery
of the relevant sectors. However, many questions remain
unanswered. The induction turns out to be non-surjective
in several cases; this is apparently closely related to
multiplicities in the mass matrix $Z$, but a good understanding
of this non-surjectivity (which can even lead to
non-commutativity of the induced sector algebra)
is still missing. It might be possible to extract more
information about the structure of the induced fusion algebra
from the $\SUn_k$ data than our results in \cite{boev1}
like the main reducibilty formula or $\alpha\sigma$-reciprocity
provide. In fact, the observation $\tr Z = D\equiv |\cV|$
is still awaiting a good explanation.

It will certainly be worth looking also at the
block-diagonal $\SUn$ modular invariants that come neither
from conformal nor from orbifold embeddings. Moreover,
it is not clear at the moment how to incorporate the type II
modular invariants in our framework. Another challanging
question, suggested by the treatment of $\SUz$, concerns
a better understanding of
the relation between the appearance of modular invariants
of $\SUn$ WZW models and the existence of
sub-(equivalent)-paragroups of the paragroups arising
from the relevant $\cA$-type subfactors (\cite{kaw4, ocn4}).
Of course it will also be interesting to construct the
associated fusion graphs also for modular invariants
of other Lie groups, e.g. $\mathit{Sp}(n)$.

\subsection*{Acknowledgement}

We would like to thank F.\ Goodman for providing us a
Pascal program for the computation of the $\SUn_k$ fusion
coefficients as well as to thank J.\ Fuchs for providing
us a printout of certain fusion matrices. We are also grateful
to T.\ Gannon for a helpful correspondence by e-mail, and it is
a pleasure to thank K.-H.\ Rehren for many useful comments
on an earlier version of the manuscript.

This project is supported by the EU TMR Network in
Non-Commutative Geometry.



\newcommand\biba[7]   {\bibitem{#1} {\sc #2:} {\sl #3.} {\rm #4} {\bf #5}
                      { (#6) } {#7}} 
\newcommand\bibb[4]   {\bibitem{#1} {\sc #2:} {\it #3.} {\rm #4}}
\newcommand\bibp[4]   {\bibitem{#1} {\sc #2:} {\sl #3.} {\rm Preprint #4}}
\newcommand\bibx[4]   {\bibitem{#1} {\sc #2:} {\sl #3} {\rm #4}}
\def\AAM              {Acta Appl.\ Math.}
\def\CMP              {Com\-mun.\ Math.\ Phys.}
\def\IJM              {Intern.\ J. Math.}
\def\JFA              {J.\ Funct.\ Anal.}
\def\JMP              {J.\ Math.\ Phys.}
\def\LMP              {Lett.\ Math.\ Phys.}
\def\RMP              {Rev.\ Math.\ Phys.}
\def\Inv              {Invent.\ Math.}
\def\npbp             {Nucl.\ Phys.\ {\bf B} (Proc.\ Suppl.)}
\def\nupb             {Nucl.\ Phys.\ {\bf B}}
\def\adma             {Adv.\ Math.}
\def\coma             {Con\-temp.\ Math.}
\def\physa            {Physica {\bf A}}
\def\ijmp             {Int.\ J.\ Mod.\ Phys.\ {\bf A}}
\def\jpa              {J.\ Phys.\ {\bf A}}
\def\FdP              {Fortschr.\ Phys.}
\def\PLB              {Phys.\ Lett.\ {\bf B}}
\def\RIMS             {Publ.\ RIMS, Kyoto Univ.}



\end{document}